%% file: microlev.tex
\documentclass[12pt,oneside,a4paper]{amsart}

\usepackage{amssymb}
\usepackage{epsfig}

\numberwithin{equation}{section}

\newcommand{\dd}{{\,\,\!\mathrm d}}
\newcommand{\ibe}{improved {Boussinesq} equation}
\newcommand{\fg}{\boldsymbol}
\newcommand{\bal}{\begin{align}}
\newcommand{\eal}{\end{align}}

\newcommand{\rand}[1] {} 
\newcommand{\labell}[1]{\label{#1}\rand{#1}}
\newcommand{\citel}[1]{\cite{#1}\rand{#1}}
\newcommand{\beq}{\begin{equation}}
\newcommand{\beql}[1]{\rand{#1}\begin{equation}\label{#1}}
\newcommand{\eeq}{\end{equation}}

\newcommand{\beqa}{\begin{eqnarray}}
\newcommand{\eeqa}{\end{eqnarray}}

\newcommand{\beqas}{\begin{eqnarray*}}
\newcommand{\eeqas}{\end{eqnarray*}}

\newcommand{\ba}{\begin{array}}
\newcommand{\ea}{\end{array}}

\newcommand{\bt}{\begin{tabular}}
\newcommand{\et}{\end{tabular}}
\newcommand{\bd}{\begin{description}}
\newcommand{\ed}{\end{description}}
\newcommand{\be}{\begin{enumerate}}
\newcommand{\ee}{\end{enumerate}}
\newcommand{\bc}{\begin{center}}
\newcommand{\ec}{\end{center}}

\newcommand{\wt}{\widetilde}

\newcommand{\fsg}{\fg \sigma}
\newcommand{\fsgm}{\fg \sigma_m}
\newcommand{\fvep}{\fg \veps}

\newcommand{\R}{\mathbb{R}} 
\newcommand{\N}{\mathbb{N}} 

\newcommand{\Z}{\mathbb{Z}}

\newcommand{\pl}{\partial}

\newcommand{\fr}{\frac}

\newcommand{\veps}{\varepsilon}
\newcommand{\reff}[1]{{\rm (\ref{#1})}}

\newcommand{\imp}{\Rightarrow}
\newcommand{\aeq}{\Leftrightarrow}
\newcommand{\wto}{\rightharpoonup}

\newcommand{\id}{I\!d}
\newcommand{\aint}{-\negthickspace\negthickspace\negthickspace
\negthickspace\int}
\newcommand{\saint}{-\negthickspace\negthickspace\negthickspace\int}

\newcommand{\lan}{\left\langle}
\newcommand{\ran}{\right\rangle}

\renewcommand{\d}{\mathrm d}
\newcommand{\nke}{\frac{1}{\rho}K_{\mathrm{mac}}(t)}

\DeclareMathOperator{\lip}{lip}

\DeclareMathOperator{\vol}{vol}

\DeclareMathOperator{\diam}{diam}

\DeclareSymbolFontAlphabet{\mathbb}{AMSb}

\theoremstyle{plain}
\newtheorem{theorem}{Theorem}[section]
\newtheorem{lemma}[theorem]{Lemma}
\newtheorem{proposition}[theorem]{Proposition}

\theoremstyle{definition}

\theoremstyle{remark}

\newcommand{\B}{{\mathcal B}}

\newcommand{\Y}{\mathrm{Y}}
\newcommand{\vepsp}{p}
\renewcommand{\Diamond}{\mbox{\circle{5}}}

\begin{document}
\thispagestyle{empty}
\vfill

\begin{center}
{\bf \sc \Large A Study of a Hamiltonian Model for\\
Phase Transformations Including Microkinetic Energy}\\[1cm]
Florian Theil\\
{\footnotesize Institute for Applied Mathematics,
University of Hannover\\
Welfengarten~1, 30167~Hannover, Germany\\
Email: theil@ifam.uni-hannover.de\\[0.3cm]}
and\\[0.3cm]
Valery I. Levitas\\
{\footnotesize Institute for Structural and Computational
Mechanics, University of Hannover\\
Appelstra\ss{}e~9A, 30167 Hannover, Germany\\
Email: levitas@ibnm.uni-hannover.de\\}
\vspace{0.5cm}
\today \\[1cm]
\end{center}

\begin{quote}\small
{\bf Abstract.} \footnotesize
How can a system in a macroscopically stable state explore
energetically more favorable states, which are far away from the
current equilibrium state?
Based on continuum mechanical considerations we derive a {\sc
  Boussinesq}-type equation
\[\rho \ddot u = \pl_x \sigma(\pl_x u) + \beta \pl_x^2 \ddot u, \quad
x \in (0,1), \; \beta>0,\]
which models the dynamics of martensitic phase transformations. Here
$\rho>0$ is the mass density, $\beta \pl_x^2 \ddot u$ is a
regularization term which models the inertial forces of oscillations
within a representative volume of length $\sqrt{\beta}$ and $\sigma$
is a nonmonotone stress-strain relation.
The solutions of the system, which we refer to as 
{\it microkinetically} regularized
wave equation, exhibit strong oscillations after times of order
$\sqrt{\beta}$, thermalization can be confirmed. That means that
macroscopic fluctuations of the solutions decay at the benefit of
microscopic fluctuations.

The mathematical analysis for the microkinetically regularized wave
equation consists in two parts. First we present
some analytical and numerical results on the propagation of phase
boundaries and thermalization effects. 
Despite the fact that model is conservative,
it exhibits the hysteretic behavior. Such a behavior is usually
interpreted in macroscopic models in terms of dissipative threshold 
which the driving force has to overcome to ensure that the phase
transformation proceeds. The threshold value
depends on the amount of the transformed phase as it is observed in 
known experiments.

Secondly we investigate the dynamics of oscillatory
solutions. Our mathematical tool are Young measures,
which describe the one-point statistics of the fluctuations.
We present a formalism which allows us to describe the effective
dynamics of rapidly fluctuating solutions. The extended system
has nontrivial equilibria which are only visible when 
oscillatory solutions are considered. The new method enables us to
derive a numerical scheme for oscillatory solutions based on particle
methods.
\end{quote}
\vfill

\mbox{ }
\newpage

\section{Introduction}
In this work we study the onset and propagation of oscillations in a
model for martensitic phase transformations.
Consider the simple tension of a rod  of nonlinear elastic material  with
a diagram $\sigma = \sigma(\veps)$, shown in Figure 1, where $\sigma$
and $\veps$ are tensile stress and strain, respectively.
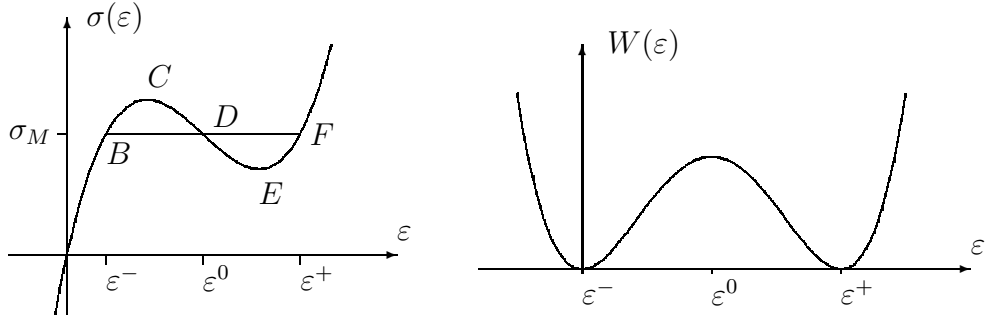
\begin{figure}
\input{sigma.tex}
\input{w.tex}
\caption{Nonmonotone stress strain relation and nonconvex elastic energy}
\labell{sigmaandw.fig}
\end{figure}
The diagram shown in Figure~\ref{sigmaandw.fig} corresponds to a nonconvex
free energy $W(\veps)$
and is typical for materials with martensitic phase
transformations (PT).  We will label the phases corresponding to the
branches $BC$ and $EF$ (-) and (+)
respectively; branch $CDE$ exhibits an unstable intermediate state.
The principle of minimum of free energy for such materials results in
formation of a phase mixture and in a macroscopic deformation along
the line $BDF$ at a constant stress $\sigma_M$ (Maxwell line),
rather than along the lines $BCDEF$.
Considering a quasi-static experiment
at some stress $\sigma_0$, a macroscopic portion of the
material is deformed from $\veps^-$ to $\veps^+$. At the next total
strain increment $\Delta \veps$, the next portion of material
undergoes PT. During the whole process of PT, the strains in the
phases (-) and (+) are fixed and equal to $\veps^-$ and $\veps^+$,
the stress is constant $\sigma = \sigma_0$ and $\Delta \veps$ is
related to the increment $\Delta c$ of the volume fraction of a new
phase.

It is clear, that at quasi-static treatments we do not follow
the original stress-strain curve $\veps\to\sigma(\veps)$, but we
rather introduce
a strain jump from $\veps^-$ to $\veps^+$ at constant stress for each
transforming particle.
To be consistent with the original stress-strain relation, the concept of
fluctuating stresses was introduced in \citel{1,2}.
Fluctuating stress $\sigma_{\mathrm{fluct}}$ is defined
as $\sigma_{\mathrm{fluct}} = \sigma(\varepsilon) - \sigma_M$ in the
interval $[\veps^-, \veps^+]$, it varies
in accordance with the line $BCDEF$, relative to the line $BDF$.
This fluctuating stress overcomes the energy barrier $BCD$, but
due to the very small time interval of its action, it is not taken into
account in the continuum relations. In the equilibrium process, the time
averaged value of $\sigma_{\mathrm{fluct}}$ is equal to zero.

This concept is useful for the development of some kinetic models \citel{1, 2}
which only use some important properties of the  stress-strain
curve $\sigma = \sigma(\veps)$ but not the precise course of the
unstable part.\par
One of the main unsolved problems is to find a proper physical model
for fluctuations and to include it in the continuum
description of PT. A
possible way is to take into account microinertia in constitutive
equations, i.e. to consider a constitutive equation
\begin{equation}
 \sigma_m = \sigma (\veps) + \beta \ddot{\veps},
\label{eq:0}
\end{equation}
where $\sigma_m$ is the macroscopic stress.
%
The constitutive assumption \reff{eq:0} and the conservation of
linear momentum lead us to the following 
evolution equation:
\begin{equation}
\labell{mw.eq} \rho \ddot u= \pl_x(\sigma(\pl_x u) + \beta \pl_x \ddot
u), \quad x\in (0,1),\; \rho \geq 0,\;\beta > 0,
\end{equation}
where $u:\R^+\times\Omega\to \R$ is the deformation of an elastic body
$\Omega =(0,1) \subset \R$ at time $t$ and $\rho$ is the mass density. The
coefficient $\beta$ describes the inertia of microscopic
fluctuations, for a derivation see Chapter~\ref{conseq.sec}.
For reasons which will become clear there,
we will refer to system \reff{mw.eq} as {\it microkinetically}
regularized wave equation. If appropriate boundary conditions are
imposed, the generalized energy $\int_0^1(\frac{\rho}{2} \dot u^2 +
W(\pl_x u) + \frac{\beta}{2}(\pl_x \dot u)^2)\,\dd x$ is conserved. 
The aim of this paper is to present the
first mathematical results for such a model.\par
Equation \reff{mw.eq} constitutes a well posed Cauchy problem.
In the following section
we present a continuum mechanical derivation of the microkinetically
regularized wave equation based on an
approach developed in \citel{10, 09, 060} to describe the dynamics of phase
transformations. Since in the derivation finite
wavelengths are assumed, it is not possible to show that \reff{mw.eq}
approximates the phase transformation problem if the strain oscillates
at a wavelength of order $\sqrt{\beta}$.
However the obviously very interesting
behavior of the solutions motivates us to analyze the dynamics of
oscillations with modern mathematical tools like Young measures.
Many of our results are also valid for the \ibe
\[ \ddot u = u_{xx}+(u^p)_{xx} + \ddot u_{xx},\]
see Chapter \ref{conseq.sec} for references.

In Chapter \ref{eqmo.sec}
important phenomena of the dynamics are investigated
numerically and analytically. We show that the system possesses
solitary-wave solutions. In the numerical simulations quasi-dissipative
behavior can be observed.
Due to the nonmonotonicity of $\sigma$ the solutions exhibit high
frequency oscillations in space after times, which are proportional to
$\sqrt{\beta}$ even if the initial state is smooth. These
spatial fluctuations can be interpreted as heat. The introductory
problem motivates the development of mathematical tools which simplify
the study of oscillatory solutions. We expect that the results will be
useful for future work.
We find a critical threshold value for the
minimal driving force of the transformation process. The threshold
value depends on the amount of transformed energy.

There exists a rich literature concerning regularizations of
quasilinear wave-equations with nonmonotone stress-strain relation,
see e.g. \citel{sl89} for a starting point. The problem lies in the
ill-posedness of the Cauchy problem for the quasilinear
wave-equation
$\rho \ddot u = \pl_x \sigma(\pl_x u)$. The hope is that the
regularization selects the physically correct solution if the strength
of the regularization tends to 0. Another approach consists
in the assumption of a kinetic relation to determine the velocity of a
moving phase boundary. Using this additional constitutive law one can
show that the Riemann problem has a unique solution, see \citel{ak91a,
  tr94}. Indeed the solutions of the Riemann problem with viscosity
and Van-der-Waals forces generate the kinetic relation,
\citel{ak91b}.\par

We are not so much interested in the singular limit $\beta\to 0$ but
more in the qualitative behavior of the solutions if $\beta$ is
nonzero.
The system distinguishes
in a natural way between two different scales (microscale and macroscale).
This phenomenon allows us to characterize the dynamics of microstructure
in a unique fashion. To this end we introduce the notion of Young
measures which can be viewed as generalized limits of sequences of highly
oscillatory functions. One major disadvantage of Young measures is
that they do not work well together with pseudo differential operators
of degree 0 like the Helmholtz projection which maps a vector
field onto the gradient part. But this is
essential if space dimensions greater than one are considered. Therefore
the analysis is restricted to the one dimensional case.

Young measures have been used successfully for phase transformation
problems. In \citel{bj87} they are applied to the three dimensional
crystallographic theory.
Using methods from calculus of variations
{\sc Kinderlehrer} and {\sc Pedregal} showed in \citel{kp92}
that dynamical models for phase transformations admit Young measure
solutions.
The approach outlined in this work has been applied
before to the viscoelastic system with nonmonotone stress-strain
relation $\rho \ddot u = \pl_x(\sigma(\pl_x u) + \beta \pl_x \dot u)$,
see \citel{th98a}.\par

The mathematical analysis for oscillatory solutions of the microkinetic model
consists of two parts: short-time behavior and long-time behavior.

In Chapter \ref{propos.sec}
we derive an effective equation for Young-measure solutions.
These generalized solutions are unique and they approximate the
dynamics of highly oscillatory solutions over finite times.
This approach can be viewed
as an analogy to homogenization theory. There the aim is the derivation
of a macroscopic law if the the material properties are
oscillatory. In our case it turns out that the dynamics of oscillatory
solutions can be reduced to a transport equation which has structural
similarity to the Vlasov-Poisson system. The Vlasov-Poisson system
describes the evolution of electrically charged plasmas (see
e.g. \citel{bmr95}).

The microkinetic system exhibits some kind of ergodic
behavior. Even if we start with a smooth initial condition the
solutions soon become oscillatory, cf. Figure~\ref{tw1.fig}. Especially
in the region $x \in [0.1, 0.2]$ the
solution looks very much like a Young measure.\par
In Chapter 5 we demonstrate that
the extended system has a large number or equilibria
which are not generalized limits of sequences of stationary solutions
of \reff{mw.eq}.
The physical interpretation of this result is as follows. Despite the
very dynamical oscillatory behavior  of each material point,
macroscopic statistical characteristics, like volume fraction of new
phase are stationary for equilibrium solutions of the extended
system. Thus Young measure are an important concept for the finding
of macroscopic
information using nonstationary oscillatory solutions.
This observation has been made earlier for the
equations of ideal incompressible magneto hydrodynamics, see
\citel{jo94, jt96}. There
the authors analyze so called relaxed equilibrium states
which correspond to highly oscillatory nonstationary solutions of the
evolution equations.\par
If we consider the nonphysical limit $\rho\to 0$
it can be shown, that almost every
solution tends to a stationary Young measure as time goes to infinity.
This is a genuinely infinite dimensional result since the system is
Hamiltonian. In finite dimensional Hamiltonian systems the
phase-space volume is conserved, thus generic solutions do not converge
to an equilibrium.\par
In \citel{th98a} the long-time behavior of oscillatory solutions has
been investigated for the viscoelastically damped system. There it is
found that the microstructure region shrinks when $t$ tends to
infinity. This contrasts to the microkinetic system where microstructure
can be generated in the long-time limit.\par
The convergence towards an equilibrium Young measure can be understood as
{\it thermalization}, i.e. the fluctuations which are
at the beginning only visible at the macroscopic level
are transformed into microscopic fluctuations. Hence on the
macroscopic level the solution becomes stationary.
This behavior does not
correspond to the numerical results of {\sc E.~Fermi, J.~Pasta} and
{\sc S.~Ulam} in
\citel{fpu55} where a discretized nonlinear wave equation
\[ \rho \ddot \veps_i = \sigma(\veps_{i+1}) - \sigma(\veps_{i-1})\]
is investigated. They observed a recurrence phenomenon, i.e. after fixed
time intervals the solution returns very close to the initial
deformation.\par
The study of the microkinetic system is concluded by a systematic
numerical investigation of the thermalization behavior in the case
$\rho>0$.
In Chapter~\ref{numres.sec} a numerical scheme based on
particle methods is developed and simulation results concerning the long-time
behavior of solutions for values of $\rho$ greater than 0 are presented.
It turns out that the spatial coupling due to the nonvanishing of $\rho$
complexifies the long-time behavior considerably. The solutions
develop microstructure, but there are no indications that
the solutions tend toward a stationary Young measure if $\rho>0$.
However, a major portion of the initial macroscopic kinetic energy
is transformed into microscopic fluctuations.
%
\section{Constitutive equation for material with microscopic inertia}
\labell{conseq.sec}
For a formal derivation of the constitutive equation we consider the
$d$-dimensional setting, $d=1,2,3$, although the subsequent analysis
is restricted to the one dimensional case.
For simplicity geometrically linear elasticity theory is used, we insist
however that the derivation is completely analogous
if geometrically nonlinear
theory is used. In this case we can use the deformation
gradient and the first nonsymmetric Piola-Kirchhoff stress tensor as
conjugate variables.
For the convenience of readers which are
not familiar with the notations in continuum mechanics
we repeat the basic definitions. Let $\Omega\subset\R^d$ be the reference
configuration of an elastically deformable body.
The superscript $t$ denotes transposition
and $\nabla$ is the gradient operator. We use bold letters to
indicate tensors and vectors,
the coefficients are written with normal letters.
The dot ``$\,\bf{\cdot}\,$'' and ``$\,\bf{:}\,$''
denotes the contraction of tensors over one and two indices:
$({\fg A \cdot \fg B})_{ik} = \sum_{j} A_{ij}B_{jk} = A_{ij}B_{jk}$ and
$\fg A : \fg B = \mathrm{tr}({\fg A}\cdot {\fg B}^t)= A_{ij}B_{ij}$,
where Einstein's sum convention is adopted for the last identities.
By ${\fg A} \otimes {\fg B}$ we denote the fourth
order tensor with components $(A_{ij} B_{kl})$.\par
Let us consider the representative volume $\omega\subset \Omega$
of the microheterogeneous
material bounded by a surface $\pl \omega$  with the unit normal vector
$\, {\fg n} \,$. We assume that $\omega$ is much smaller than
$\Omega$ and that the center of mass is located at the origin.
Let $\, {\fg r} \,$ be the particle position, $\fg u$
the continuous displacement field, i.e. for
every material point ${\boldsymbol r}$ in the reference configuration the
position at time $t$ is given by $\wt{\fg u}(t,{\boldsymbol
  r})$. By
$\; \widetilde{\fsg} \;$ we denote the stress tensor and $\;
\widetilde{\fvep} = \nabla \wt{\fg u}$ is
the deformation gradient tensor, where {\Large $ \, _{\widetilde{}}
  \,$ } means the local value of parameters as opposed to the macroscopic
counterpart. Let us  introduce in the usual way  the  macroscopic
stress $\, \fsgm \,$ and deformation gradient $\fvep_m$ tensors
\begin{equation}
\fsgm^t \, = \vol(\omega)^{-1} \int_{\pl \omega} {\fg r} \otimes
\widetilde{\fsg} \, {\fg \cdot} \,
{\fg n} \, \d S \; ; \qquad\qquad\qquad
\fvep_m = \vol(\omega)^{-1} \int_{\pl \omega} \wt{\fg u}\otimes {\fg n}
\, \d S.
\label{eq:1}
\end{equation}
The macroscopic stress tensor is the critical quantity which allows to
analyze the macroscopic behavior.
Definition \reff{eq:1} can be motivated as follows. Let
$E_\omega$ be the total energy contained in the representative
volume, then
the power of the external forces can be computed with the formula
\begin{equation} \label{work.eq}
\frac{\dd}{\dd t} E_\omega = \int_{\pl \omega} (\widetilde{\boldsymbol
\sigma}^t \cdot \dot{\wt{\fg u}}) \cdot \fg n \, \dd S.
\end{equation}
Assuming that the velocity distribution corresponds to a
macroscopically homogeneous deformation on
the boundary, we find that $\dot{ \wt{\fg u}}(\fg r) =
\dot{\fg u}_0 + \dot{ \fg \veps}_m \cdot \fg r$ holds for all
$\fg r\in \pl \omega$.
Plugging the Ansatz in \reff{work.eq} one finds that
\begin{align*}
&\frac{\dd }{\dd t} E_\omega = \int_{\pl \omega} (\wt{\fg\sigma}^t \cdot
(\dot{\fg\veps}_m \cdot \fg r))\cdot \fg n\, \dd S
+ \int_{\pl \omega} (\wt{\fg \sigma}\cdot \fg n) \cdot \dot{\fg u}_0\, \dd S\\
= &\dot {\fg \veps}^t_m : \int_{\pl \omega} \fg r \otimes {\wt{ \fg
\sigma}}\cdot \fg n\, \dd S + \dot{\fg u}_0\cdot \int_{\pl \omega}(\wt{\fg
\sigma} \cdot \fg n) \, \dd S\\
=& \vol(\omega)\cdot\dot{\fg{\veps}}_m:\fg \sigma_m+ \dot{\fg u}_0\cdot
\fg F_m,
\end{align*}
where $\fg F_m=\int_{\pl \omega} (\wt{\fg \sigma}\cdot \fg n) \, \dd S$ is the
macroscopic force.
Thus in absence of macroscopic translations
the power of the external stresses, through which the
representative volume interacts with the neighborhood, is equal to
$\fg{\sigma}_m: \dot{\fg{\veps}}_m$, as for material points.
The symmetry of the stress tensor in the geometrically linear theory
is corresponds to the symmetry of the Cauchy stress
tensor in the geometrically nonlinear theory. With respect to both theories
the macroscopic stress tensor is in general not symmetric, it is
possible to transfer energy by rotations.
The macroscopic forces will be neglected in the further
considerations.
Using the  Gauss theorem and the equations of motion without body
forces
\newcommand{\tirho}{\widetilde \rho}
\begin{eqnarray}
\nabla \, {\fg \cdot} \, \widetilde{\fsg} \, = \, \tirho \ddot{\fg u} \; ,
\label{eq:2}
\end{eqnarray}
where $\tirho$ is the mass density, we obtain from eqs.~(\ref{eq:1})
and (\ref{eq:2})
\begin{equation}
\fsgm \, = \, \langle \widetilde{\fsg} \rangle \, + \,
\langle \tirho \,  \ddot{\fg u}  \otimes {\fg r} \rangle \; ;
\qquad\qquad\qquad
\fvep_m \, = \, \langle \widetilde{\fvep} \rangle \; ,
\label{eq:3}
\end{equation}
where $\; \langle ... \rangle \, = \, \vol(\omega)^{-1} \int_{\omega}
(...) \; \d x\,$
is the averaging over the volume $\omega$.
When the macroscopic stress tensor is approximated by the averaged
stress tensor the resulting errors can be expressed as accelerations
or inertial stresses. The averaged value $\langle \wt{\fg \sigma} \rangle$
neglects rapid oscillations in the interior of the representative
volume. The kinetic energy of these
oscillations is responsible for the inertial stresses.
It is crucial that the inertial stresses have a
regularizing effect. They enlarge the inertia of the material in
such a way that a well posed Cauchy problem can be derived even if for the
averaged stress tensor a nonmonotone dependency of the macroscopic
deformation gradient $\fg{\veps}_m$ is assumed.\par
If the diameter of $\omega$
tends to 0 then in (\ref{eq:3}) the second term vanishes and
\begin{equation}
\fsgm \, = \, \langle \widetilde{\fsg} \rangle \; ,
\label{eq:4}
\end{equation}
i.e. the macroscopic stress $\, \fsgm \,$ is equal to microscopic stresses
averaged over the infinitesimal volume $\, \fsg_v \, := \, \langle
\widetilde{\fsg} \rangle \,$. The same holds if the body is in an
equilibrium state. The next step consists in prescribing the
constitutive equations for $\, \widetilde{\fsg} \,$ (in the micromechanical
approach) or directly for $\, {\fsg}_v \,$ at phenomenological
consideration. In order to describe martensitic PT in elastic materials
a nonconvex elastic potential is used, which results in nonmonotone
stress-strain relation  $\; \fsg_v \, = \, {\fg f} \left(\fvep_m \right) \,$.
Due to the ill posedness of the dynamic boundary-value problem for such type
of materials, a number of mathematical and physical regularization methods
have been suggested. One of them consists in the introduction of
viscosity, i.e.
in considering a viscoelastic material. In the given paper we would
like to limit ourselves to nondissipative materials. Another approach is
related to introduction of characteristic material size, e.g. by
introduction of higher strain gradients $\, \nabla \, \fvep_m \,$,
$\;\, \nabla \, \nabla \, \fvep_m \,$, \ldots or
nonlocal stress-strain relations. As it is shown in \citel{1}, in a framework
of micromechanical consideration, the macroscopic deformation gradient
$\, {\fg B} \,$ and the work-conjugated third-order tensors of the
hyperstress $\, {\fg M} \,$ can be defined by the formulae
\begin{equation}
{\fg B} \, = \, \vol(\omega)^{-1} \int_{\pl \omega} \widetilde{\fvep}
\otimes {\fg n} \, \d S \; ;
\qquad
{\fg M} \, = \, \frac{\vol(\omega)^{-1}}{2} \int_{\pl \omega} {\fg r}
\otimes {\fg r} \,
\otimes\widetilde{\fsg} {\fg \cdot} \, {\fg n} \, \d S \; .
\label{eq:5}
\end{equation}
With the help of the Gauss theorem and the equations of motion for
static case we obtain
\begin{equation}
{\fg B} \, = \, \langle \nabla \, \nabla \, {\fg u} \rangle \, = \,
\langle \nabla \, \widetilde{\fvep} \rangle \; ;
\qquad\qquad\qquad
{\fg M} = \langle {\fg r} \otimes \widetilde{\fsg}
\rangle.
\label{eq:6}
\end{equation}
The tensor $\, {\fg M} \,$ is nonzero only if the diameter of the
representative volume $\omega$ is finite,
this follows from equation~(\ref{eq:6})$_2$ by considering the limit
$\diam(\omega) \to 0$. At macroscopically homogeneous
strains $\; {\fg u} \, = \, {\fg u}_0 \, + \, \fvep_m \, {\fg \cdot} \,
{\fg r} \;$ on  $\pl \omega$ we have $\, \widetilde{\fvep} \, = \,
\fvep_m \,$ on the surface $\pl \omega$, and according to equation~(\ref{eq:5})
$\, {\fg B} \, = \, {\fg 0} \,$. If the macroscopic stresses on $\pl
\omega$ are homogeneous
$\; \widetilde{\fsg} \, {\fg \cdot} \, {\fg n} \, = \, \fsgm \, {\fg
  \cdot} \,  {\fg n} \,$ holds and according to equation~(\ref{eq:6}) $\,
{\fg M} \, = \,  { \fg 0} \,$ (using $\, \langle {\fg r} \rangle = 0)$.
Consequently, the tensors $\, {\fg B} \,$ and $\, {\fg M} \,$ characterize
the heterogeneous distribution of strains and stresses inside the
representative volume. Both vanish for macroscopically homogeneous
conditions even if the representative volume is finite.
\par
At the same time for finite representative volume, the inertial stresses
$\, \fsg_i \, := \, \langle \tirho \, {\fg r} \otimes \ddot{\fg u} \rangle \,$
are nonzero for macroscopically homogeneous stress state, but high
accelerations $\, \ddot{\fg u} \,$. As PT represent a highly dynamic
process, the contribution of inertial stresses can be in fact very important
for finite representative volume.\par
Our next aim is to approximate the compact integral operator $\langle
\rho \fg r \otimes \ddot{\fg u} \rangle$ by an unbounded differential
operator. Expanding $\ddot{\wt{\fg u}}$ around $\boldsymbol 0$ to the first
order we have $\ddot{\wt{\fg u}}(\boldsymbol r) = \ddot{\wt{\fg{u}}}(\fg 0) +
\fg{\ddot{\widetilde \veps}} (\fg 0)\cdot \boldsymbol r + \mathcal
O(\diam(\omega)^2\|\nabla^2 \ddot{\wt{\fg u}}\|_{L^\infty(\Omega)})$. Thus
\begin{align*}
& \langle \rho \boldsymbol r \otimes \ddot{\wt{\fg u}}(t,\cdot) \rangle_{ij}
= \tfrac{\rho}{\vol(\omega)} \biggl\{ \underbrace{\int_{\omega}
   x_i\ddot u_j(\fg 0) \, \d x}_{=0}+ \sum_{k=1}^{d}
  \int_{\omega}\ddot{\widetilde\veps}_{jk}(t,\fg 0)
    x_i x_k\, \d x
\biggr\} + \mathcal O(\diam(\omega)^3\|\nabla^2 \ddot{\wt{\fg
    u}}\|_{L^\infty(\Omega)})\\
= & \rho \sum_{k=1}^d\ddot{\widetilde \veps}_{jk}(t,\fg 0)
\tfrac{1}{\vol(\omega)}\int_\omega x_ix_k \, \d x
+ \mathcal O(\diam(\omega)^3\|\nabla^2 \ddot{\wt{\fg u}}\|_{L^\infty(\Omega)}).
\end{align*}
Due to the linearity of the integral operator the estimates are stable
under weak convergence, i.e. we may replace $\wt{\boldsymbol \veps}$ by
$\fg\veps_m$. Dropping the error term we obtain
\begin{equation} \label{eq:7}
\fg \sigma_m = \fg \sigma_v + \langle \widetilde \rho \boldsymbol r \otimes
\boldsymbol r \rangle \boldsymbol \cdot \boldsymbol{\ddot{\fg
    \veps}}_m \end{equation}
It is clear that the approximation looses validity if
the oscillations of $\ddot{\fg\veps}_m$ occur on a length scale of order
$\diam(\omega)$.\par
The final step in our derivation consists in taking equation~\reff{eq:3}
as a definition for $\widetilde{\fg{\sigma}}_m$. This
introduces a correction term, which ensures that
the averaged stress contains enough information to predict the
macroscopic dynamics. This new stress is
conservative and related to the acceleration, therefore we refer to it
as microinertial stress. It has a similar
effect as a viscoelastic damping, namely the spatial regularization of
the velocity.\par
At one dimensional tension, when only tensile stress $\, \sigma \,$ and
strain $\, \varepsilon \,$ are important, equation~(\ref{eq:7}) results in
\begin{equation}
\sigma_m \, = \, {\sigma}_v \, + \, \beta \, \ddot{\varepsilon} \; ;
\qquad\qquad\qquad
\beta \, = \, \langle \rho \, r^2 \rangle \; ,
\labell{eq:8}
\end{equation}
where the coefficient $\, \beta \,$ characterizes the inertia of the
representative volume at the fixed center of mass, i.e. microscopic
inertia. Equation~(\ref{eq:8}) resembles to a Lagrange equation in mechanics of
discrete systems with the generalized coordinate $\, {\varepsilon} \,$
under the action of external $\sigma_m$ and conservative
$\sigma_v$ generalized forces. For this reason we will denote
the kinetic energy of the representative volume
$\, \frac{1}{2} \, \beta \, \dot{\varepsilon}^2 \,$ as
{\it microkinetic} energy.
If the representative volume is a cube with side length $2a$ and
the mass density is homogeneous, then $\beta = \frac{\rho}{3} \,
a^2$.\par
Equation \reff{mw.eq} has been derived in the case
$\sigma(\veps)=E\veps$ ($E$ is the Young modulus)
to describe longitudinal waves in a slender elastic rod
\cite[p. 428]{love}, the equation is known as {\sc Love's}
modified wave equation. The inertia of radial deformations are taken
into account; this
allows in contrast to the classical one-dimensional wave equation
to describe wave propagation problems in rods where the size of the
diameter is
comparable to the wavelength. In the case $\sigma(\veps) =
\veps + \veps^p$,
$p=2$ the equation is known as improved {\sc Boussinesq} equation,
it has been derived in \citel{bo76} to describe ion-sound waves. The
classical {Boussinesq} equation
\begin{equation} \labell{be.eq} \ddot u = \pl_x^2 (u +u^2 +\pl_x^2u)
\end{equation}
which has been derived in 1872 to describe shallow water waves
\citel{boussinesq} has the shortcoming that the Cauchy problem is ill
posed. Thus it cannot be used to analyze wave propagation problems
numerically. The dispersion relation of both {Boussinesq}
equations
\begin{align*}
\omega^2 &= k^2(1-k^2) \quad \text{(classical)},\\
\omega^2 &= k^2(1+k^2)^{-1} \quad \text{(improved)}.\\
\end{align*}
are equivalent to the fourth order for $|k| \ll 1$, this motivates the
nomenclature. We will focus on the short wave-length
limit $|k|\to \infty$. In this limit the improved and the classical
{Boussinesq} equation are not related.\par
For odd numbers $p$ the mapping $\veps \to \sigma(\veps)$
is monotone, for even numbers the
solutions can blowup in finite times, see, e.g. \citel{scl84}. We are
interested in the case that $\sigma$ is nonmonotone but the
stored elastic energy function $W$ ($W'=\sigma$) is coercive,
i.e. $W(\veps) \to \infty$ as $\veps \to \infty$.
To our best knowledge this case has not been studied in
the literature yet.\par
Note, that an inertial term in stress-strain relations was introduced
and made consistent with thermodynamics
in \citel{10,09}. Earlier {\sc Valanis} \citel{061} suggested inertial
contribution in
constitutive equations for internal variables, and a way to satisfy
the second law of thermodynamics in this case. A general theory,
which includes generalized inertial forces for each thermodynamic variable,
was developed in \citel{10,09}. It was shown, that in particular cases
thermodynamic inertial forces were well known for thermal conductivity,
moisture propagation in a colloid capillary body and turbulent transfer,
but the equations were not consistent with the second law of thermodynamics.
%
\section{First discussion of the equations of motion, solitary waves}
\labell{eqmo.sec}
In this section we state the precise evolution equations
and explain central properties. To motivate the consideration of
oscillatory solutions, simulation results of special solutions are
presented.\par
We set $\Omega = (0,1)\subset \R$.
Using the constitutive equation~\reff{eq:8} where the
subscript of $\sigma_m$  is dropped, we obtain the
following nonlinear partial differential equation which governs the
time evolution of $u$:
\begin{align}
\labell{mwex.eq}
&\rho \ddot u = \pl_x(\sigma(\pl_x u) + \beta \pl_x \ddot u), \qquad
\beta >0, \rho \geq 0,\\
\nonumber
& u|_{t=0} = u_0,\; \dot u|_{t=0} = v_0 \qquad \text{(initial
  conditions)},\\ \nonumber
& u|_{x=0} = 0,\; (\sigma(\pl_x u) + \beta \pl_x \ddot u)|_{x=1} =0
\qquad \text{(Neumann-boundary condition)}.
\end{align}
The chosen boundary conditions model an experiment where the
rod is free at the right boundary and fixed at the left boundary
in order to exclude translations.\\
We assume that $\sigma(\cdot)$ is a nonmonotone function with three
zeros $\veps^- < \veps^0 < \veps^+$ which satisfies $\lim_{|\veps| \to
  \infty} \veps \cdot \sigma(\veps) = \infty$. For mathematical
simplicity we do not consider boundary forces to control the
experiment. Instead the shape of the mapping
$\veps \mapsto \sigma(\veps)$ is varied with a scalar parameter $\mu$ 
in order to
stabilize or to destabilize a certain phase, therefore no assumption
that $\int_{\veps^-}^{\veps^+} \sigma(\veps)\, \d \veps =0$ is made.
The nonconvex function $W(\veps) = \int_0^\veps \sigma(\gamma) \, \d
\gamma$ is usually called double-well potential, the
two local minima correspond to two different phases which
are stable with respect to small perturbations. Global existence
and uniqueness of solutions to \reff{mwex.eq} is clear, although a
formal proof is not provided in this work.\par
If $\beta=0$ system~\reff{mwex.eq} describes the purely elastic
behavior of the rod where no microkinetic forces are taken into
account. It is clear that the initial-boundary value problem is well posed
only if $\sigma$ is monotone and thus equation~\reff{mwex.eq} is hyperbolic.
By assuming a nonmonotone stress-strain relation we have
introduced a Hadamard-instability which causes the solutions to start
oscillating heavily and blow up in finite time.
The term $\beta
\pl_x^2 \ddot u$ is of higher order thus for $\beta>0$ global existence and
uniqueness of solutions can be established.\par
System~\reff{mwex.eq} is the evolution equation associated to a Hamiltonian
system. The Hamiltonian (energy) which is conserved by the dynamics is given
by
\[ H(q,p) = \int_{x\in\Omega} \frac{1}{2}p\cdot \B p+ W(q)\, \d x,\]
where $q(t,x) = \pl_x u(t,x) \in L^2(\Omega)$,
$p(t,x) = (\rho - \beta \pl_x^2)\dot u(t,x)\in H^{-1}(\Omega)$ and $\B p \in
H^1(\Omega)$ is the solution of the elliptic equation
\[ (\rho - \beta \pl_x^2) v = p, \quad \pl_x v|_{x=0} = 0, \; v|_{x=1} = 0\]
e.g. $v= \B p$. The symplectic structure is given by the
skew-symmetric operator
$J=\left(\genfrac{}{}{0pt}{}{0}{\pl_x}
  \genfrac{}{}{0pt}{}{\pl_x}{0}\right)$.
In $u, \dot
u$-coordinates the energy takes the form
\[E(u,\dot u)=\int_{x\in\Omega} \left(\tfrac{\rho}{2}\dot u ^2 + W(\pl_x u) +
\tfrac{\beta}{2}(\pl_x \dot u)^2\right)\, \d x.\]
The expression
$K_{\mathrm{mac}}=\frac{\rho}{2} \int_{x\in\Omega}\dot u^2 \, \d x$ is called
{\it macroscopic} kinetic energy, $E= \int_{x\in\Omega} W(\pl_x u) \,\d x$
{\it elastic} energy and $K_{\mathrm{mic}}= \frac{\beta}{2} \int_{x\in\Omega}
(\pl_x \dot
u)^2 \, \d x$ {\it microscopic} kinetic energy.
\subsection{Qualitative and quantitative properties of typical solutions}
The term $\beta \pl_x^2 \ddot u$ in \reff{mwex.eq} models the inertia of
microscopic
structures, e.g. phase boundaries. To give an illustration, we consider a phase
boundary which is at time $t=0$ at the position $x=0$ and moves to the right
with velocity $c$. This corresponds to the initial value
\begin{align} \labell{ptf.eq}
u_0(x) &= \left\{ \ba{rl} \frac{1}{\kappa}x^2-x; & x \in [0,\kappa],\\
  x-\kappa; & x\in (\kappa, 1],
\end{array} \right. , \qquad
v_0(x) = \left\{ \ba{rl} -\frac{2c}{\kappa}x; & x \in [0,\kappa],\\
-2c; & x\in(\kappa,1].
\end{array} \right.
\end{align}
The variable $\kappa\ll 1$ describes the thickness of the transition layer
between the two phases. With respect to the stored energy function
$W(\veps) = \frac{1}{4}(\veps^2-1)^2$ we obtain the energy
\[ E(u_0,v_0) = (2-\tfrac{4}{3}\kappa)\rho c^2 + \tfrac{2}{15}\kappa+ 2
\tfrac{\beta c^2}{\kappa}.\]
This example shows heuristically that the velocity $c$ of the phase
boundary has vanishing influence on the macroscopic kinetic energy and the
elastic energy, if $\kappa$ is close to 0.
Only the microscopic kinetic energy is able to measure the
velocity of the phase boundary.\par
The solution of system~\reff{mwex.eq} generated
by the above initial condition is by no means a traveling front in
the variable $\pl_x u(t,\cdot)$ as it
can be seen in Figure~\ref{tw1.fig} where the
graph of $\pl_x u(t,\cdot)$ is plotted for different values of $t$.
\begin{figure}
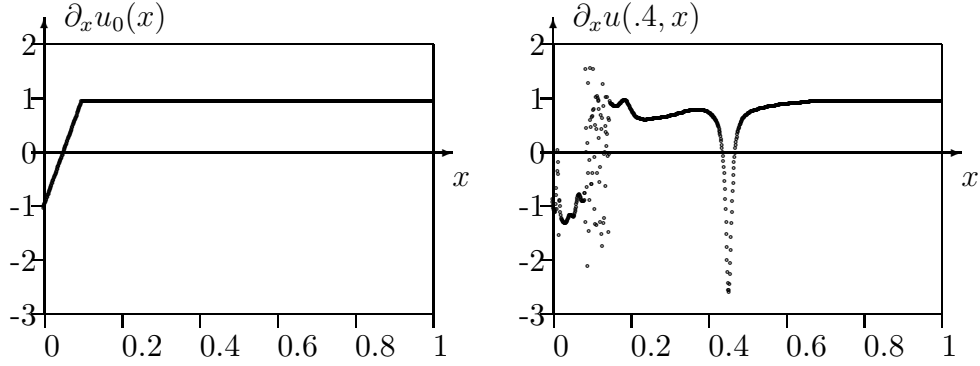

\input{iv.tex}
%
\input{pulse.tex}
\vspace{2mm}
\caption{Initial value $\pl_x u_0(\cdot)$, $c=2$, $\kappa=0.1$
and a snapshot of the strain $\pl_x u$ at time $t=0.4$ ($\rho=1$,
$\beta=10^{-4}$).}
\labell{tw1.fig}
\end{figure}
We have used $\sigma(\veps) = E(\veps ^3 - \veps)$, the Young modulus
$E=1$ MPa and the density $\rho=1\frac{\mathrm{g}}{\mathrm{cm}^3}$
in the numerical simulation, for a precise description of our
numerical scheme, error estimates and
additional results on the long-time behavior see
Chapter~\ref{numres.sec}.
The solution depicted in Figure~\ref{tw1.fig} has been computed
with mirrored boundary conditions i.e. Dirichlet-boundary condition at
$x=1$ and Neumann at $x=0$.
It is obvious that the transition region with thickness $\kappa$
generates a strongly oscillatory region which is located in the
interval $[0,\; 0.2]$. In addition one can identify a pulse at $x=0.45$
which is moving at constant velocity to the right.\par
Solitary waves have been found and analyzed for modifications of
the {Boussinesq} equation \reff{be.eq} as well as for the \ibe. Formal
calculations show that the former equation is completely integrable
and the latter is not since it does not pass the {\sc Painlev{\'e}{}}
test. This implies that solitary waves may exist but the interaction
of two pulses is not elastic, i.e. energy is transferred into
uncorrelated fluctuations, see \citel{cls86, scl84}. The analysis in
\citel{cls86} shows that also for the microkinetically regularized
wave equation it cannot be expected that it is completely integrable,
this is confirmed by the results of the numerical simulations.\par
We find two kinds of solitary waves which we denote by {\it
  nucleation} pulse and {\it elastic} pulse. The amplitudes of the
nucleation pulses are bounded from below, they traverse the nonmonotone
sector of $\sigma$ completely. The amplitude of the elastic pulses may
be arbitrarily small. Elastic pulses have nothing to do with the
nonmonotonicity of $\sigma$, they are closely related to the solitary
waves which have been analyzed in the {Boussinesq} equation.\par
\subsection{Nucleation waves}
To discuss traveling waves
we make the usual Ansatz $\pl_x u(t,x) = \vepsp(x-c t)$
where $\vepsp(\cdot)$ is
the initial strain and $c$ is the velocity of the propagation.
Plugging $\vepsp$ into
\reff{mwex.eq} we obtain the ordinary differential equation
\[
\rho c^2 \vepsp'' = \sigma(\vepsp)'' + \beta c^2 \vepsp''''.
\]
Integrating the above equation twice yields
\begin{equation} \labell{ode2.eq}
\rho c^2 \vepsp = \sigma(\vepsp) + \beta c^2 \vepsp'' + \lambda+
\wt\lambda x,
\end{equation}
where $\lambda$ and $\wt \lambda$ are the constants of integration.
We are interested in solutions $\vepsp$ satisfying $\lim_{|x|\to \infty}
\vepsp(x) = \veps^+$,
i.e. nucleations within the phase which is given by the local minimum
$\veps = \veps^+$ of $W$, cf. Figure~\ref{sigmaandw.fig}. Using
equation~\reff{ode2.eq} we
find that $\lambda = \rho c^2 \veps^+$ and $\wt \lambda=0$
and we end up with the second order equation
\begin{equation} \labell{ode3.eq}
\beta c^2 \vepsp'' = - \sigma(\vepsp) + \rho c^2 \vepsp - \rho c^2 \veps^+.
\end{equation}
The solutions of \reff{ode3.eq} move along the equipotential lines
\[ \tfrac{1}{2}\beta c^2 (\vepsp')^2+ W(\vepsp) - \tfrac{1}{2}
\rho c^2 \vepsp^2 + \rho c^2 \veps^+ \vepsp = \mathrm{const}.\]
In order to find a homoclinic orbit attached to the point $y^\infty = (\veps^+,0)
\in \R^2$ the point $\veps^+$ has to be a local maximum of the function
$(\veps \mapsto W(\veps) - \rho c^2 \veps (\frac{1}{2} \veps - \veps^+))$,
hence $\rho c^2 \geq \sigma'(\veps^+)$. The amplitude $A$ of the pulse is
$|\veps^+-\gamma|$ where $\gamma$ is the smallest respectively greatest zero
of
\begin{equation} \labell{ode4.eq}
\frac{\beta c^2}{2} (\veps')^2+ W(\veps) - \frac{1}{2} \rho c^2 \veps^2
+ \rho c^2 \veps^+ \veps - W(\veps^+) - \frac{1}{2} \rho c^2 (\veps^+)^2.
\end{equation}
For the choice $W(\veps) = \frac{1}{4}(\veps^2 - 1)^2 + \mu(\veps-
\frac{1}{3}\veps^3)$,
which has been used in the numerical simulations, the shape of the pulse
can be computed explicitly:
\[ \vepsp(x) = 1+ \frac{\rho c^2-2(1-\mu)}{1-\frac{\mu}{3} \pm
  \sqrt{\frac{1}{2}\rho c^2+ \frac{1}{9}\mu(\mu+3)}
  \cosh\left( \sqrt{\frac{\rho c^2 -2(1-\mu)}{\beta c^2}}x \right)}.\]
The role of $\mu$ will be explained in the following section.
Thus we have obtained two families of pulse solutions
($\vepsp_{\mathrm{el}}$ and $\vepsp_{\mathrm{nucl}}$) moving at velocity
$c$. If $\mu= 0$ one finds the simple
velocity-amplitude relation $A(c)=\max_{x\in\R}|\veps^+-\vepsp(x)| =
c\sqrt{2\rho}\pm 2$.\par
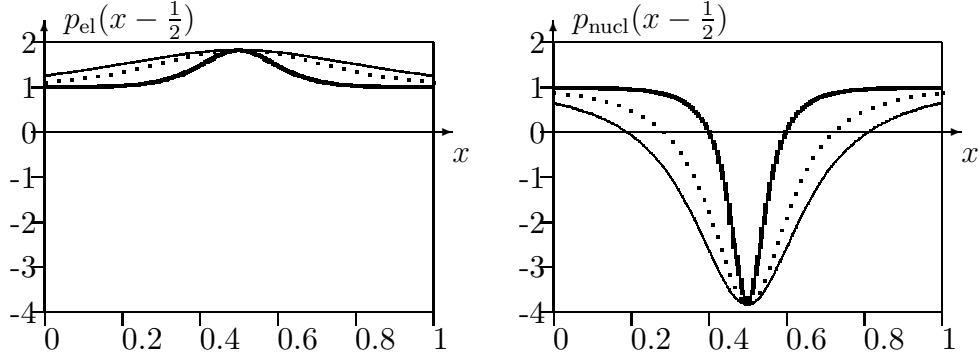
\begin{figure}
\input{Elastpls.tex}
\input{Nuclpls.tex}
\vspace{2mm}
\caption{The shape of elastic and nucleation waves for $\rho =1$, $\mu=0$,
  $c=2$ and $\beta =$ 0.025 (--), 0.0125 ($\cdots$) and 0.0025 ({\bf --})}
\labell{tws.fig}
\end{figure}
For $\rho=1$, $c=\sqrt{2}$, $\mu=0$ the amplitude $A=4$ is
approximated well by the numerically computed pulse. Note, that in the
case $\rho=0$ pulse solutions satisfying $\lim_{|x|\to \infty}
\vepsp(x)= \veps^+$ do not exist. Of course these results also hold
true with minor modifications if
the stored elastic energy $W$ is given by a more general function than
a polynomial of fourth order.\par
\subsection{Propagation of a phase transformation front}
To discuss the simulation results concerning the
considerations in the introduction, we would like to study the
propagation of a phase transformation front. From the preceding
discussion of the pulse solutions it is clear that front solution
can not exist. Front-like solutions require that
the energy $W(\veps) - \frac{1}{2} \rho c^2 \veps^2 + \lambda \veps$
(cf.~\reff{ode2.eq}) has two local maxima of same height at the
positions $\veps=\veps^-$ and $\veps=\veps^+$, but this is
only possible in degenerate cases which we will not consider in the
present work.\par
The model is intended to describe
the dynamics of phase transformations when no dissipation is present.
Since the solutions do not have the possibility to dissipate kinetic
energy, the fluctuations due to the phase transformation remain present
after the transformation region has propagated. Therefore we expect to
observe wild oscillations in the wake of the nucleation pulse. These
oscillations should be interpreted as heat since they do not give a
contribution to the kinetic energy but only to the microkinetic
energy.\par
We consider a scenario where phase 2 is slightly destabilized i.e. the
boundary force $\Sigma$ is below the Maxwell-line. It is unpleasant
that the local minima of the Gibbs' energy
$W(\veps) - \Sigma\cdot \veps$ are only given implicitly, this
makes the algebraic manipulations intricate. Therefore we model the
effect of $\Sigma \not= 0$ by taking
$\sigma(\veps) = \veps^3-\veps + \mu\cdot(1-\veps^2)$ where
$\mu \in \R$ is a control parameter, which allows us to destabilize
($\mu>0$) or to stabilize ($\mu<0$) phase~2. For
$|\mu| \ll 1$ we obtain the relation
\[ \mu \sim  -\frac{3}{2} \Sigma.\]
The simulation of the solution
defined by the initial datum~\reff{ptf.eq}
shows exactly the expected behavior, see Figure~\ref{tw2.fig}.
\begin{figure}[h]
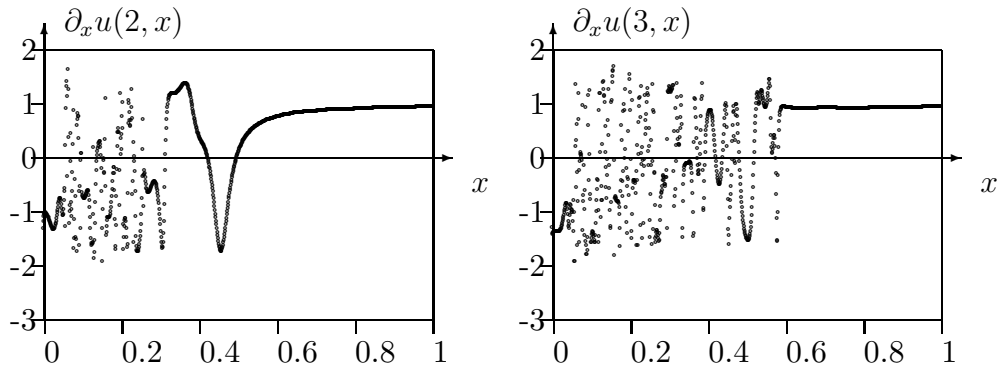

\vspace{2mm}
%
\input{movpls.tex}
\input{stckpls.tex}
\vspace*{2mm}
\caption{Two snapshots at time $t=2$ and $t=3$ show, how a propagating
phase-transformation region gets stuck at $x_{\mathrm{term}}=0.6$
($\rho=1$, $\beta=10^{-2}$, $\mu=0.5$).}
\labell{tw2.fig}
\end{figure}
One can observe a phase transformation front characterized by the the
tip of the pulse at $x= 0.45$ moving from left to right through the
material, leaving a
wake of oscillations behind. It is interesting that the phase
transformation front comes to a stop when the gained energy is transformed
too quickly into microscopic fluctuations. This obstructs the
activation of the stable phase and the phase transformation may
terminate, see Figure~\ref{tw2.fig}. This astounding effect demonstrates
that the study of the qualitative dynamic behavior requires a good
understanding of the high frequency oscillations in space.
If the sign of $\mu$ is reversed, phase~2 is stabilized and the
phase transformation does not even start, i.e. hysteresis is
observed. It is very remarkable that allowance for microkinetic
energy results in this phenomenon. In macroscopic models it is
interpreted in terms of dissipative threshold which the driving
force has to overcome in order to phase transformation proceeds
(see e.g. \cite{ak88,l94a}). Moreover in the numerical simulations
this threshold depends on volume fraction of transformed
phase, which corresponds to macroscopic description of known
experimental results \cite{fhm93, l94a}.
In Figure~\ref{vvmu.fig} the
position where the phase transformations stopped is plotted over different
values of $\mu$.
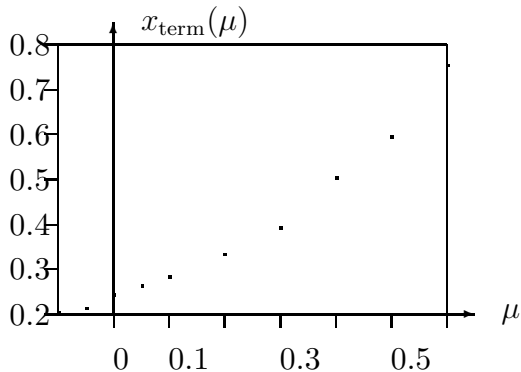
\begin{figure}[h]
\vspace{2mm}
\input{treg.tex}
\vspace{2mm}
\caption{Position of the stuck PT-front for different values of $\mu$.}
\labell{vvmu.fig}
\end{figure}
\section{The propagation of oscillations} \labell{propos.sec}
The Hadamard instability which is present in equation~\reff{mw.eq} if
$\beta =0$ is not completely removed by the regularization term $\beta
\pl_x^2 \ddot u$. The onset of the fluctuations is delayed, we expect
that after time $T$ the oscillations will occur on a length scale $h=
\sqrt{\beta}/ T$. Their presence makes a rigorous discussion of the
qualitative dynamical behavior very difficult.\par
Our approach is to study the oscillations which are brought via
oscillatory initial states into the system. We do not let $\beta$ tend
to 0, instead we consider for fixed $\beta>0$ a sequence of
oscillatory initial states and ask: ``How do the oscillations propagate
as the time increases?''. It turns out that they remain stationary in
$\Omega$, but the stochastic properties change in a well defined
fashion. To give a motivation one could think of taking for $\beta \ll
1$ the solution at time $t=1$ (which should look very similar to the
simulation result depicted in Figure~\ref{tw1.fig}) as a new initial
value.\par
In the derivation of the constitutive
law~\reff{eq:8} we have assumed that local stresses and strains $\tilde
\sigma$ and $\tilde \veps$ can be
replaced by the homogenized counterparts $\sigma_m$ and
$\veps_m$. Thus the constitutive law~\reff{eq:8} looses its validity
if oscillations occur which have a wavelength shorter than
$\sqrt{\beta}$, the size of a characteristic volume.
On the other hand there
are only few systems known, which support oscillatory solutions.
The authors are not aware of any model where the dynamic creation of
microstructure can be studied rigorously.
We will ignore the
modeling restriction and concentrate on the discussion of of
oscillatory solutions to the partial
differential equation~\reff{mwex.eq}.

The mathematical analysis of oscillatory solutions consists in several
steps. First an integro-differential equation, which
allows a clearer view to the structure of the equation is
derived. Then we define Young measures and oscillatory solutions in a
mathematical rigorous way. The propagation of oscillations is
characterized by the extended system \reff{yms.eq}, see
Theorem~\ref{app.thm}.
In Chapter~\ref{statsol.sec} the stationary solutions of the extended
system are characterized and discussed. If $\rho=0$ the long-time
behavior or the microkinetic system can be understood
completely. Almost every solution generates a nontrivial stationary
Young measure as time tends to infinity. The deformation $u(t)$ becomes
stationary but the long-time limit is not an equilibrium solution of 
\reff{mw.eq} (Theorem~\ref{ymltb.thm}).
 
In order to simplify the analysis for system~\reff{mwex.eq} we
transform it into an integro-differential equation which does not
contain spatial derivatives. The new variables are given by
$(\veps,\eta)=\mathcal T(u,\dot u) = (\pl_x u, \pl_x \dot u)$.
By differentiating equation~\reff{mwex.eq} once by $x$ one obtains
\[ (\rho-\beta \pl_x^2) \ddot \veps = \pl_x^2 \sigma(\veps).\]
Using the operator $\mathcal B$ which has been defined in
section~\ref{conseq.sec} we find that
\begin{align*}
&\ddot \veps -\ddot \veps|_{x=1}  = \rho\B\left( \pl_x^2
  \fr1\rho\sigma(\veps) - \ddot \veps|_{x=1} \right)\\
=& - \fr1\beta\rho\B \left[ \left( \rho - \beta \pl_x^2 \right) \fr1\rho
  (\sigma(\veps) - \sigma(\veps|_{x=1}))\right]
+ \fr1\beta\rho\B \biggl[ \underbrace{-\sigma(\veps|_{x=1}) - \beta
    \ddot \veps|_{x=1}}_{=0}+ \sigma(\veps) \biggr]\\
=& -\fr1\beta(\sigma(\veps) - \sigma(\veps|_{x=1})) + \fr1\beta \rho\B
\sigma(\veps)\imp \ddot \veps=-\fr1\beta (\sigma(\veps) - \rho\B\sigma(\veps)).
\end{align*}
Thus $(\veps,\eta)$ satisfy the following integro-differential equation
\begin{subequations}
\labell{ev}
\begin{align}
\labell{mko1} \dot \veps & = \eta, && \veps|_{t=0} = \veps_0\in
L^2(\Omega), \\
\labell{mko2} \dot \eta & = - \tfrac1\beta(\sigma(\veps) - k), &&
\eta|_{t=0}= \eta_0 \in L^2(\Omega),\\
k &=  \rho \B \sigma(\veps),
\end{align}
\end{subequations}
which defines us a dynamics in $L^2(\Omega,\R^2)$. The evolution preserves
the regularity of the solutions i.e. if $(\veps_0,\eta_0)\in
C^1(\Omega,\R^2)$ then $(\veps,\eta)(t)\in C^1(\Omega,\R^2)$ for all $t\geq
0$. The most important property of this system is that the equations
for $\veps$ and $\eta$ are hyperbolic with wave speed 0, thus
the solutions cannot develop shocks. This transport
equation allows the system to carry solutions oscillating rapidly
in $\veps$ and $\eta$. Since the spatial coupling via $\rho\B$ is compact, $k$
cannot fluctuate in space.
\subsection{Definition of Young measures and oscillatory solutions}
A relevant tool for the study of oscillations are Young
measures (or parameterized measures) which have
gained of importance during the last twenty years.
A Young measure $\Psi\in \Y(\Omega,\R^d)$, $d\in\N$,
is a mapping from $\Omega$ into the set of probability measures on $\R^d$,
which is denoted by $PM(\R^d)$. For each $x\in \Omega$ the probability
density $\Psi(x)(\cdot)$ measures the probability that the strain and
the strain velocity $(\pl_x u^{(n)}, \pl_x \dot u^{(n)})$
take the value $(\veps, \eta)$.\par
In the context of oscillations Young measures
are conceived as a family of one point distributions which are usually
generated by a sequence of rapidly oscillating functions. In order to clarify
the term ``generated'' we have to become a little bit more technical and
introduce notions from probability theory. Let $f^{(n)}:\Omega\to \R$
be a bounded sequence of functions. For a given point $x_0 \in \Omega$
we choose a positive number $r$, a positive integer $n$ and a step length
$h$. We intend to consider the limits $h\to 0$, $n \to \infty$ and $r\to 0$ in
this order. The triple $(r, n, h)$ defines a finite sample consisting of
the values $f^{(n)}(x_0+hk)$, $k\in\{\eta \in \Z | x_0 + \eta h \in
[x_0-r,x_0+r]\}$. Now
we let $h$ tend to $0$. This generates a unique family of probability measures
$\widetilde \nu(x_0,r,n,\cdot)\in PM(\R^d)$
which is parameterized over $(x_0,r,n) \in \Omega\times \mathcal \R^+ \times
\N$. If for almost every $x_0\in\Omega$
the limit $\nu(x_0) = \lim_{r\to 0} \lim_{n\to\infty}
\widetilde \nu(x_0,r,n)$ exists (in the vague sense) then we say that $f^{(n)}$
generates the Young measure $\nu:\Omega \to PM(\R^d):x_0\mapsto\nu(x_0)$.\par
An alternative characterization, which is more suitable in the context
of partial differential equations, is given by the duality of
probability measures and continuous functions.
A sequence $f^{(n)}$ generates a Young measure $\Psi$ if and only if
for every continuous function $g:\R^d\to\R$
\[\lim_{n\to\infty} g(f^{(n)}(x))\wto \lan \Psi(x),g\ran_{y\in\R^d}
\text{ weakly in } L^2(\Omega).\]
By ``$\lan \cdot, \cdot\ran$'' the duality between probability
measures and continuous functions is denoted. To obtain a clear
distinction between the variables the integration variable is used as an
index, i.e.\[ \lan \Psi(x), g\ran_{y\in\R^d}= \int_{y\in\R^d} g(y) \cdot
\Psi(x,y)\, \d y\]
if the probability measure $\Psi(x)$ has a density $\Psi(x,y)$. By slight
abuse of notation we will adhere to this function notation even if the
measure is singular. It is our aim to emphasize the geometric viewpoint by
this notation.\par
We say that the sequence of solutions $u^{(n)}$ defines an oscillatory
solution if the initial values $(\pl_x u_0^{(n)}, \pl_x v_0^{(n)})$
generate a nontrivial Young measure $\Psi_0$.
\subsection{Effective evolution equations}
Spatial oscillations (microstructure) are not damped away by the
dynamics of the microkinetic model. A possible characterization of the
time evolution of oscillations is given by the following problem:

\vspace{1em}
\fbox{\parbox{\linewidth}{\vspace{2mm}
Given a sequence of initial states to the microkinetic system
$(u_0^{(n)}, v_0^{(n)})$ so that $(\pl_x u_0^{(n)}, \pl_x v_0^{(n)})$ 
generates a unique Young measure $\Psi_0$ as $n\to \infty$. Does
$y^{(n)}(t,x)=(\pl_x u^{(n)},\pl_x \dot u^{(n)})(t,x)$ generate also
for $t>0$ a unique time-dependent Young measure $\Psi(t)$?
\vspace{2mm} 
}}\vspace{1em}

It can be shown that the answer is positive, the time
depending Young measure $\Psi(t,x,y)$ is a distributional solution of
the transport equation
\begin{subequations}
\labell{yms.eq}
\begin{align}
\labell{ymsa.eq}
&\pl_t \Psi(t,x,y)+ \nabla_y\cdot\{\Psi(t,x,y)\cdot(y_2,
-\tfrac{1}{\beta}(\sigma(y_1) - k(t,x)))^t \} = 0, \; \Psi|_{t=0} =
\Psi_0,\\
&(\rho-\beta\pl_x^2)k(t,x)= \rho\lan \Psi(t,x,y),\sigma(y_1) \ran_{y\in\R^2},\;
\pl_x k|_{x=0} = 0, \; k|_{x=1} = 0.
\end{align}
\end{subequations}
The transport term is exactly the right-hand side of system~\reff{ev}.
Equation~\reff{ymsa.eq} has to be read in the weak sense, i.e.
for every test function $g\in C^1(\R^2)$
\begin{align}
\labell{ymws}
\frac{\d}{\d t} \lan \Psi, g \ran-\lan \Psi(t,x,y), \nabla g(y) \cdot
(y_2, -\tfrac{1}{\beta}(\sigma(y_1)-k(t,x)))^t\ran_{y\in\R^2}= 0
\end{align}
for every $x\in\Omega$. System~\reff{yms.eq} is a extension of
\reff{mwex.eq},
i.e. every solution of \reff{mwex.eq} is a special solution of \reff{ymws}. We
formulate this fundamental assertion in a
\begin{proposition} \labell{consist.prop}
Let $u(t,x)$ be a solution of \reff{mwex.eq} and $\veps=\pl_x u$,
$\eta=\pl_x \dot u$. Then the Young measure
\[\Psi(t,x)= \delta_{(\veps, \eta)(t,x)}\]
where $\delta_y$ is the Dirac distribution with mass at $y\in\R^2$, is a
distributional solution of \reff{yms.eq}.
\end{proposition}
\begin{proof}
Since the equivalence of \reff{mwex.eq} and \reff{ev} has already been
established, we set $y(t,x) = \mathcal T(u(t,x),\dot u(t,x))$.
By inserting $\delta_{y(t,x)}$ into \reff{ymws} we obtain
\begin{align*}
\frac{\d}{\d t} g(y(t))- \nabla g(y(t)) \cdot ( y_2, -
\tfrac{1}{\beta}(\sigma(y_1) - k(t,x)))^t = 0
\end{align*}
by \reff{ev}.
\end{proof}
It can be shown under reasonable assumptions that the solutions of
\reff{yms.eq} are unique and exist globally in time.
The central assertion which justifies \reff{yms.eq} as an effective
equation for oscillatory solutions is the following continuity
result. If the oscillations of the initial state can be captured with
a Young measure $\Psi_0$ it suffices to take $\Psi_0$ as an initial
value for \reff{yms.eq}. The unique solution at time $t$ is a Young
measure which describes the oscillations at time $t$.
\begin{theorem} \labell{app.thm}
Let $(\pl_x u_0^{(n)},\pl_x v_0^{(n)})\in L^2(\Omega,\R^2)$ generate the
Young measure $\Psi_0\in \Y(\Omega,\R^2)$ and
$u^{(n)}(t)$ be a sequence of solutions to the microkinetically
regularized wave equation
\reff{mwex.eq},
so that $(\pl_x u^{(n)},\pl_x \dot u^{(n)})(t=0) = (\pl_x u_0^{(n)},
\pl_x v_0^{(n)})$. Then $(\pl_x u^{(n)}, \pl_x \dot u^{(n)})(t)$
generates a Young measure $\Psi(t)$, which is the unique solution of 
\reff{yms.eq} satisfying the initial condition $\Psi(0) = \Psi_0$.
\end{theorem}
\begin{proof}
For a proof see \citel{th97,th98b}.
\end{proof}
It is not at all clear that Young measures are always
the right objects to capture oscillatory behavior since they ignore
spatial correlations. In higher dimensional settings it can be shown
that mathematically more elaborate objects than one-point statistics
are needed. However for this prototypical one-dimensional equation
they suffice.
A general discussion on the relation between Young-measure solutions
and classical solutions of generalizations of system~\reff{ev} see
\citel{mi98}.\par
We summarize the results of this section. Replacing the original
coordinates displacement-velocity $(u,\dot u)$ by the new coordinates
strain-strain rate $(\veps,\eta)$ it was possible to transform
system~\reff{mwex.eq} into the equivalent, mathematically better accessible
system~\reff{ev}. The system supports oscillating solutions and one
can show that the dynamics of oscillatory solution can be described
with time dependent Young measures which are unique solutions of
\reff{yms.eq}. Thus the latter equation is a homogenized version
of \reff{ev}, it describes the macroscopic behavior of the
microscopically fluctuating strain.

\section{Long-time behavior} \labell{statsol.sec}
The discussion of the dynamics of the microkinetically
regularized wave equation \reff{mw.eq} shows that
the stationary solutions of \reff{yms.eq}
are central for the understanding of the
qualitative behavior. They represent a kind of
oscillatory ground state, which is to a large extend
independent from the macroscopic dynamics.\par
In the numerical simulation results in Chapter~\ref{eqmo.sec}
one can see that the phase transformation induces
strong oscillations to the solutions. The macroscopic dynamics of
these oscillatory solutions is very slow, this suggests the
supposition that they approximate stationary Young measures. This
supposition is partly supported by the results of
Section~\ref{ltb.sec}. In the case $\rho=0$ generic
solutions thermalize, i.e. $(\pl_x u,\pl_x \dot u)(t)$ generates a
nontrivial stationary Young measure as time tends to infinity.
For nonzero $\rho$ thermalization can be confirmed neither
analytically nor numerically. The
solutions still develop arbitrarily fine oscillations but a certain
amount of macroscopic dynamics remains in the simulations.\par
Since the microkinetic system \reff{mwex.eq} is Hamiltonian it
might be surprising at the first glance, that generic solutions may
converge to an equilibrium for $t\to \infty$.
This is only possible by developing fine oscillations so that
the macroscopically visible kinetic energy vanishes into uncorrelated
microscopic fluctuations. This phenomenon is denoted thermalization,
it is
the usual behavior of solids in the absence of an exciting force. The
fundamental laws of physics are reversible, thus the introduction of
additional quantities like temperature, which are necessary to describe
irreversible processes, only give a model like, statistical description
of this behavior.\par
\subsection{Generalized stationary states}
To clarify the role of system~\reff{yms.eq} we demonstrate
that the system is a nontrivial extension of
\reff{ev}. This is done by
comparing the set of equilibrium points $\mathcal C$ of \reff{mwex.eq} with
the equilibria $\mathcal Y$ of \reff{yms.eq}. It turns out that $\mathcal Y$
is much larger than $\mathcal C$, the latter set is not even dense in the
former.
The stationary points of \reff{mwex.eq} are characterized by the
equation $\pl_x \sigma(\pl_x u) = 0$.
Using the Neumann-boundary condition this equation can be integrated once
and one obtains $\sigma(\pl_x u)=0$. We denote the three zeros of $\sigma$ by
$z_i$ where $\veps^- < z_0 < \veps^+$. The set of equilibrium points can be
characterized as follows: For every stationary solution $u^*$
of \reff{mwex.eq} there exists a partition $(\Omega_1,\Omega_0,\Omega_2)$ of
$\Omega$ so that $\pl_x u^*(x) = \sum_{i=0}^2 \chi_{\Omega_i}(x)z_i$ almost
everywhere.\par
The equilibrium points of the extended system~\reff{yms.eq} are defined by the
equation
\begin{align}
\labell{statym} \lan \Psi(x,y), \nabla g(y)\cdot(y_2,-\tfrac{1}{\beta}
(\sigma(y_1)-k(x)))^t\ran_{y\in\R^2} = 0,\\
\labell{statcoup}
(\rho - \beta\pl_x^2) k = \rho\lan \Psi(x,y),\sigma(y_1) \ran_{y\in\R^2}
\end{align}
for every test function $g\in C^1(\R^2)$. By choosing $g(y) = y_2$ we obtain
the relation
\[ \lan \Psi(x,y), -\tfrac{1}{\beta}(\sigma(y_1) - k(x)) \ran_{y\in\R^2} =0\]
for every $x\in\Omega$. Hence \reff{statcoup} implies $k=0$. Thus it suffices
to analyze the equation
\begin{align} \nonumber
\nabla_y\cdot\{\Psi\cdot(\beta y_2,-\sigma(y_1))^t\}&=0\\
\labell{dpjnh}
\aeq \nabla \Psi(x,y) \cdot J\cdot
 \nabla H(y)&=0,\quad J=\left(\ba{rr} 0 & 1 \\ -1 & 0\ea \right)
\end{align}
where $H(y)= \frac{\beta}{2} y_2^2 + W(y_1)$. Equation~\reff{dpjnh} expresses
the fact that for fixed $x\in\Omega$ the function $\Psi(x,y)$ is constant
on the connected components of the equipotential lines of $H$. For the
convenience of the reader we give a short proof of this well known fact.
\begin{lemma} \labell{courhil.lem} Let $\Psi,H\in C^1(\R^2,\R)$ and
\begin{equation} \labell{orth}
\pl_{y_1} \Psi \cdot \pl_{y_2} H= \pl_{y_2} \Psi \cdot\pl_{y_1} H
\end{equation}
for every $y\in\R^2$. Let in addition be $y^1,
  y^2\in \R^2$ and $\varphi\in C^1([0,1],\R^2)$ satisfying
$\varphi(0)= y^{(1)}$, $\varphi(1)=y^{(2)}$ and $\fr{\d}{\d t} H(\varphi(t))
=0$ for every $t\in[0,1]$. Then
$\Psi(y^{(1)})=\Psi(y^{(2)})$ holds.
\end{lemma}
\begin{proof}
We have that
$\Psi(y_2)-\Psi(y_1) = \int_0^1\nabla\Psi(y(t))\cdot\dot y(t)\, \d t=0$ because
of \reff{orth} and since $\fr{\d }{\d t} H(y(t))= \nabla H(y(t)) \cdot \dot y$.
\end{proof}
For double well-potentials like $W(\veps) = \frac{1}{4}(\veps^2-1)^2$
we have to split the phase space $\R^2$ into three domains in oder to
represent arbitrary phase distributions:
\begin{equation} \label{pssplit.eq} \begin{array}{rl}
P_1&=\{y\in\R^2 | H(y) < W(z_0) \text{ and } y_1 < z_0 \},\\
P_2&=\{y\in\R^2 | H(y) < W(z_0) \text{ and } y_1 > z_0 \},\\
P_0&=\R^2\setminus(P_1 \cup P_2). \end{array}
\end{equation}
The sets $P_1$ and $P_2$ correspond approximately to the two stable phases
$\veps^-$ and $\veps^+$. By Lemma~\ref{courhil.lem} every stationary solution
$\Psi^*(x,y)$ of
\reff{yms.eq} can be expressed using three functions
\begin{theorem}
Let $\Psi$ be a stationary solution of \reff{yms.eq} which has a
density function $\Psi^*:\Omega\times \R^2\to \R_\geq$. Then there
exist three functions $\alpha_i:\Omega\times \R^+ \to \R^+$, $i=0,1,2$
so that
\[ \Psi^*(x,y) = \sum_{i=0}^2\chi_{P_i}(x)\cdot \alpha_i(x,H(y)),\]
where $P_0$, $P_1$, $P_2$ is the partition of $\R^2$ defined by
\reff{pssplit.eq}.
\end{theorem}
The converse of the theorem also holds true if $\alpha_i$ satisfy
normality constraints like
$\lan \Psi^*(x,y),1\ran_{y\in\R^2}=1$ for every $x\in
\Omega$.
By taking suitable sequences $(\alpha_i^k)_{k\in\N}$ we also obtain singular
equilibria. The notation of the singular equilibrium Young measures
requires the use of geometric measure theory, so we have
decided not to characterize the full set of stationary solutions in this
work.\par
Very similar results concerning stationary Young measures for a two
dimensional magneto fluid have been obtained in
\citel{jo94, jt96}. There
the most probable stationary states are determined by the maximum of an
entropy functional. 
In our case the importance of the stationary Young
measures is not clear if $\rho>0$, cf. the results of the long-time
integrations in Chapter~\ref{numres.sec}.\par
The striking difference between the stationary solutions of \reff{mwex.eq} and
stationary Young measures is that the latter may have nonvanishing velocity
components, classical equilibria are restricted to the line $\eta=0$.
Since the velocity distribution of stationary Young measures
is symmetric with respect to $\eta=0$ the average velocity $\lan \Psi(x,y),
y_2\ran_{y\in\R^2}$
vanishes for every $x\in\Omega$. One should also note that the displacement
$u^*(x)$ associated to a stationary Young measure $\Psi^*(x)$ via
\[ u^*(x) = \int_0^x \lan \Psi^*(x',y), y_1\ran_{y\in\R^2} \, \d x'\]
is in general not a stationary solution of \reff{mwex.eq}. This is no
contradiction to the fact that $\Psi^*$ is an equilibrium since the
microscopic fluctuations are not visible from the macroscopic level.
The discussion of the stationary solutions shows, that the
continuation of equation~\reff{mwex.eq} in the space of Young measures reveals
many stationary solutions which have been formerly hidden.\par
\subsection{Thermalization}
\labell{ltb.sec}
The first attempt to observe the rise of uncorrelated microscopic
fluctuations  is described in the well-known paper of
{\sc E.~Fermi, J.~Pasta}
and {\sc S.~Ulam} \citel{fpu55}. The result, which is obtained by numerical
integration of several discretized one-dimensional quasilinear wave equations
is truly opposite of the expected behavior. It is found that the motion of
the rod is
almost periodic, especially the initial state is approximated up to a very
high accuracy in fixed time intervals.\par
We consider equation~\reff{mwex.eq} to be an example for
thermalization. If
$\rho=0$ almost every solution $u(t,x)$ of \reff{mwex.eq} converges to a
stationary
function $u^*(x)$ in $L^\infty(\Omega)$ at $t\to \infty$,
i.e. \reff{mwex.eq} is a good candidate
to study the transformation of macroscopic kinetic energy into heat. But
not only $u(t)$ converges but also $(\pl_x u(t,x), \pl_x \dot u(t,x))$
generates a unique nontrivial equilibrium Young measure $\Psi^*$
as time goes to infinity. Thus equation~\reff{mwex.eq} is also an example for
the dynamic creation of microstructure. The case $\rho=0$ is an important
limit since the equations of motion for $\rho>0$ emerge from the
unperturbed system via a compact perturbation. Thus one may hope that the
essential dynamics can be understood by analyzing the case $\rho = 0$.\par
The convergence of
the solutions if $\rho =0$ is due to the fact that  the system is completely
integrable, hence it can be analyzed easily using canonical variables.
\begin{theorem} \labell{ymltb.thm}
Let $\rho=0$ and $u(t,x)$ be a solution of system~\reff{mwex.eq} generated
by the initial condition $(u,\dot u)|_{t=0}=(u_0,v_0)$ and let the genericity
assumption $\mathrm{closure}(G) = \Omega$ be fulfilled, where the
subset $G\subset \Omega$ is defined as follows:
\[ G = \left\{ x \in \Omega \quad \Bigl| \quad
  \frac{\d}{\d x} \int_{\veps \in \R}
    \Re \left(\tfrac{\beta}{2}(\pl_x v_0(x))^2+W(\pl_x
          u_0(x)) -W(\veps) \right)^{-\frac{1}{2}}\, \d \veps\not=0
  \right\} .\]
Then there exists a unique nontrivial stationary Young measure $\Psi^*\in
\Y(\Omega,\R^2)$ so
that $(\pl_x u(t,x), \pl_x \dot u(t,x))$ generates $\Psi^*$ for
$t\to \infty$.
\end{theorem}
Using this abstract result one can easily deduce the following theorem
which describes the long-time behavior of solutions more graphically.
\begin{theorem} \labell{ultb.thm}
Let $u(t,x)$ be a solution of \reff{mwex.eq} and let the assumptions of
Theorem~\ref{ymltb.thm} be satisfied. Then there exists a function $u^*(x)$ so
that $\lim_{t\to\infty}\|u(t,x)-u^*(x)\|_{L^\infty(\Omega)}+
\lim_{t\to\infty} \|\dot u(t,x) \|_{L^\infty(\Omega)}=0$. If in addition
$G=\Omega$ then there is a constant $C>0$ so that
$\|u(t,x)-u^*(x)\|_{L^\infty(\Omega)}+\|\dot u(t,x)\|_{L^\infty(\Omega)}
\leq C/t$.
\end{theorem}
For the proofs of both theorems see \citel{th97, th98b}.\par
Crucial for this result is the nonmonotonicity of $\sigma$. If
$\sigma$ is linear,
the set $G$ in Theorem~\ref{ymltb.thm} is empty, the
integral is constantly equal to $2 \pi \sqrt{\beta/\sigma'}$.
Thus the thermalization process
is strongly related to underlying phase transformation problem.\par
The discussion of the stationary solutions of the microkinetic system
reveals that there exists a great variety of Young measure equilibria,
which cannot be approximated with classical stationary solutions.
The set of the
generalized equilibria is very important for the dynamics, it can be shown
that it attracts generic solutions in the case $\rho=0$. The simulation
results show that also in the coupled case $\rho>0$  the solutions stay
close to the stationary Young measures.
\section{Numerical scheme and simulation results}
\labell{numres.sec}
Since there are no rigorous results concerning the long-time behavior
if $\rho>0$ is considered, we present some numerical results which should
illuminate the qualitative behavior in the coupled case. The results indicate
that the asymptotics for $\rho>0$ differ from the uncoupled
case. Before we comment
the individual long-time integrations we explain the notion of convergence of
the numerical scheme. For simplicity we set $\beta=1$.\par
A Young-measure solution is approximated by aggregated Young measures:
\begin{align*}
  \Y^{h,N}(\Omega,\R^2) = \left\{\Psi\in\Y(\Omega,\R^2) \left|\;
      \Psi(x) = \tfrac{1}{N} \sum_{k=0}^{\frac{1}{h}-1}
      \sum_{i=1}^N \chi_{[hk,h(k+1))}(x)
      \delta_{y^{ik}} \text{ where } y \in \R^{N\times\fr{1}{h}\times
        2}\right. \right\}.
\end{align*}
Aggregated Young measures $\Psi^{h,N} \in \Y^{h,N}(\Omega,\R^2)$ are
constant on
intervals of length $h$. For $x\in (h k, h(k+1))$ the probability measure
$\Psi^{h,N}(x)$ is a convex combination of $N$ Dirac masses located at
$\{y^{ik}, i=1\ldots N\}$.
We remark, that there are various possibilities of approximating
nontrivial Young measures by simpler objects. For a discussion of some
methods, see {\sc Roubicek's} book \citel{ro97}, Chapter 4.\par
In order to obtain useful estimates, we introduce a metric
which has the property that $\cup_{M,N=1}^\infty
\Y^{\frac{1}{M}, N}(\Omega,\R^2)$ is dense in $\Y(\Omega,\R^2)$. We define
\[ d_2(\Psi_1,\Psi_2) = \left\| \sup_{\|\nabla
    g\|_{L^\infty(\R^2)}\leq 1} \lan \Psi_1(x)-\Psi_2(x),
  g\ran\right\|_{L^2_x(\Omega)},\]
where the subscript $x$ denotes the integration variable of the $L^2$-norm.
The metric $d_2(\cdot,\cdot)$ is an extension of the Wasserstein distance
which metrizes
vague convergence in the space of probability measures with first moment,
see, e.g. \citel{r91}. Of
course for given $\Psi\in\Y(\Omega,\R^2)$
there is no a priori estimate for $\inf_{\Psi^{h,N} \in
  \Y^{h,N}(\Omega,\R^2)} d_2(\Psi^{h,N},\Psi)$ independent of $\Psi$ which can
be used to compute in advance how $h$ and $N$ have to be chosen in order
to obtain a prescribed accuracy. The propagation of this initial
error can be controlled by using the continuity of the solution with respect
to the initial value.\par
$\Y^{h,N}$ is invariant under the evolution defined by \reff{yms.eq}
only if $\rho=0$. For nonvanishing $\rho$ the right hand side has to
be projected on  $\Y^{h,N}$.
This motivates the consideration of the modified evolution equation
\begin{align} \labell{pyms}
&\dot \Psi^h = -\nabla_{(\veps,\eta)} \cdot \left\{ \Psi^h\cdot (\eta,
  -(\sigma(\veps)-\pi_h \rho\B \lan \Psi^h, \sigma\ran))^t\right\},
\quad \Psi^h|_{t=0} =\Psi_0 \in \Y(\Omega,\R^2).
\end{align}
The linear mapping $\pi_h^k$ is the projection onto the space of piecewise
constant functions: $\pi_h^kf(x) =
-\frac{1}{h}\int_{x\in[kh,(k+1)h]} f(x')\, \d x'$
for $x\in [kh, (k+1) h)$. The dynamics defined by \reff{pyms} leaves
$\Y^{h,N}$ invariant.
For initial values $\Psi_0^{h,N} \in \Y^{h,N}(\Omega,\R^2)$ we obtain
\[ \Psi^{h,N}(t,x) = \frac{1}{N} \sum_{k,i} \chi_{[kh,(k+1)h)}
\delta_{y^{ki}(t)},\]
where the $(\frac{1}{h}\times N\times 2)$-vector $y^{ki}(t)$ solves the
ordinary differential equation
\begin{equation} \labell{node}
\dot y^{ki}= \left( \ba{c} y_2^{ki}\\
- \sigma(y^{ki}_1)+ \pi_h^k \mathcal B \Bigl[ 
\tfrac{1}{N}\sum\limits_{k,i}
  \chi_{[kh,(k+1)h)} \sigma(y^{ki}_1)\Bigr] \ea \right),
\quad y^{ki}(0) = y^{ki}_0.
\end{equation}
This system can be integrated using an arbitrary integration scheme for
ordinary differential equations. For the error accumulated due to the
modification of the spatial coupling an exponential bound can be derived.
\begin{theorem} \label{cns.thm}
Let $\beta=1$ and $\Psi(t)$ and $\Psi^{h,N}(t)$  solutions of
\reff{yms.eq} and \reff{pyms} satisfying the initial condition
$\Psi(0) = \Psi^{h,N}(0) = \Psi^{h,N}_0 \in
\Y^{h,N}(\Omega,\R^2)$. Then there exist numbers $C_1(\Psi_0)$ and
$C_2(\rho,\|\sigma'\|_{L^\infty(\R)})$ so that
\[ d_2(\Psi(t),\Psi^{h,N}(t)) \leq C_1 \sqrt{\rho}h^2\,e^{C_2t} \]
holds.
\end{theorem}
\begin{proof} See Appendix. \end{proof}
To control the error, when arbitrary initial states are considered, one
has to use Gronwall estimates for the Young-measure semiflow. This
does not contribute to the understanding of the numerical scheme,
therefore this step is omitted here, see however \citel{th98b}.\par
In the numerical simulations in this work we used $N=1$, i.e. only
functions, not general Young measures. The reason we proved the
convergence for the generalized scheme is, that we are convinced that
the numerical results give a good picture of the qualitative dynamics
although $h$ and the time increment $\tau$ are too large to see the
asymptotic regime of the scheme. Theorem~\ref{cns.thm} shows that it
is possible to follow solutions which are strongly oscillating.
For a proper implementation of these ideas, more sophisticated
numerical techniques like adaptive schemes have to used, but this is
beyond the scope of this work.
\subsection{Numerical results}
In order to check the decay of macroscopic kinetic energy long-time
integrations were performed. This might seem dubious since the error grows
exponentially in time. At first it seems that the numerically computed
solution has nothing to do with the exact solution for large times.
The same problem
occurs when integration schemes for ordinary differential equations are
analyzed; only estimates which behave exponentially in time can be
derived. This situation greatly improves if the viewpoint is slightly
changed: For integration schemes for ordinary differential equations it
can be shown that the numerically computed solution corresponds to a
high precision to a slightly perturbed vector field, more precisely it
can be shown that the error remains exponentially small over a time interval
$T=\mathcal O(\frac{1}{\triangle t})$ (see e.g. \citel{re97}).
Since we are only
interested in qualitative properties of the dynamics generated by certain
vector fields, this weakened result is satisfactory. It is not our aim to
give rigorous estimates for the long-time integrations, we just point out
the ideas which are relevant if a rigorous justification of the validity
of the results of the long-time integrations is attempted.\par
As integration method for the system of ordinary differential equations
the implicit midpoint rule with step length $\tau=0.1$ is used.
Equation~\reff{node} is integrated for
different values of $\rho$ over the interval $[0,400]$, the parameter
$\beta$ is constantly kept to $1$. The solutions which are computed in the
described fashion conserve the total energy not exactly but up to a precision
of 1\% over the whole time interval. For test purposes we reduced $\tau$ to
the value $0.01$, now the total energy was conserved up to an error of
0.01\% and the qualitative results were practically unchanged.\par
In order to be able to compare the results for different values of $\rho$ the
initial value is  $u_0(x) = \sqrt{2} x + \frac{1}{2} x^2$,
$v_0(x)=0$ in every experiment.\par
The simulations have been performed with different
values of $h$ ($h_1 = 0.005$, $h_2=0.002$, $h_3=0.001$, $h_4=0.0005$). The
results of runs using $h=h_3$ and $h=h_4$ differ very little from each
other, thus
it can be expected that the numerical results for $h=h_4$
are in agreement with the exact behavior.\par
The normalized macroscopic kinetic energy
$M(t)=\nke=\fr12\int_{x\in\Omega} \dot u^2\, \d x$ serves as an
indicator for the qualitative long-time dynamics. In the
figures~\ref{rho0.0.fig} - \ref{rho0.2.fig} the value $M(t)$ is plotted
over the time in double logarithmic scale. Time $t$ and normalized kinetic
energy $M(t)$ are plotted horizontally respectively vertically.
 The first experiment is the
uncoupled case $\rho=0$. One can see clearly that $M(t)$
converges to 0 for $t\to\infty$. The convergence rate of the computed
solution is in agreement with the analytical result:
$M(t) \leq C/t$ for some constant $C>0$. Since
$M(t)$ is oscillating strongly within the large
integration interval the graph $\{(t,M(t))|\; t\in[0,400]\}$
resembles more to a region than a line.\par
In the coupled case ($\rho>0$) the behavior of the solutions is
more complex that in the uncoupled case. On the one hand in the beginning
the solutions develop fine structure, otherwise $M(t)$ would
not decay. On the other hand this process soon is hidden by chaotic
oscillations which no longer decay. The nonvanishing of
$M(t)$ as $t\to \infty$ is due to the nontrivial coupling. With growing
values of $\rho$ the amplitude of the chaotic oscillations becomes larger.
This can be seen in the results for $\rho=0.02$ (Figure~\ref{rho0.02.fig}),
$\rho=0.05$ (Figure~\ref{rho0.05.fig}), $\rho=0.1$ (Figure~\ref{rho0.1.fig}) and
$\rho = 0.2$ (Figure~\ref{rho0.2.fig}). These results indicate that the
long-time behavior of the solutions in the coupled case cannot be understood
by an analysis which is only based on considerations made for the uncoupled
case.\par
\begin{figure}
\begin{minipage}[t]{.43\linewidth}
\setlength{\unitlength}{0.240900pt}
\ifx\plotpoint\undefined\newsavebox{\plotpoint}\fi
\begin{picture}(750,450)(0,0)
\epsfig{height=4cm,file=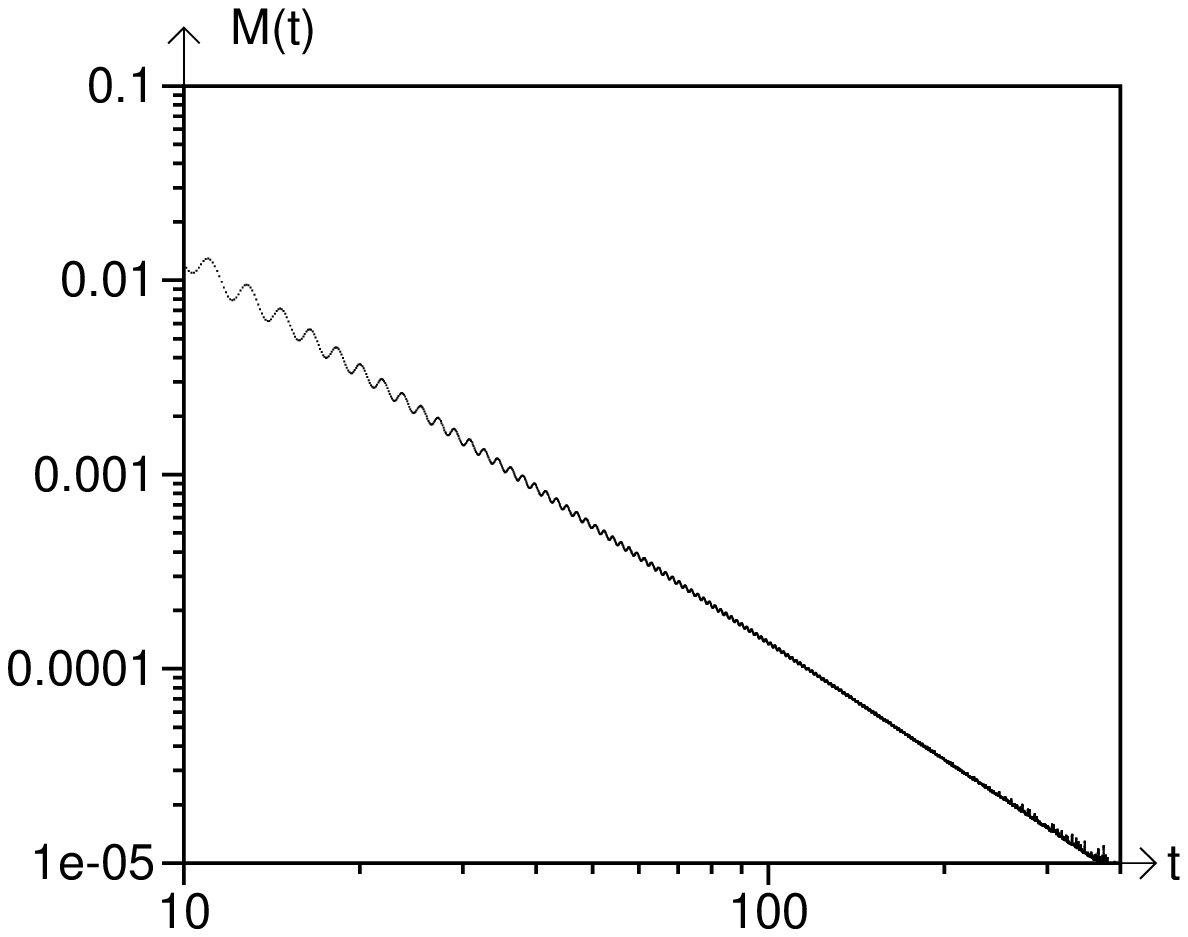}
\end{picture}
\caption{Macroscopic kinetic energy over time, $\rho=0$,
$n=2000$} \labell{rho0.0.fig}
\end{minipage}
\begin{minipage}[t]{.43\linewidth}
\setlength{\unitlength}{0.240900pt}
\ifx\plotpoint\undefined\newsavebox{\plotpoint}\fi
\begin{picture}(750,450)(0,0)
\epsfig{height=4cm,file=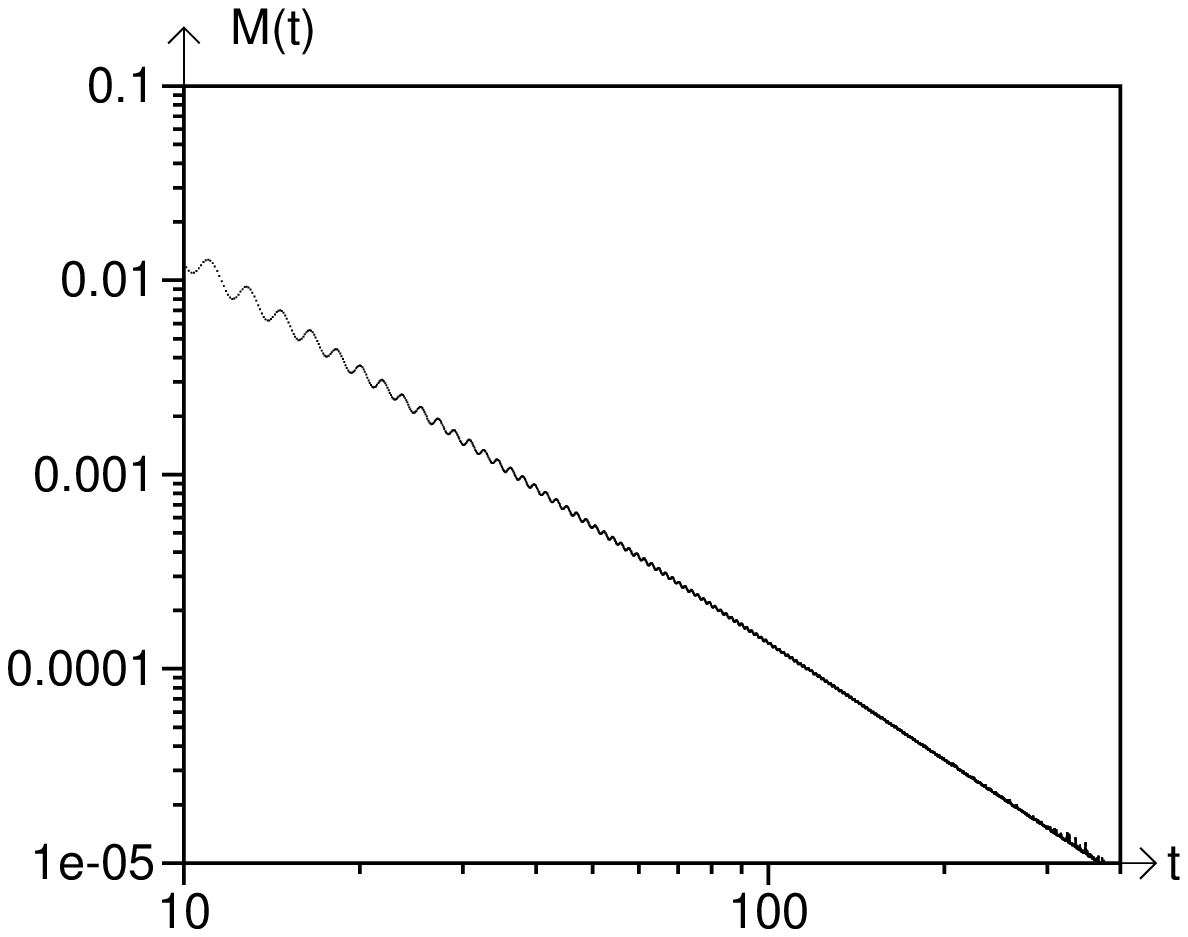}
\end{picture}
\caption{$\rho=0.02$, $n=2000$} \labell{rho0.02.fig}
\end{minipage}
\end{figure}
%
%
\begin{figure}
\vspace*{5mm}
\begin{minipage}[t]{.43\linewidth}
\setlength{\unitlength}{0.240900pt}
\ifx\plotpoint\undefined\newsavebox{\plotpoint}\fi
\begin{picture}(750,450)(0,0)
\epsfig{height=4cm,file=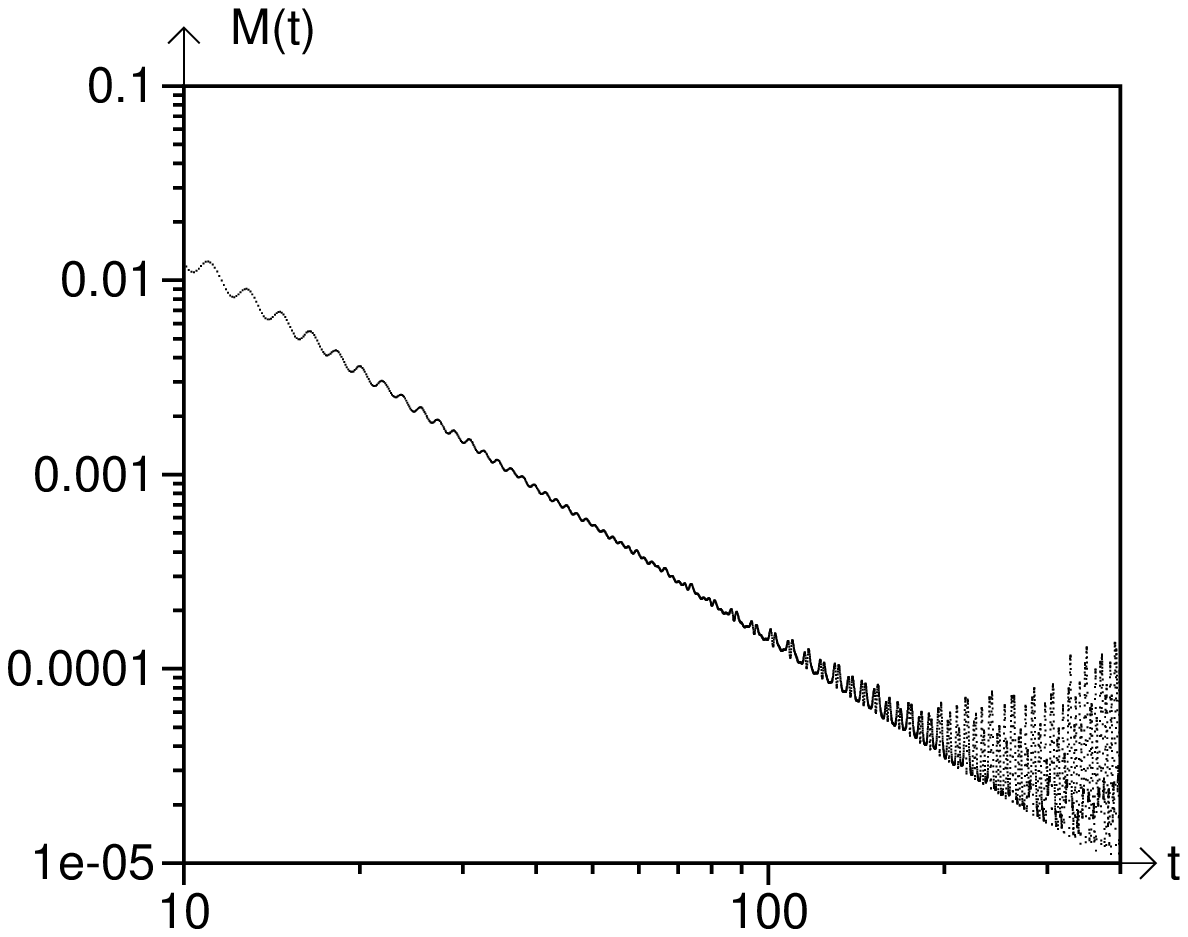}
\end{picture}
\caption{$\rho=0.05$, $n=2000$} \labell{rho0.05.fig}
\end{minipage}
%
\begin{minipage}[t]{.43\linewidth}
\setlength{\unitlength}{0.240900pt}
\ifx\plotpoint\undefined\newsavebox{\plotpoint}\fi
\begin{picture}(750,450)(0,0)
\epsfig{height=4cm,file=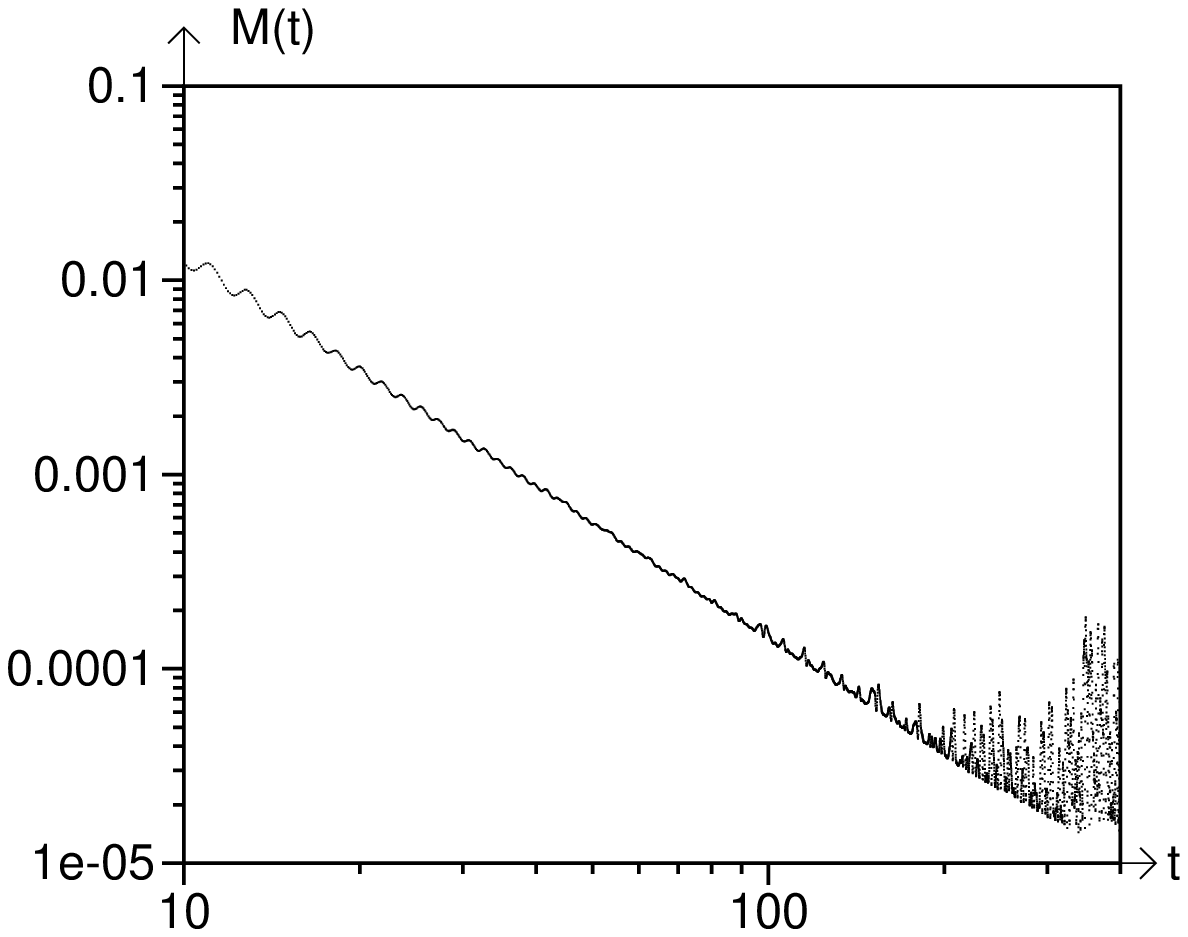}
\end{picture}
\caption{$\rho=0.1$, $n=2000$} \labell{rho0.1.fig}
\end{minipage}
\end{figure}
%
%
\begin{figure}
%
\begin{minipage}[t]{.43\linewidth}
\setlength{\unitlength}{0.240900pt}
\ifx\plotpoint\undefined\newsavebox{\plotpoint}\fi
\begin{picture}(750,450)(0,0)
\epsfig{height=4cm,file=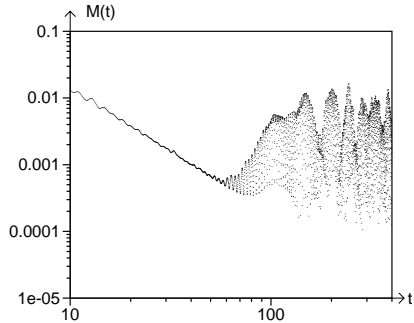}
\end{picture}
\caption{$\rho=0.2$, $n=2000$} \labell{rho0.2.fig}
\end{minipage}
\end{figure}
%
\section{Concluding remarks}
The high frequency oscillations due to inertial stresses are not observable
at the macro level and represent the continuum mechanical counterpart of
thermal fluctuations. Such stress fluctuations can overcome the stress
and energy barrier $(BCD)$ in Figure~1 when macroscopic stress-strain
curve coincides with the Maxwell line.

The possible scenario is as follows.
After some macroscopic perturbation (e.g. prescribed velocity at the
boundary), macroscopic kinetic energy transforms into microscopic
kinetic energy, or energy of stress fluctuations. Such energy exists
in each specimen before experiment. When under prescribed growing
displacement macroscopic stress reaches the Maxwell stress and stress
fluctuations are sufficient to overcome the barrier $(BCD)$ in Figure~1,
then PT will occur at constant stress.

If for conservative systems the
macroscopic kinetic energy transforms into microscopic kinetic energy,
which is not observable macroscopically, then this can be considered as
a macroscopic dissipation mechanism.
Indeed, the numerical simulations exhibit the hysteretic behavior. 
Such a behavior is usually interpreted
in macroscopic models in terms of dissipative threshold which the driving
force has to overcome in order to phase transformation proceeds. The
threshold value 
depends on the amount of the transformed phase as it is observed in
known experiments. 

The transformation of the energetically less favorable phase $\veps^+$
into $\veps^-$ is initiated by a nucleation pulse. The macroscopic
kinetic energy of the system is transformed into microkinetic energy
by the onset of strong oscillations.
This process decreases the macroscopic
kinetic energy which is necessary to overcome the energy barrier
between the phases and the transformation process stops. To
quantify this picture the notion of Young-measure solutions has been
introduced. It has been shown that the extended system, which describes the
dynamics of oscillatory solutions, has nontrivial equilibria which can
store the microkinetic energy.\par
We conjecture that the chaotic fluctuations are very close
to a unique
``most probable'' Young-measure equilibrium in the big set of stationary
Young measures. The notion of a probable state is meant in the sense
of statistical physics where the Gibbs distributions are identified as
maximizers of the entropy.

Most questions concerning the dynamics of the microkinetically
regularized wave equation are still open:

It should be possible to obtain a bound $C$ depending on $\rho$ so
that the macroscopic kinetic energy is bounded by $C(\rho)$ as time
goes to infinity i.e.
\[ \limsup_{t\to \infty} K_{\mathrm{mac}}(t) \leq C(\rho) .\]
What happens if the more physical limit $\rho=1$, $\beta\to 0$ is
taken? Do the solution $u^\beta$ converge to solutions of the
convexified equation
\[ \rho \ddot u = \pl_x\sigma^M(\pl_x u)\]
where $\sigma^M$ is obtained from $\sigma$ by replacing the
nonmonotone part by the Maxwell line? This is only meaningful for
smooth solutions. In the case of shocks the question is complicated by
the nonuniqueness of weak solutions.
\section{Appendix}
\begin{proof}[Proof of Theorem~\ref{cns.thm}]
We follow the usual procedure to establish the convergence of a numerical
scheme: Consistency and stability imply convergence.\par
For the given exact Young measure solutions
$t\mapsto \Psi(t,x)$ and $t\mapsto \Psi^{h,N}(t,x)$ of \reff{yms.eq} and
\reff{pyms} we define $\phi(t,x):\R^2\to\R^2:y\mapsto \phi(t,x,y)$
as the family of diffeomorphisms generated by the ordinary differential
equation:
\[ \dot \phi_1(t,x) = \phi_2(t,x),
\quad \dot \phi_2(t,x) = - (\sigma(\phi_1(t,x)) - \rho\B \lan
\Psi(t,x,y), \sigma(\phi_1(t,x,y)) \ran_{y\in\R^2} ),\]
Correspondingly the solution flow of the discretized system
\[ \dot q^h =  p^h, \quad \dot p^h = - (\sigma(q^h) - \pi_h\rho\B
\lan \Psi^{h,N}(t,x), \sigma \ran ),\]
is denoted by $\phi^h(t,x)$. Note, that from equation~\reff{ymws} it follows
immediately that $\Psi$ and $\Psi^{h,N}$ satisfy
\[ \lan \Psi(t), g \ran_{y\in\R^2} = \lan \Psi_0, g \circ \phi(t,x)
\ran_{y\in\R^2},\]
and
\[ \lan \Psi^{h,N}(t), g \ran_{y\in\R^2} = \lan \Psi^{h,N}_0, g \circ
\phi^h(t,x)\ran_{y\in\R^2},\]
for every test function $g\in C(\R^2)$.
\subsection*{Consistency of $\rho\B$ with $\pi_h \rho\B$:}
We establish the existence of a constant $C>0$ so that
\[ \|(\id-\pi_h)\rho\B\|_{\mathrm{Lin}(L^2(\Omega),L^2(\Omega))} \leq Ch^2\]
holds. This results mainly from the smoothing property of $\rho\B$. We set
$\omega_k = [hk,h(k+1)) \subset \R$. Firstly $\|\pi_h u-u\|_{L^2(\Omega)}$ is
estimated by $\|\pl_x u\|_{L^2(\Omega)}$.
Using the intermediate value theorem for every
$k\in \{0,\ldots 1/h-1\}$ one obtains the existence of $\xi_k\in\omega_k$
so that $\saint_{x\in\omega_k} u(x)\, \d x = u(\xi_k)$ holds. Hence
\begin{align*}
\left| \aint_{x'\in \omega_k} u(x')\, \d x-u(x)\right|
= |u(\xi_k) -u(x)|\leq \int_{\xi_k}^x|u'(s)| \d s \leq
\sqrt{|x-\xi_k|}\left(\int_{\xi_k}^x|u'(s)|^2 \d s \right)^{\frac{1}{2}}.
\end{align*}
Using this estimate we obtain
\begin{align} \nonumber
& \left( \int_\Omega |\pi_h u(x) - u(x)|^2\d x \right)^{\frac{1}{2}}
= \left( \sum_k\int_{x\in\omega_k}  \left| \aint_{x'\in\omega_k} u(x') \, \d x'
    - u(x)\right|^2 \,\d x\right)^{\frac{1}{2}}\\
\leq & \biggl( \sum_k \int_{\omega_k}\biggl( \underbrace{\sqrt{
 \diam(\omega_k)}}_{\leq \sqrt{h}}\| u'\|_{L^2(\omega_k)} \biggr)^2 \d x
  \biggr)^\frac{1}{2}
\leq \sqrt{h} \biggl(\sum_k \underbrace{\vol(\omega_k)}_{\leq h} \| u'
  \|^2_{L^2(\omega_k)} \biggr)^\frac{1}{2}\\
\labell{kons} \leq & h \| u'\|_{L^2(\Omega)}.
\end{align}
The $L^2(\Omega)$-norm of $\pl_x \rho\B u$ can be estimated by the
$L^2(\Omega)$-Norm of $u$ by expanding $u$ into a Fourier
series and estimating the coefficients of $\pl_x \rho\B u$. Since $\rho\B$ is
selfadjoint, the supremum of the modulus of the spectrum of $\pl_x\rho\B$ is
the
operator norm in $\mathrm{Lin}(L^2(\Omega),L^2(\Omega))$. Hence $\|\pl_x
\rho\B\| \leq \sup_{k\in\R}\frac{\rho \pi k}{\rho + k^2 \pi^2}=
\sqrt{\frac{\rho}{2}}$. By estimate \reff{kons}, the inequality $\|(\id-\pi_h)
  \rho \B\|_{\mathrm{Lin}(L^2(\Omega),L^2(\Omega))}\leq C h$ where $C=
  \sqrt{\frac{\rho}{2}}$ is established. Repeating the last step gives
  us the desired estimate $\|(\id-\pi_h)
  \rho \B\|_{\mathrm{Lin}(L^2(\Omega),L^2(\Omega))}\leq C h^2$.
\subsection*{Stability:}
The next step is to control $d_2(\Psi_0,\Psi(t))$ using Gronwall's
inequality. This result is a purely analytical estimate for the exact
system~\reff{yms.eq}.
\begin{align}
\nonumber
&\| \lan \Psi_0(y), | \phi_t(y)-y |\ran_{y\in\R^2} \|_{L^2(\Omega)}\\
= & \left\|\lan \Psi_0(y), \left| \int_0^t \left( \ba{c}
\phi_{s,2}(y) - y_2 + y_2\\
\nonumber
- \sigma \circ\phi_{s,1}(y) + \rho\B \lan \Psi_0, \sigma \circ \phi_{s,1}
\ran
\ea \right) \d s \right| \ran_{y\in\R^2} \right\|_{L^2(\Omega)}\\
\nonumber
\leq & \int_0^t (\ell + \| \rho\B \| \|\sigma'\|_{L^\infty(\R)}) \cdot \|\lan \Psi_0(y),
|\phi_s(y) - y| \ran \|\, \d s\\
\nonumber
&+ \int_0^t \left\| \lan\Psi_0(y), \left|-\sigma(y_2)+\sigma_0+ \rho\B
\lan \Psi_0(d\widetilde y), \sigma(\widetilde y_2)-\sigma_0\ran \right|
\ran\right\|_{L^2(\Omega)} \d s + t \sigma_0(1+\|\rho\B\|)\\
\label{stabil}
\imp & \left\|\lan \Psi_0(y), |\phi_t(y)-y|\ran\right\|_{L^2(\Omega)} \leq
\widetilde{C} e^{(2\ell + \|\rho\B\|\|\sigma'\|_{L^\infty(\R)})t}.
\end{align}
where
$\widetilde{C}= (1+ \|\rho\B \|) (\|\sigma'\|_{L^\infty(\R)}
\lan\Psi_0(y), |y|\ran +\sigma_0)$ and $\sigma_0 = |\sigma(0)|$.
Using this we can estimate the error due to the discretization:
\begin{align*}
&d_2( \Psi(t), \Psi^{h,N}(t)) =
\left\| \sup_{\lip(g) \leq 1} \lan \Psi_0, g \circ \phi_t - g \circ
  \phi^h_t \ran \right\|_{L^2(\Omega)} \leq \left\| \lan \Psi_0, |\phi_t
  -\phi_t^h | \ran \right\|_{L^2(\Omega)}\\
\leq & \int_0^t \left\| \lan \Psi_0, \left( \ba{c} \phi_{s,2} -
\phi^h_{s,2}\\
-\sigma\circ \phi_{s,1} + \sigma\circ\phi^h_{s,1} \ea \right)
\ran\right\|_{L^2(\Omega)}+ \left\| \rho\B \lan \Psi_0, \sigma \circ
  \phi_{s,1}\ran - \pi_h\rho\B\lan \Psi_0, \sigma\circ \phi^h_{s,1} \ran
\right\|_{L^2(\Omega)}\,\d s \\
\leq & \int_0^t \ell \left\| \lan \Psi_0, |\phi_s-\phi_s^h|\ran
\right\|_{L^2(\Omega)}
+ \|(\id-\pi_h)\rho\B\|\,
\| \lan \Psi_0, \sigma\circ \phi_{s,1}\ran\|_{L^2(\Omega)}\\
&+ \|\pi_h \rho\B\| \, \|\lan\Psi_0, |\sigma\circ\phi_{s,1}-\sigma\circ
\phi^h_{s,1}|\ran \|_{L^2(\Omega)}\, \d s.
\end{align*}
The second term can be controlled using the stability
inequality~\reff{stabil}
\begin{align*}
&\| \lan \Psi_0, \sigma \circ \phi_{t,1}\ran \| \leq \| \lan \Psi_0,
|\sigma\circ\phi_{t,1}-\sigma|\ran\|+ \|\lan \Psi_0, \sigma\ran \|\\
\leq & \|\sigma'\|_{L^\infty(\R)} \left( \| \lan \Psi_0(y), | \phi_t(y)-y|\ran \|
+\|\lan\Psi_0(y),|y|\ran \|_{L^2(\Omega)}\right)+\sigma_0\\
\leq & \|\sigma'\|_{L^\infty(\R)} \widetilde{C} e^{(2\ell + \| \rho\B \| \|\sigma'\|_{L^\infty(\R)}) t}
+ \|\sigma'\|_{L^\infty(\R)} \lan \Psi_0(y), |y|\ran + \sigma_0.
\end{align*}
Using Gronwall's inequality again, one gets
\begin{align*}
&d_2(\Psi(t), \Psi^{h,N}(t))\\
\leq & C h^2 e^{(\ell + \|\sigma'\|_{L^\infty(\R)} \| \rho\B\|)t}
\cdot \left\{ \widetilde{C} e^{(2\ell+\|\rho\B \|\|\sigma'\|_{L^\infty(\R)})t} +
  \|\sigma'\|_{L^\infty(\R)} \lan\Psi_0(y),|y|\ran+ \sigma_0 \right\}.
\end{align*}
Setting $\widetilde{\widetilde{C}}:= C(\|\sigma'\|_{L^\infty(\R)}\widetilde{C} +
\|\sigma'\|_{L^\infty(\R)}\lan \Psi_0(y), |y|\ran + \sigma_0)$ yields
the desired estimate
\begin{equation}
\nonumber
d_2( \Psi(t)), \Psi^{h,N}(t)) \leq
\widetilde{\widetilde{C}} h e^{3(\ell+ \|\sigma'\|_{L^\infty(\R)}\| \rho\B \|)t}.
\end{equation}
\end{proof}
\section*{Acknowledgments}
F.T. wishes to thank Prof. Alexander Mielke for his constant
encouragement and many helpful suggestions. He is also grateful to
Prof. Felix Otto for pointing him to \citel{jo94}. V.I.L. acknowledges
discussions with Prof. Erwin Stein.
The support by the VW-Foundation under the project I/70 284
and I/70 282 ``Stress and strain induced phase transformations in
engineering materials'' is greatly acknowledged.

\end{document}

%% file: sigma.tex
\setlength{\unitlength}{0.240900pt}
\ifx\plotpoint\undefined\newsavebox{\plotpoint}\fi
\begin{picture}(750,450)(0,0)
\font\gnuplot=cmr10 at 12pt
\gnuplot
\sbox{\plotpoint}{\rule[-0.200pt]{0.400pt}{0.400pt}}%
\put(706,146){\makebox(0,0)[l]{$\varepsilon$}}
\put(218,485){\makebox(0,0)[l]{$\sigma(\varepsilon)$}}
\put(249,273){\makebox(0,0)[l]{$B$}}
\put(314,394){\makebox(0,0)[l]{$C$}}
\put(416,330){\makebox(0,0)[l]{$D$}}
\put(488,210){\makebox(0,0)[l]{$E$}}
\put(569,302){\makebox(0,0)[l]{$F$}}
\put(96,302){\makebox(0,0)[l]{$\sigma_M$}}
\put(249,76){\makebox(0,0)[l]{$\varepsilon^-$}}
\put(401,76){\makebox(0,0)[l]{$\varepsilon^0$}}
\put(554,76){\makebox(0,0)[l]{$\varepsilon^+$}}
\put(96,112){\vector(1,0){610}}
\put(188,19){\vector(0,1){466}}
\put(249,302){\line(1,0){305}}
\put(172,302){\line(1,0){16}}
\put(249,112){\line(0,-1){14}}
\put(401,112){\line(0,-1){14}}
\put(554,112){\line(0,-1){14}}
\put(168.67,19){\rule{0.400pt}{1.204pt}}
\multiput(168.17,19.00)(1.000,2.500){2}{\rule{0.400pt}{0.602pt}}
\multiput(170.59,24.00)(0.482,2.932){9}{\rule{0.116pt}{2.300pt}}
\multiput(169.17,24.00)(6.000,28.226){2}{\rule{0.400pt}{1.150pt}}
\multiput(176.59,57.00)(0.482,2.660){9}{\rule{0.116pt}{2.100pt}}
\multiput(175.17,57.00)(6.000,25.641){2}{\rule{0.400pt}{1.050pt}}
\multiput(182.59,87.00)(0.482,2.570){9}{\rule{0.116pt}{2.033pt}}
\multiput(181.17,87.00)(6.000,24.780){2}{\rule{0.400pt}{1.017pt}}
\multiput(188.59,116.00)(0.485,1.942){11}{\rule{0.117pt}{1.586pt}}
\multiput(187.17,116.00)(7.000,22.709){2}{\rule{0.400pt}{0.793pt}}
\multiput(195.59,142.00)(0.482,2.208){9}{\rule{0.116pt}{1.767pt}}
\multiput(194.17,142.00)(6.000,21.333){2}{\rule{0.400pt}{0.883pt}}
\multiput(201.59,167.00)(0.482,2.027){9}{\rule{0.116pt}{1.633pt}}
\multiput(200.17,167.00)(6.000,19.610){2}{\rule{0.400pt}{0.817pt}}
\multiput(207.59,190.00)(0.482,1.847){9}{\rule{0.116pt}{1.500pt}}
\multiput(206.17,190.00)(6.000,17.887){2}{\rule{0.400pt}{0.750pt}}
\multiput(213.59,211.00)(0.482,1.756){9}{\rule{0.116pt}{1.433pt}}
\multiput(212.17,211.00)(6.000,17.025){2}{\rule{0.400pt}{0.717pt}}
\multiput(219.59,231.00)(0.482,1.575){9}{\rule{0.116pt}{1.300pt}}
\multiput(218.17,231.00)(6.000,15.302){2}{\rule{0.400pt}{0.650pt}}
\multiput(225.59,249.00)(0.485,1.179){11}{\rule{0.117pt}{1.014pt}}
\multiput(224.17,249.00)(7.000,13.895){2}{\rule{0.400pt}{0.507pt}}
\multiput(232.59,265.00)(0.482,1.304){9}{\rule{0.116pt}{1.100pt}}
\multiput(231.17,265.00)(6.000,12.717){2}{\rule{0.400pt}{0.550pt}}
\multiput(238.59,280.00)(0.482,1.123){9}{\rule{0.116pt}{0.967pt}}
\multiput(237.17,280.00)(6.000,10.994){2}{\rule{0.400pt}{0.483pt}}
\multiput(244.59,293.00)(0.482,0.943){9}{\rule{0.116pt}{0.833pt}}
\multiput(243.17,293.00)(6.000,9.270){2}{\rule{0.400pt}{0.417pt}}
\multiput(250.59,304.00)(0.482,0.943){9}{\rule{0.116pt}{0.833pt}}
\multiput(249.17,304.00)(6.000,9.270){2}{\rule{0.400pt}{0.417pt}}
\multiput(256.59,315.00)(0.482,0.762){9}{\rule{0.116pt}{0.700pt}}
\multiput(255.17,315.00)(6.000,7.547){2}{\rule{0.400pt}{0.350pt}}
\multiput(262.59,324.00)(0.485,0.569){11}{\rule{0.117pt}{0.557pt}}
\multiput(261.17,324.00)(7.000,6.844){2}{\rule{0.400pt}{0.279pt}}
\multiput(269.00,332.59)(0.491,0.482){9}{\rule{0.500pt}{0.116pt}}
\multiput(269.00,331.17)(4.962,6.000){2}{\rule{0.250pt}{0.400pt}}
\multiput(275.00,338.59)(0.491,0.482){9}{\rule{0.500pt}{0.116pt}}
\multiput(275.00,337.17)(4.962,6.000){2}{\rule{0.250pt}{0.400pt}}
\multiput(281.00,344.60)(0.774,0.468){5}{\rule{0.700pt}{0.113pt}}
\multiput(281.00,343.17)(4.547,4.000){2}{\rule{0.350pt}{0.400pt}}
\multiput(287.00,348.60)(0.774,0.468){5}{\rule{0.700pt}{0.113pt}}
\multiput(287.00,347.17)(4.547,4.000){2}{\rule{0.350pt}{0.400pt}}
\put(293,352.17){\rule{1.300pt}{0.400pt}}
\multiput(293.00,351.17)(3.302,2.000){2}{\rule{0.650pt}{0.400pt}}
\put(299,353.67){\rule{1.445pt}{0.400pt}}
\multiput(299.00,353.17)(3.000,1.000){2}{\rule{0.723pt}{0.400pt}}
\put(305,354.67){\rule{1.686pt}{0.400pt}}
\multiput(305.00,354.17)(3.500,1.000){2}{\rule{0.843pt}{0.400pt}}
\put(318,354.67){\rule{1.445pt}{0.400pt}}
\multiput(318.00,355.17)(3.000,-1.000){2}{\rule{0.723pt}{0.400pt}}
\put(324,353.17){\rule{1.300pt}{0.400pt}}
\multiput(324.00,354.17)(3.302,-2.000){2}{\rule{0.650pt}{0.400pt}}
\put(330,351.17){\rule{1.300pt}{0.400pt}}
\multiput(330.00,352.17)(3.302,-2.000){2}{\rule{0.650pt}{0.400pt}}
\multiput(336.00,349.95)(1.132,-0.447){3}{\rule{0.900pt}{0.108pt}}
\multiput(336.00,350.17)(4.132,-3.000){2}{\rule{0.450pt}{0.400pt}}
\multiput(342.00,346.94)(0.920,-0.468){5}{\rule{0.800pt}{0.113pt}}
\multiput(342.00,347.17)(5.340,-4.000){2}{\rule{0.400pt}{0.400pt}}
\multiput(349.00,342.95)(1.132,-0.447){3}{\rule{0.900pt}{0.108pt}}
\multiput(349.00,343.17)(4.132,-3.000){2}{\rule{0.450pt}{0.400pt}}
\multiput(355.00,339.93)(0.599,-0.477){7}{\rule{0.580pt}{0.115pt}}
\multiput(355.00,340.17)(4.796,-5.000){2}{\rule{0.290pt}{0.400pt}}
\multiput(361.00,334.94)(0.774,-0.468){5}{\rule{0.700pt}{0.113pt}}
\multiput(361.00,335.17)(4.547,-4.000){2}{\rule{0.350pt}{0.400pt}}
\multiput(367.00,330.93)(0.599,-0.477){7}{\rule{0.580pt}{0.115pt}}
\multiput(367.00,331.17)(4.796,-5.000){2}{\rule{0.290pt}{0.400pt}}
\multiput(373.00,325.93)(0.491,-0.482){9}{\rule{0.500pt}{0.116pt}}
\multiput(373.00,326.17)(4.962,-6.000){2}{\rule{0.250pt}{0.400pt}}
\multiput(379.00,319.93)(0.710,-0.477){7}{\rule{0.660pt}{0.115pt}}
\multiput(379.00,320.17)(5.630,-5.000){2}{\rule{0.330pt}{0.400pt}}
\multiput(386.00,314.93)(0.491,-0.482){9}{\rule{0.500pt}{0.116pt}}
\multiput(386.00,315.17)(4.962,-6.000){2}{\rule{0.250pt}{0.400pt}}
\multiput(392.00,308.93)(0.599,-0.477){7}{\rule{0.580pt}{0.115pt}}
\multiput(392.00,309.17)(4.796,-5.000){2}{\rule{0.290pt}{0.400pt}}
\multiput(398.00,303.93)(0.491,-0.482){9}{\rule{0.500pt}{0.116pt}}
\multiput(398.00,304.17)(4.962,-6.000){2}{\rule{0.250pt}{0.400pt}}
\multiput(404.00,297.93)(0.491,-0.482){9}{\rule{0.500pt}{0.116pt}}
\multiput(404.00,298.17)(4.962,-6.000){2}{\rule{0.250pt}{0.400pt}}
\multiput(410.00,291.93)(0.599,-0.477){7}{\rule{0.580pt}{0.115pt}}
\multiput(410.00,292.17)(4.796,-5.000){2}{\rule{0.290pt}{0.400pt}}
\multiput(416.00,286.93)(0.581,-0.482){9}{\rule{0.567pt}{0.116pt}}
\multiput(416.00,287.17)(5.824,-6.000){2}{\rule{0.283pt}{0.400pt}}
\multiput(423.00,280.93)(0.599,-0.477){7}{\rule{0.580pt}{0.115pt}}
\multiput(423.00,281.17)(4.796,-5.000){2}{\rule{0.290pt}{0.400pt}}
\multiput(429.00,275.93)(0.599,-0.477){7}{\rule{0.580pt}{0.115pt}}
\multiput(429.00,276.17)(4.796,-5.000){2}{\rule{0.290pt}{0.400pt}}
\multiput(435.00,270.93)(0.599,-0.477){7}{\rule{0.580pt}{0.115pt}}
\multiput(435.00,271.17)(4.796,-5.000){2}{\rule{0.290pt}{0.400pt}}
\multiput(441.00,265.94)(0.774,-0.468){5}{\rule{0.700pt}{0.113pt}}
\multiput(441.00,266.17)(4.547,-4.000){2}{\rule{0.350pt}{0.400pt}}
\multiput(447.00,261.94)(0.774,-0.468){5}{\rule{0.700pt}{0.113pt}}
\multiput(447.00,262.17)(4.547,-4.000){2}{\rule{0.350pt}{0.400pt}}
\multiput(453.00,257.94)(0.920,-0.468){5}{\rule{0.800pt}{0.113pt}}
\multiput(453.00,258.17)(5.340,-4.000){2}{\rule{0.400pt}{0.400pt}}
\multiput(460.00,253.95)(1.132,-0.447){3}{\rule{0.900pt}{0.108pt}}
\multiput(460.00,254.17)(4.132,-3.000){2}{\rule{0.450pt}{0.400pt}}
\put(466,250.17){\rule{1.300pt}{0.400pt}}
\multiput(466.00,251.17)(3.302,-2.000){2}{\rule{0.650pt}{0.400pt}}
\put(472,248.17){\rule{1.300pt}{0.400pt}}
\multiput(472.00,249.17)(3.302,-2.000){2}{\rule{0.650pt}{0.400pt}}
\put(312.0,356.0){\rule[-0.200pt]{1.445pt}{0.400pt}}
\put(484,246.67){\rule{1.445pt}{0.400pt}}
\multiput(484.00,247.17)(3.000,-1.000){2}{\rule{0.723pt}{0.400pt}}
\put(490,246.67){\rule{1.686pt}{0.400pt}}
\multiput(490.00,246.17)(3.500,1.000){2}{\rule{0.843pt}{0.400pt}}
\put(497,247.67){\rule{1.445pt}{0.400pt}}
\multiput(497.00,247.17)(3.000,1.000){2}{\rule{0.723pt}{0.400pt}}
\multiput(503.00,249.61)(1.132,0.447){3}{\rule{0.900pt}{0.108pt}}
\multiput(503.00,248.17)(4.132,3.000){2}{\rule{0.450pt}{0.400pt}}
\multiput(509.00,252.61)(1.132,0.447){3}{\rule{0.900pt}{0.108pt}}
\multiput(509.00,251.17)(4.132,3.000){2}{\rule{0.450pt}{0.400pt}}
\multiput(515.00,255.60)(0.774,0.468){5}{\rule{0.700pt}{0.113pt}}
\multiput(515.00,254.17)(4.547,4.000){2}{\rule{0.350pt}{0.400pt}}
\multiput(521.00,259.59)(0.491,0.482){9}{\rule{0.500pt}{0.116pt}}
\multiput(521.00,258.17)(4.962,6.000){2}{\rule{0.250pt}{0.400pt}}
\multiput(527.59,265.00)(0.482,0.581){9}{\rule{0.116pt}{0.567pt}}
\multiput(526.17,265.00)(6.000,5.824){2}{\rule{0.400pt}{0.283pt}}
\multiput(533.00,272.59)(0.492,0.485){11}{\rule{0.500pt}{0.117pt}}
\multiput(533.00,271.17)(5.962,7.000){2}{\rule{0.250pt}{0.400pt}}
\multiput(540.59,279.00)(0.482,0.762){9}{\rule{0.116pt}{0.700pt}}
\multiput(539.17,279.00)(6.000,7.547){2}{\rule{0.400pt}{0.350pt}}
\multiput(546.59,288.00)(0.482,0.943){9}{\rule{0.116pt}{0.833pt}}
\multiput(545.17,288.00)(6.000,9.270){2}{\rule{0.400pt}{0.417pt}}
\multiput(552.59,299.00)(0.482,1.033){9}{\rule{0.116pt}{0.900pt}}
\multiput(551.17,299.00)(6.000,10.132){2}{\rule{0.400pt}{0.450pt}}
\multiput(558.59,311.00)(0.482,1.123){9}{\rule{0.116pt}{0.967pt}}
\multiput(557.17,311.00)(6.000,10.994){2}{\rule{0.400pt}{0.483pt}}
\multiput(564.59,324.00)(0.482,1.304){9}{\rule{0.116pt}{1.100pt}}
\multiput(563.17,324.00)(6.000,12.717){2}{\rule{0.400pt}{0.550pt}}
\multiput(570.59,339.00)(0.485,1.179){11}{\rule{0.117pt}{1.014pt}}
\multiput(569.17,339.00)(7.000,13.895){2}{\rule{0.400pt}{0.507pt}}
\multiput(577.59,355.00)(0.482,1.575){9}{\rule{0.116pt}{1.300pt}}
\multiput(576.17,355.00)(6.000,15.302){2}{\rule{0.400pt}{0.650pt}}
\multiput(583.59,373.00)(0.482,1.666){9}{\rule{0.116pt}{1.367pt}}
\multiput(582.17,373.00)(6.000,16.163){2}{\rule{0.400pt}{0.683pt}}
\multiput(589.59,392.00)(0.482,1.847){9}{\rule{0.116pt}{1.500pt}}
\multiput(588.17,392.00)(6.000,17.887){2}{\rule{0.400pt}{0.750pt}}
\multiput(595.59,413.00)(0.482,2.027){9}{\rule{0.116pt}{1.633pt}}
\multiput(594.17,413.00)(6.000,19.610){2}{\rule{0.400pt}{0.817pt}}
\put(601.17,436){\rule{0.400pt}{1.500pt}}
\multiput(600.17,436.00)(2.000,3.887){2}{\rule{0.400pt}{0.750pt}}
\put(478.0,248.0){\rule[-0.200pt]{1.445pt}{0.400pt}}
\end{picture}

%% file: w.tex
\setlength{\unitlength}{0.240900pt}
\ifx\plotpoint\undefined\newsavebox{\plotpoint}\fi
\begin{picture}(750,450)(0,0)
\font\gnuplot=cmr10 at 12pt
\gnuplot
\sbox{\plotpoint}{\rule[-0.200pt]{0.400pt}{0.400pt}}%
\put(238,443){\makebox(0,0)[l]{$W(\varepsilon)$}}
\put(808,125){\makebox(0,0)[l]{$\varepsilon$}}
\put(198,54){\makebox(0,0)[l]{$\varepsilon^-$}}
\put(401,54){\makebox(0,0)[l]{$\varepsilon^0$}}
\put(604,54){\makebox(0,0)[l]{$\varepsilon^+$}}
\put(35,90){\vector(1,0){773}}
\put(198,90){\vector(0,1){353}}
\put(198,90){\line(0,-1){7}}
\put(401,90){\line(0,-1){7}}
\put(604,90){\line(0,-1){7}}
\put(96,366){\usebox{\plotpoint}}
\multiput(96.59,354.79)(0.482,-3.474){9}{\rule{0.116pt}{2.700pt}}
\multiput(95.17,360.40)(6.000,-33.396){2}{\rule{0.400pt}{1.350pt}}
\multiput(102.59,317.18)(0.482,-3.022){9}{\rule{0.116pt}{2.367pt}}
\multiput(101.17,322.09)(6.000,-29.088){2}{\rule{0.400pt}{1.183pt}}
\multiput(108.59,283.73)(0.482,-2.841){9}{\rule{0.116pt}{2.233pt}}
\multiput(107.17,288.36)(6.000,-27.365){2}{\rule{0.400pt}{1.117pt}}
\multiput(114.59,253.94)(0.485,-2.094){11}{\rule{0.117pt}{1.700pt}}
\multiput(113.17,257.47)(7.000,-24.472){2}{\rule{0.400pt}{0.850pt}}
\multiput(121.59,225.67)(0.482,-2.208){9}{\rule{0.116pt}{1.767pt}}
\multiput(120.17,229.33)(6.000,-21.333){2}{\rule{0.400pt}{0.883pt}}
\multiput(127.59,201.50)(0.482,-1.937){9}{\rule{0.116pt}{1.567pt}}
\multiput(126.17,204.75)(6.000,-18.748){2}{\rule{0.400pt}{0.783pt}}
\multiput(133.59,180.05)(0.482,-1.756){9}{\rule{0.116pt}{1.433pt}}
\multiput(132.17,183.03)(6.000,-17.025){2}{\rule{0.400pt}{0.717pt}}
\multiput(139.59,160.88)(0.482,-1.485){9}{\rule{0.116pt}{1.233pt}}
\multiput(138.17,163.44)(6.000,-14.440){2}{\rule{0.400pt}{0.617pt}}
\multiput(145.59,144.71)(0.482,-1.214){9}{\rule{0.116pt}{1.033pt}}
\multiput(144.17,146.86)(6.000,-11.855){2}{\rule{0.400pt}{0.517pt}}
\multiput(151.59,131.74)(0.485,-0.874){11}{\rule{0.117pt}{0.786pt}}
\multiput(150.17,133.37)(7.000,-10.369){2}{\rule{0.400pt}{0.393pt}}
\multiput(158.59,119.82)(0.482,-0.852){9}{\rule{0.116pt}{0.767pt}}
\multiput(157.17,121.41)(6.000,-8.409){2}{\rule{0.400pt}{0.383pt}}
\multiput(164.59,110.37)(0.482,-0.671){9}{\rule{0.116pt}{0.633pt}}
\multiput(163.17,111.69)(6.000,-6.685){2}{\rule{0.400pt}{0.317pt}}
\multiput(170.59,102.65)(0.482,-0.581){9}{\rule{0.116pt}{0.567pt}}
\multiput(169.17,103.82)(6.000,-5.824){2}{\rule{0.400pt}{0.283pt}}
\multiput(176.00,96.94)(0.774,-0.468){5}{\rule{0.700pt}{0.113pt}}
\multiput(176.00,97.17)(4.547,-4.000){2}{\rule{0.350pt}{0.400pt}}
\multiput(182.00,92.95)(1.132,-0.447){3}{\rule{0.900pt}{0.108pt}}
\multiput(182.00,93.17)(4.132,-3.000){2}{\rule{0.450pt}{0.400pt}}
\put(188,89.67){\rule{1.686pt}{0.400pt}}
\multiput(188.00,90.17)(3.500,-1.000){2}{\rule{0.843pt}{0.400pt}}
\put(201,89.67){\rule{1.445pt}{0.400pt}}
\multiput(201.00,89.17)(3.000,1.000){2}{\rule{0.723pt}{0.400pt}}
\put(207,91.17){\rule{1.300pt}{0.400pt}}
\multiput(207.00,90.17)(3.302,2.000){2}{\rule{0.650pt}{0.400pt}}
\multiput(213.00,93.60)(0.774,0.468){5}{\rule{0.700pt}{0.113pt}}
\multiput(213.00,92.17)(4.547,4.000){2}{\rule{0.350pt}{0.400pt}}
\multiput(219.00,97.60)(0.774,0.468){5}{\rule{0.700pt}{0.113pt}}
\multiput(219.00,96.17)(4.547,4.000){2}{\rule{0.350pt}{0.400pt}}
\multiput(225.00,101.59)(0.710,0.477){7}{\rule{0.660pt}{0.115pt}}
\multiput(225.00,100.17)(5.630,5.000){2}{\rule{0.330pt}{0.400pt}}
\multiput(232.00,106.59)(0.491,0.482){9}{\rule{0.500pt}{0.116pt}}
\multiput(232.00,105.17)(4.962,6.000){2}{\rule{0.250pt}{0.400pt}}
\multiput(238.00,112.59)(0.491,0.482){9}{\rule{0.500pt}{0.116pt}}
\multiput(238.00,111.17)(4.962,6.000){2}{\rule{0.250pt}{0.400pt}}
\multiput(244.59,118.00)(0.482,0.581){9}{\rule{0.116pt}{0.567pt}}
\multiput(243.17,118.00)(6.000,5.824){2}{\rule{0.400pt}{0.283pt}}
\multiput(250.59,125.00)(0.482,0.671){9}{\rule{0.116pt}{0.633pt}}
\multiput(249.17,125.00)(6.000,6.685){2}{\rule{0.400pt}{0.317pt}}
\multiput(256.59,133.00)(0.482,0.581){9}{\rule{0.116pt}{0.567pt}}
\multiput(255.17,133.00)(6.000,5.824){2}{\rule{0.400pt}{0.283pt}}
\multiput(262.59,140.00)(0.485,0.569){11}{\rule{0.117pt}{0.557pt}}
\multiput(261.17,140.00)(7.000,6.844){2}{\rule{0.400pt}{0.279pt}}
\multiput(269.59,148.00)(0.482,0.671){9}{\rule{0.116pt}{0.633pt}}
\multiput(268.17,148.00)(6.000,6.685){2}{\rule{0.400pt}{0.317pt}}
\multiput(275.59,156.00)(0.482,0.671){9}{\rule{0.116pt}{0.633pt}}
\multiput(274.17,156.00)(6.000,6.685){2}{\rule{0.400pt}{0.317pt}}
\multiput(281.59,164.00)(0.482,0.762){9}{\rule{0.116pt}{0.700pt}}
\multiput(280.17,164.00)(6.000,7.547){2}{\rule{0.400pt}{0.350pt}}
\multiput(287.59,173.00)(0.482,0.671){9}{\rule{0.116pt}{0.633pt}}
\multiput(286.17,173.00)(6.000,6.685){2}{\rule{0.400pt}{0.317pt}}
\multiput(293.59,181.00)(0.482,0.671){9}{\rule{0.116pt}{0.633pt}}
\multiput(292.17,181.00)(6.000,6.685){2}{\rule{0.400pt}{0.317pt}}
\multiput(299.59,189.00)(0.482,0.671){9}{\rule{0.116pt}{0.633pt}}
\multiput(298.17,189.00)(6.000,6.685){2}{\rule{0.400pt}{0.317pt}}
\multiput(305.59,197.00)(0.485,0.569){11}{\rule{0.117pt}{0.557pt}}
\multiput(304.17,197.00)(7.000,6.844){2}{\rule{0.400pt}{0.279pt}}
\multiput(312.59,205.00)(0.482,0.581){9}{\rule{0.116pt}{0.567pt}}
\multiput(311.17,205.00)(6.000,5.824){2}{\rule{0.400pt}{0.283pt}}
\multiput(318.59,212.00)(0.482,0.581){9}{\rule{0.116pt}{0.567pt}}
\multiput(317.17,212.00)(6.000,5.824){2}{\rule{0.400pt}{0.283pt}}
\multiput(324.59,219.00)(0.482,0.581){9}{\rule{0.116pt}{0.567pt}}
\multiput(323.17,219.00)(6.000,5.824){2}{\rule{0.400pt}{0.283pt}}
\multiput(330.00,226.59)(0.491,0.482){9}{\rule{0.500pt}{0.116pt}}
\multiput(330.00,225.17)(4.962,6.000){2}{\rule{0.250pt}{0.400pt}}
\multiput(336.00,232.59)(0.491,0.482){9}{\rule{0.500pt}{0.116pt}}
\multiput(336.00,231.17)(4.962,6.000){2}{\rule{0.250pt}{0.400pt}}
\multiput(342.00,238.59)(0.581,0.482){9}{\rule{0.567pt}{0.116pt}}
\multiput(342.00,237.17)(5.824,6.000){2}{\rule{0.283pt}{0.400pt}}
\multiput(349.00,244.59)(0.599,0.477){7}{\rule{0.580pt}{0.115pt}}
\multiput(349.00,243.17)(4.796,5.000){2}{\rule{0.290pt}{0.400pt}}
\multiput(355.00,249.60)(0.774,0.468){5}{\rule{0.700pt}{0.113pt}}
\multiput(355.00,248.17)(4.547,4.000){2}{\rule{0.350pt}{0.400pt}}
\multiput(361.00,253.60)(0.774,0.468){5}{\rule{0.700pt}{0.113pt}}
\multiput(361.00,252.17)(4.547,4.000){2}{\rule{0.350pt}{0.400pt}}
\multiput(367.00,257.61)(1.132,0.447){3}{\rule{0.900pt}{0.108pt}}
\multiput(367.00,256.17)(4.132,3.000){2}{\rule{0.450pt}{0.400pt}}
\put(373,260.17){\rule{1.300pt}{0.400pt}}
\multiput(373.00,259.17)(3.302,2.000){2}{\rule{0.650pt}{0.400pt}}
\put(379,262.17){\rule{1.500pt}{0.400pt}}
\multiput(379.00,261.17)(3.887,2.000){2}{\rule{0.750pt}{0.400pt}}
\put(386,264.17){\rule{1.300pt}{0.400pt}}
\multiput(386.00,263.17)(3.302,2.000){2}{\rule{0.650pt}{0.400pt}}
\put(195.0,90.0){\rule[-0.200pt]{1.445pt}{0.400pt}}
\put(410,264.17){\rule{1.300pt}{0.400pt}}
\multiput(410.00,265.17)(3.302,-2.000){2}{\rule{0.650pt}{0.400pt}}
\put(416,262.17){\rule{1.500pt}{0.400pt}}
\multiput(416.00,263.17)(3.887,-2.000){2}{\rule{0.750pt}{0.400pt}}
\put(423,260.17){\rule{1.300pt}{0.400pt}}
\multiput(423.00,261.17)(3.302,-2.000){2}{\rule{0.650pt}{0.400pt}}
\multiput(429.00,258.95)(1.132,-0.447){3}{\rule{0.900pt}{0.108pt}}
\multiput(429.00,259.17)(4.132,-3.000){2}{\rule{0.450pt}{0.400pt}}
\multiput(435.00,255.94)(0.774,-0.468){5}{\rule{0.700pt}{0.113pt}}
\multiput(435.00,256.17)(4.547,-4.000){2}{\rule{0.350pt}{0.400pt}}
\multiput(441.00,251.94)(0.774,-0.468){5}{\rule{0.700pt}{0.113pt}}
\multiput(441.00,252.17)(4.547,-4.000){2}{\rule{0.350pt}{0.400pt}}
\multiput(447.00,247.93)(0.599,-0.477){7}{\rule{0.580pt}{0.115pt}}
\multiput(447.00,248.17)(4.796,-5.000){2}{\rule{0.290pt}{0.400pt}}
\multiput(453.00,242.93)(0.581,-0.482){9}{\rule{0.567pt}{0.116pt}}
\multiput(453.00,243.17)(5.824,-6.000){2}{\rule{0.283pt}{0.400pt}}
\multiput(460.00,236.93)(0.491,-0.482){9}{\rule{0.500pt}{0.116pt}}
\multiput(460.00,237.17)(4.962,-6.000){2}{\rule{0.250pt}{0.400pt}}
\multiput(466.00,230.93)(0.491,-0.482){9}{\rule{0.500pt}{0.116pt}}
\multiput(466.00,231.17)(4.962,-6.000){2}{\rule{0.250pt}{0.400pt}}
\multiput(472.59,223.65)(0.482,-0.581){9}{\rule{0.116pt}{0.567pt}}
\multiput(471.17,224.82)(6.000,-5.824){2}{\rule{0.400pt}{0.283pt}}
\multiput(478.59,216.65)(0.482,-0.581){9}{\rule{0.116pt}{0.567pt}}
\multiput(477.17,217.82)(6.000,-5.824){2}{\rule{0.400pt}{0.283pt}}
\multiput(484.59,209.65)(0.482,-0.581){9}{\rule{0.116pt}{0.567pt}}
\multiput(483.17,210.82)(6.000,-5.824){2}{\rule{0.400pt}{0.283pt}}
\multiput(490.59,202.69)(0.485,-0.569){11}{\rule{0.117pt}{0.557pt}}
\multiput(489.17,203.84)(7.000,-6.844){2}{\rule{0.400pt}{0.279pt}}
\multiput(497.59,194.37)(0.482,-0.671){9}{\rule{0.116pt}{0.633pt}}
\multiput(496.17,195.69)(6.000,-6.685){2}{\rule{0.400pt}{0.317pt}}
\multiput(503.59,186.37)(0.482,-0.671){9}{\rule{0.116pt}{0.633pt}}
\multiput(502.17,187.69)(6.000,-6.685){2}{\rule{0.400pt}{0.317pt}}
\multiput(509.59,178.37)(0.482,-0.671){9}{\rule{0.116pt}{0.633pt}}
\multiput(508.17,179.69)(6.000,-6.685){2}{\rule{0.400pt}{0.317pt}}
\multiput(515.59,170.09)(0.482,-0.762){9}{\rule{0.116pt}{0.700pt}}
\multiput(514.17,171.55)(6.000,-7.547){2}{\rule{0.400pt}{0.350pt}}
\multiput(521.59,161.37)(0.482,-0.671){9}{\rule{0.116pt}{0.633pt}}
\multiput(520.17,162.69)(6.000,-6.685){2}{\rule{0.400pt}{0.317pt}}
\multiput(527.59,153.37)(0.482,-0.671){9}{\rule{0.116pt}{0.633pt}}
\multiput(526.17,154.69)(6.000,-6.685){2}{\rule{0.400pt}{0.317pt}}
\multiput(533.59,145.69)(0.485,-0.569){11}{\rule{0.117pt}{0.557pt}}
\multiput(532.17,146.84)(7.000,-6.844){2}{\rule{0.400pt}{0.279pt}}
\multiput(540.59,137.65)(0.482,-0.581){9}{\rule{0.116pt}{0.567pt}}
\multiput(539.17,138.82)(6.000,-5.824){2}{\rule{0.400pt}{0.283pt}}
\multiput(546.59,130.37)(0.482,-0.671){9}{\rule{0.116pt}{0.633pt}}
\multiput(545.17,131.69)(6.000,-6.685){2}{\rule{0.400pt}{0.317pt}}
\multiput(552.59,122.65)(0.482,-0.581){9}{\rule{0.116pt}{0.567pt}}
\multiput(551.17,123.82)(6.000,-5.824){2}{\rule{0.400pt}{0.283pt}}
\multiput(558.00,116.93)(0.491,-0.482){9}{\rule{0.500pt}{0.116pt}}
\multiput(558.00,117.17)(4.962,-6.000){2}{\rule{0.250pt}{0.400pt}}
\multiput(564.00,110.93)(0.491,-0.482){9}{\rule{0.500pt}{0.116pt}}
\multiput(564.00,111.17)(4.962,-6.000){2}{\rule{0.250pt}{0.400pt}}
\multiput(570.00,104.93)(0.710,-0.477){7}{\rule{0.660pt}{0.115pt}}
\multiput(570.00,105.17)(5.630,-5.000){2}{\rule{0.330pt}{0.400pt}}
\multiput(577.00,99.94)(0.774,-0.468){5}{\rule{0.700pt}{0.113pt}}
\multiput(577.00,100.17)(4.547,-4.000){2}{\rule{0.350pt}{0.400pt}}
\multiput(583.00,95.94)(0.774,-0.468){5}{\rule{0.700pt}{0.113pt}}
\multiput(583.00,96.17)(4.547,-4.000){2}{\rule{0.350pt}{0.400pt}}
\put(589,91.17){\rule{1.300pt}{0.400pt}}
\multiput(589.00,92.17)(3.302,-2.000){2}{\rule{0.650pt}{0.400pt}}
\put(595,89.67){\rule{1.445pt}{0.400pt}}
\multiput(595.00,90.17)(3.000,-1.000){2}{\rule{0.723pt}{0.400pt}}
\put(392.0,266.0){\rule[-0.200pt]{4.336pt}{0.400pt}}
\put(607,89.67){\rule{1.686pt}{0.400pt}}
\multiput(607.00,89.17)(3.500,1.000){2}{\rule{0.843pt}{0.400pt}}
\multiput(614.00,91.61)(1.132,0.447){3}{\rule{0.900pt}{0.108pt}}
\multiput(614.00,90.17)(4.132,3.000){2}{\rule{0.450pt}{0.400pt}}
\multiput(620.00,94.60)(0.774,0.468){5}{\rule{0.700pt}{0.113pt}}
\multiput(620.00,93.17)(4.547,4.000){2}{\rule{0.350pt}{0.400pt}}
\multiput(626.59,98.00)(0.482,0.581){9}{\rule{0.116pt}{0.567pt}}
\multiput(625.17,98.00)(6.000,5.824){2}{\rule{0.400pt}{0.283pt}}
\multiput(632.59,105.00)(0.482,0.671){9}{\rule{0.116pt}{0.633pt}}
\multiput(631.17,105.00)(6.000,6.685){2}{\rule{0.400pt}{0.317pt}}
\multiput(638.59,113.00)(0.482,0.852){9}{\rule{0.116pt}{0.767pt}}
\multiput(637.17,113.00)(6.000,8.409){2}{\rule{0.400pt}{0.383pt}}
\multiput(644.59,123.00)(0.485,0.874){11}{\rule{0.117pt}{0.786pt}}
\multiput(643.17,123.00)(7.000,10.369){2}{\rule{0.400pt}{0.393pt}}
\multiput(651.59,135.00)(0.482,1.214){9}{\rule{0.116pt}{1.033pt}}
\multiput(650.17,135.00)(6.000,11.855){2}{\rule{0.400pt}{0.517pt}}
\multiput(657.59,149.00)(0.482,1.485){9}{\rule{0.116pt}{1.233pt}}
\multiput(656.17,149.00)(6.000,14.440){2}{\rule{0.400pt}{0.617pt}}
\multiput(663.59,166.00)(0.482,1.756){9}{\rule{0.116pt}{1.433pt}}
\multiput(662.17,166.00)(6.000,17.025){2}{\rule{0.400pt}{0.717pt}}
\multiput(669.59,186.00)(0.482,1.937){9}{\rule{0.116pt}{1.567pt}}
\multiput(668.17,186.00)(6.000,18.748){2}{\rule{0.400pt}{0.783pt}}
\multiput(675.59,208.00)(0.482,2.208){9}{\rule{0.116pt}{1.767pt}}
\multiput(674.17,208.00)(6.000,21.333){2}{\rule{0.400pt}{0.883pt}}
\multiput(681.59,233.00)(0.485,2.094){11}{\rule{0.117pt}{1.700pt}}
\multiput(680.17,233.00)(7.000,24.472){2}{\rule{0.400pt}{0.850pt}}
\multiput(688.59,261.00)(0.482,2.841){9}{\rule{0.116pt}{2.233pt}}
\multiput(687.17,261.00)(6.000,27.365){2}{\rule{0.400pt}{1.117pt}}
\multiput(694.59,293.00)(0.482,3.022){9}{\rule{0.116pt}{2.367pt}}
\multiput(693.17,293.00)(6.000,29.088){2}{\rule{0.400pt}{1.183pt}}
\multiput(700.59,327.00)(0.482,3.474){9}{\rule{0.116pt}{2.700pt}}
\multiput(699.17,327.00)(6.000,33.396){2}{\rule{0.400pt}{1.350pt}}
\put(601.0,90.0){\rule[-0.200pt]{1.445pt}{0.400pt}}
\end{picture}

%% file: Elastpls.tex
\setlength{\unitlength}{0.240900pt}
\ifx\plotpoint\undefined\newsavebox{\plotpoint}\fi
\begin{picture}(750,450)(0,0)
\font\gnuplot=cmr10 at 12pt
\gnuplot
\sbox{\plotpoint}{\rule[-0.200pt]{0.400pt}{0.400pt}}%
\put(84,19){\makebox(0,0)[r]{-4}}
\put(76.0,19.0){\rule[-0.200pt]{4.818pt}{0.400pt}}
\put(84,90){\makebox(0,0)[r]{-3}}
\put(76.0,90.0){\rule[-0.200pt]{4.818pt}{0.400pt}}
\put(84,160){\makebox(0,0)[r]{-2}}
\put(76.0,160.0){\rule[-0.200pt]{4.818pt}{0.400pt}}
\put(84,231){\makebox(0,0)[r]{-1}}
\put(76.0,231.0){\rule[-0.200pt]{4.818pt}{0.400pt}}
\put(84,302){\makebox(0,0)[r]{0}}
\put(76.0,302.0){\rule[-0.200pt]{4.818pt}{0.400pt}}
\put(84,372){\makebox(0,0)[r]{1}}
\put(76.0,372.0){\rule[-0.200pt]{4.818pt}{0.400pt}}
\put(84,443){\makebox(0,0)[r]{2}}
\put(76.0,443.0){\rule[-0.200pt]{4.818pt}{0.400pt}}
\put(96.0,-1.0){\rule[-0.200pt]{0.400pt}{4.818pt}}
\put(218.0,-1.0){\rule[-0.200pt]{0.400pt}{4.818pt}}
\put(340.0,-1.0){\rule[-0.200pt]{0.400pt}{4.818pt}}
\put(462.0,-1.0){\rule[-0.200pt]{0.400pt}{4.818pt}}
\put(584.0,-1.0){\rule[-0.200pt]{0.400pt}{4.818pt}}
\put(706.0,-1.0){\rule[-0.200pt]{0.400pt}{4.818pt}}
\put(96.0,19.0){\rule[-0.200pt]{146.949pt}{0.400pt}}
\put(706.0,19.0){\rule[-0.200pt]{0.400pt}{102.142pt}}
\put(96.0,443.0){\rule[-0.200pt]{146.949pt}{0.400pt}}
\put(737,266){\makebox(0,0)[l]{$x$}}
\put(127,478){\makebox(0,0)[l]{$p_{\mathrm{el}}(x-\frac{1}{2})$}}
\put(218,-22){\makebox(0,0)[l]{0.2}}
\put(340,-22){\makebox(0,0)[l]{0.4}}
\put(462,-22){\makebox(0,0)[l]{0.6}}
\put(584,-22){\makebox(0,0)[l]{0.8}}
\put(706,-22){\makebox(0,0)[l]{1}}
\put(96,-22){\makebox(0,0)[l]{0}}
\put(96.0,19.0){\rule[-0.200pt]{0.400pt}{102.142pt}}
\put(96,302){\vector(1,0){641}}
\put(96,19){\vector(0,1){459}}
\put(96,391){\usebox{\plotpoint}}
\put(102,390.67){\rule{1.445pt}{0.400pt}}
\multiput(102.00,390.17)(3.000,1.000){2}{\rule{0.723pt}{0.400pt}}
\put(108,391.67){\rule{1.445pt}{0.400pt}}
\multiput(108.00,391.17)(3.000,1.000){2}{\rule{0.723pt}{0.400pt}}
\put(114,392.67){\rule{1.686pt}{0.400pt}}
\multiput(114.00,392.17)(3.500,1.000){2}{\rule{0.843pt}{0.400pt}}
\put(121,393.67){\rule{1.445pt}{0.400pt}}
\multiput(121.00,393.17)(3.000,1.000){2}{\rule{0.723pt}{0.400pt}}
\put(96.0,391.0){\rule[-0.200pt]{1.445pt}{0.400pt}}
\put(133,394.67){\rule{1.445pt}{0.400pt}}
\multiput(133.00,394.17)(3.000,1.000){2}{\rule{0.723pt}{0.400pt}}
\put(139,395.67){\rule{1.445pt}{0.400pt}}
\multiput(139.00,395.17)(3.000,1.000){2}{\rule{0.723pt}{0.400pt}}
\put(145,396.67){\rule{1.445pt}{0.400pt}}
\multiput(145.00,396.17)(3.000,1.000){2}{\rule{0.723pt}{0.400pt}}
\put(151,397.67){\rule{1.686pt}{0.400pt}}
\multiput(151.00,397.17)(3.500,1.000){2}{\rule{0.843pt}{0.400pt}}
\put(158,398.67){\rule{1.445pt}{0.400pt}}
\multiput(158.00,398.17)(3.000,1.000){2}{\rule{0.723pt}{0.400pt}}
\put(164,399.67){\rule{1.445pt}{0.400pt}}
\multiput(164.00,399.17)(3.000,1.000){2}{\rule{0.723pt}{0.400pt}}
\put(170,400.67){\rule{1.445pt}{0.400pt}}
\multiput(170.00,400.17)(3.000,1.000){2}{\rule{0.723pt}{0.400pt}}
\put(176,401.67){\rule{1.445pt}{0.400pt}}
\multiput(176.00,401.17)(3.000,1.000){2}{\rule{0.723pt}{0.400pt}}
\put(182,402.67){\rule{1.445pt}{0.400pt}}
\multiput(182.00,402.17)(3.000,1.000){2}{\rule{0.723pt}{0.400pt}}
\put(188,403.67){\rule{1.686pt}{0.400pt}}
\multiput(188.00,403.17)(3.500,1.000){2}{\rule{0.843pt}{0.400pt}}
\put(195,404.67){\rule{1.445pt}{0.400pt}}
\multiput(195.00,404.17)(3.000,1.000){2}{\rule{0.723pt}{0.400pt}}
\put(201,405.67){\rule{1.445pt}{0.400pt}}
\multiput(201.00,405.17)(3.000,1.000){2}{\rule{0.723pt}{0.400pt}}
\put(207,406.67){\rule{1.445pt}{0.400pt}}
\multiput(207.00,406.17)(3.000,1.000){2}{\rule{0.723pt}{0.400pt}}
\put(213,407.67){\rule{1.445pt}{0.400pt}}
\multiput(213.00,407.17)(3.000,1.000){2}{\rule{0.723pt}{0.400pt}}
\put(219,408.67){\rule{1.445pt}{0.400pt}}
\multiput(219.00,408.17)(3.000,1.000){2}{\rule{0.723pt}{0.400pt}}
\put(225,409.67){\rule{1.686pt}{0.400pt}}
\multiput(225.00,409.17)(3.500,1.000){2}{\rule{0.843pt}{0.400pt}}
\put(232,410.67){\rule{1.445pt}{0.400pt}}
\multiput(232.00,410.17)(3.000,1.000){2}{\rule{0.723pt}{0.400pt}}
\put(238,411.67){\rule{1.445pt}{0.400pt}}
\multiput(238.00,411.17)(3.000,1.000){2}{\rule{0.723pt}{0.400pt}}
\put(244,412.67){\rule{1.445pt}{0.400pt}}
\multiput(244.00,412.17)(3.000,1.000){2}{\rule{0.723pt}{0.400pt}}
\put(250,413.67){\rule{1.445pt}{0.400pt}}
\multiput(250.00,413.17)(3.000,1.000){2}{\rule{0.723pt}{0.400pt}}
\put(256,414.67){\rule{1.445pt}{0.400pt}}
\multiput(256.00,414.17)(3.000,1.000){2}{\rule{0.723pt}{0.400pt}}
\put(262,415.67){\rule{1.686pt}{0.400pt}}
\multiput(262.00,415.17)(3.500,1.000){2}{\rule{0.843pt}{0.400pt}}
\put(269,416.67){\rule{1.445pt}{0.400pt}}
\multiput(269.00,416.17)(3.000,1.000){2}{\rule{0.723pt}{0.400pt}}
\put(275,417.67){\rule{1.445pt}{0.400pt}}
\multiput(275.00,417.17)(3.000,1.000){2}{\rule{0.723pt}{0.400pt}}
\put(281,418.67){\rule{1.445pt}{0.400pt}}
\multiput(281.00,418.17)(3.000,1.000){2}{\rule{0.723pt}{0.400pt}}
\put(287,419.67){\rule{1.445pt}{0.400pt}}
\multiput(287.00,419.17)(3.000,1.000){2}{\rule{0.723pt}{0.400pt}}
\put(293,420.67){\rule{1.445pt}{0.400pt}}
\multiput(293.00,420.17)(3.000,1.000){2}{\rule{0.723pt}{0.400pt}}
\put(299,421.67){\rule{1.445pt}{0.400pt}}
\multiput(299.00,421.17)(3.000,1.000){2}{\rule{0.723pt}{0.400pt}}
\put(305,422.67){\rule{1.686pt}{0.400pt}}
\multiput(305.00,422.17)(3.500,1.000){2}{\rule{0.843pt}{0.400pt}}
\put(312,423.67){\rule{1.445pt}{0.400pt}}
\multiput(312.00,423.17)(3.000,1.000){2}{\rule{0.723pt}{0.400pt}}
\put(318,424.67){\rule{1.445pt}{0.400pt}}
\multiput(318.00,424.17)(3.000,1.000){2}{\rule{0.723pt}{0.400pt}}
\put(127.0,395.0){\rule[-0.200pt]{1.445pt}{0.400pt}}
\put(330,425.67){\rule{1.445pt}{0.400pt}}
\multiput(330.00,425.17)(3.000,1.000){2}{\rule{0.723pt}{0.400pt}}
\put(336,426.67){\rule{1.445pt}{0.400pt}}
\multiput(336.00,426.17)(3.000,1.000){2}{\rule{0.723pt}{0.400pt}}
\put(324.0,426.0){\rule[-0.200pt]{1.445pt}{0.400pt}}
\put(349,427.67){\rule{1.445pt}{0.400pt}}
\multiput(349.00,427.17)(3.000,1.000){2}{\rule{0.723pt}{0.400pt}}
\put(342.0,428.0){\rule[-0.200pt]{1.686pt}{0.400pt}}
\put(361,428.67){\rule{1.445pt}{0.400pt}}
\multiput(361.00,428.17)(3.000,1.000){2}{\rule{0.723pt}{0.400pt}}
\put(355.0,429.0){\rule[-0.200pt]{1.445pt}{0.400pt}}
\put(379,429.67){\rule{1.686pt}{0.400pt}}
\multiput(379.00,429.17)(3.500,1.000){2}{\rule{0.843pt}{0.400pt}}
\put(367.0,430.0){\rule[-0.200pt]{2.891pt}{0.400pt}}
\put(416,429.67){\rule{1.686pt}{0.400pt}}
\multiput(416.00,430.17)(3.500,-1.000){2}{\rule{0.843pt}{0.400pt}}
\put(386.0,431.0){\rule[-0.200pt]{7.227pt}{0.400pt}}
\put(435,428.67){\rule{1.445pt}{0.400pt}}
\multiput(435.00,429.17)(3.000,-1.000){2}{\rule{0.723pt}{0.400pt}}
\put(423.0,430.0){\rule[-0.200pt]{2.891pt}{0.400pt}}
\put(447,427.67){\rule{1.445pt}{0.400pt}}
\multiput(447.00,428.17)(3.000,-1.000){2}{\rule{0.723pt}{0.400pt}}
\put(441.0,429.0){\rule[-0.200pt]{1.445pt}{0.400pt}}
\put(460,426.67){\rule{1.445pt}{0.400pt}}
\multiput(460.00,427.17)(3.000,-1.000){2}{\rule{0.723pt}{0.400pt}}
\put(466,425.67){\rule{1.445pt}{0.400pt}}
\multiput(466.00,426.17)(3.000,-1.000){2}{\rule{0.723pt}{0.400pt}}
\put(453.0,428.0){\rule[-0.200pt]{1.686pt}{0.400pt}}
\put(478,424.67){\rule{1.445pt}{0.400pt}}
\multiput(478.00,425.17)(3.000,-1.000){2}{\rule{0.723pt}{0.400pt}}
\put(484,423.67){\rule{1.445pt}{0.400pt}}
\multiput(484.00,424.17)(3.000,-1.000){2}{\rule{0.723pt}{0.400pt}}
\put(490,422.67){\rule{1.686pt}{0.400pt}}
\multiput(490.00,423.17)(3.500,-1.000){2}{\rule{0.843pt}{0.400pt}}
\put(497,421.67){\rule{1.445pt}{0.400pt}}
\multiput(497.00,422.17)(3.000,-1.000){2}{\rule{0.723pt}{0.400pt}}
\put(503,420.67){\rule{1.445pt}{0.400pt}}
\multiput(503.00,421.17)(3.000,-1.000){2}{\rule{0.723pt}{0.400pt}}
\put(509,419.67){\rule{1.445pt}{0.400pt}}
\multiput(509.00,420.17)(3.000,-1.000){2}{\rule{0.723pt}{0.400pt}}
\put(515,418.67){\rule{1.445pt}{0.400pt}}
\multiput(515.00,419.17)(3.000,-1.000){2}{\rule{0.723pt}{0.400pt}}
\put(521,417.67){\rule{1.445pt}{0.400pt}}
\multiput(521.00,418.17)(3.000,-1.000){2}{\rule{0.723pt}{0.400pt}}
\put(527,416.67){\rule{1.445pt}{0.400pt}}
\multiput(527.00,417.17)(3.000,-1.000){2}{\rule{0.723pt}{0.400pt}}
\put(533,415.67){\rule{1.686pt}{0.400pt}}
\multiput(533.00,416.17)(3.500,-1.000){2}{\rule{0.843pt}{0.400pt}}
\put(540,414.67){\rule{1.445pt}{0.400pt}}
\multiput(540.00,415.17)(3.000,-1.000){2}{\rule{0.723pt}{0.400pt}}
\put(546,413.67){\rule{1.445pt}{0.400pt}}
\multiput(546.00,414.17)(3.000,-1.000){2}{\rule{0.723pt}{0.400pt}}
\put(552,412.67){\rule{1.445pt}{0.400pt}}
\multiput(552.00,413.17)(3.000,-1.000){2}{\rule{0.723pt}{0.400pt}}
\put(558,411.67){\rule{1.445pt}{0.400pt}}
\multiput(558.00,412.17)(3.000,-1.000){2}{\rule{0.723pt}{0.400pt}}
\put(564,410.67){\rule{1.445pt}{0.400pt}}
\multiput(564.00,411.17)(3.000,-1.000){2}{\rule{0.723pt}{0.400pt}}
\put(570,409.67){\rule{1.686pt}{0.400pt}}
\multiput(570.00,410.17)(3.500,-1.000){2}{\rule{0.843pt}{0.400pt}}
\put(577,408.67){\rule{1.445pt}{0.400pt}}
\multiput(577.00,409.17)(3.000,-1.000){2}{\rule{0.723pt}{0.400pt}}
\put(583,407.67){\rule{1.445pt}{0.400pt}}
\multiput(583.00,408.17)(3.000,-1.000){2}{\rule{0.723pt}{0.400pt}}
\put(589,406.67){\rule{1.445pt}{0.400pt}}
\multiput(589.00,407.17)(3.000,-1.000){2}{\rule{0.723pt}{0.400pt}}
\put(595,405.67){\rule{1.445pt}{0.400pt}}
\multiput(595.00,406.17)(3.000,-1.000){2}{\rule{0.723pt}{0.400pt}}
\put(601,404.67){\rule{1.445pt}{0.400pt}}
\multiput(601.00,405.17)(3.000,-1.000){2}{\rule{0.723pt}{0.400pt}}
\put(607,403.67){\rule{1.686pt}{0.400pt}}
\multiput(607.00,404.17)(3.500,-1.000){2}{\rule{0.843pt}{0.400pt}}
\put(614,402.67){\rule{1.445pt}{0.400pt}}
\multiput(614.00,403.17)(3.000,-1.000){2}{\rule{0.723pt}{0.400pt}}
\put(620,401.67){\rule{1.445pt}{0.400pt}}
\multiput(620.00,402.17)(3.000,-1.000){2}{\rule{0.723pt}{0.400pt}}
\put(626,400.67){\rule{1.445pt}{0.400pt}}
\multiput(626.00,401.17)(3.000,-1.000){2}{\rule{0.723pt}{0.400pt}}
\put(632,399.67){\rule{1.445pt}{0.400pt}}
\multiput(632.00,400.17)(3.000,-1.000){2}{\rule{0.723pt}{0.400pt}}
\put(638,398.67){\rule{1.445pt}{0.400pt}}
\multiput(638.00,399.17)(3.000,-1.000){2}{\rule{0.723pt}{0.400pt}}
\put(644,397.67){\rule{1.686pt}{0.400pt}}
\multiput(644.00,398.17)(3.500,-1.000){2}{\rule{0.843pt}{0.400pt}}
\put(651,396.67){\rule{1.445pt}{0.400pt}}
\multiput(651.00,397.17)(3.000,-1.000){2}{\rule{0.723pt}{0.400pt}}
\put(657,395.67){\rule{1.445pt}{0.400pt}}
\multiput(657.00,396.17)(3.000,-1.000){2}{\rule{0.723pt}{0.400pt}}
\put(663,394.67){\rule{1.445pt}{0.400pt}}
\multiput(663.00,395.17)(3.000,-1.000){2}{\rule{0.723pt}{0.400pt}}
\put(472.0,426.0){\rule[-0.200pt]{1.445pt}{0.400pt}}
\put(675,393.67){\rule{1.445pt}{0.400pt}}
\multiput(675.00,394.17)(3.000,-1.000){2}{\rule{0.723pt}{0.400pt}}
\put(681,392.67){\rule{1.686pt}{0.400pt}}
\multiput(681.00,393.17)(3.500,-1.000){2}{\rule{0.843pt}{0.400pt}}
\put(688,391.67){\rule{1.445pt}{0.400pt}}
\multiput(688.00,392.17)(3.000,-1.000){2}{\rule{0.723pt}{0.400pt}}
\put(694,390.67){\rule{1.445pt}{0.400pt}}
\multiput(694.00,391.17)(3.000,-1.000){2}{\rule{0.723pt}{0.400pt}}
\put(669.0,395.0){\rule[-0.200pt]{1.445pt}{0.400pt}}
\put(700.0,391.0){\rule[-0.200pt]{1.445pt}{0.400pt}}
\sbox{\plotpoint}{\rule[-0.500pt]{1.000pt}{1.000pt}}%
\put(96,380){\usebox{\plotpoint}}
\put(96.00,380.00){\usebox{\plotpoint}}
\multiput(102,381)(20.756,0.000){0}{\usebox{\plotpoint}}
\multiput(108,381)(20.473,3.412){0}{\usebox{\plotpoint}}
\put(116.59,382.00){\usebox{\plotpoint}}
\multiput(121,382)(20.473,3.412){0}{\usebox{\plotpoint}}
\multiput(127,383)(20.473,3.412){0}{\usebox{\plotpoint}}
\put(137.18,384.00){\usebox{\plotpoint}}
\multiput(139,384)(20.473,3.412){0}{\usebox{\plotpoint}}
\multiput(145,385)(20.473,3.412){0}{\usebox{\plotpoint}}
\put(157.70,386.96){\usebox{\plotpoint}}
\multiput(158,387)(20.756,0.000){0}{\usebox{\plotpoint}}
\multiput(164,387)(20.473,3.412){0}{\usebox{\plotpoint}}
\multiput(170,388)(20.473,3.412){0}{\usebox{\plotpoint}}
\put(178.26,389.38){\usebox{\plotpoint}}
\multiput(182,390)(20.473,3.412){0}{\usebox{\plotpoint}}
\multiput(188,391)(20.547,2.935){0}{\usebox{\plotpoint}}
\put(198.76,392.63){\usebox{\plotpoint}}
\multiput(201,393)(20.473,3.412){0}{\usebox{\plotpoint}}
\multiput(207,394)(19.690,6.563){0}{\usebox{\plotpoint}}
\put(218.99,397.00){\usebox{\plotpoint}}
\multiput(219,397)(20.473,3.412){0}{\usebox{\plotpoint}}
\multiput(225,398)(20.547,2.935){0}{\usebox{\plotpoint}}
\multiput(232,399)(19.690,6.563){0}{\usebox{\plotpoint}}
\put(239.25,401.21){\usebox{\plotpoint}}
\multiput(244,402)(19.690,6.563){0}{\usebox{\plotpoint}}
\multiput(250,404)(20.473,3.412){0}{\usebox{\plotpoint}}
\put(259.48,405.58){\usebox{\plotpoint}}
\multiput(262,406)(19.957,5.702){0}{\usebox{\plotpoint}}
\multiput(269,408)(20.473,3.412){0}{\usebox{\plotpoint}}
\put(279.59,410.53){\usebox{\plotpoint}}
\multiput(281,411)(20.473,3.412){0}{\usebox{\plotpoint}}
\multiput(287,412)(19.690,6.563){0}{\usebox{\plotpoint}}
\multiput(293,414)(20.473,3.412){0}{\usebox{\plotpoint}}
\put(299.74,415.25){\usebox{\plotpoint}}
\multiput(305,417)(20.547,2.935){0}{\usebox{\plotpoint}}
\multiput(312,418)(19.690,6.563){0}{\usebox{\plotpoint}}
\put(319.79,420.30){\usebox{\plotpoint}}
\multiput(324,421)(19.690,6.563){0}{\usebox{\plotpoint}}
\multiput(330,423)(20.473,3.412){0}{\usebox{\plotpoint}}
\put(340.03,424.67){\usebox{\plotpoint}}
\multiput(342,425)(20.547,2.935){0}{\usebox{\plotpoint}}
\multiput(349,426)(20.473,3.412){0}{\usebox{\plotpoint}}
\put(360.53,427.92){\usebox{\plotpoint}}
\multiput(361,428)(20.473,3.412){0}{\usebox{\plotpoint}}
\multiput(367,429)(20.756,0.000){0}{\usebox{\plotpoint}}
\multiput(373,429)(20.473,3.412){0}{\usebox{\plotpoint}}
\put(381.11,430.00){\usebox{\plotpoint}}
\multiput(386,430)(20.473,3.412){0}{\usebox{\plotpoint}}
\multiput(392,431)(20.756,0.000){0}{\usebox{\plotpoint}}
\put(401.78,431.00){\usebox{\plotpoint}}
\multiput(404,431)(20.756,0.000){0}{\usebox{\plotpoint}}
\multiput(410,431)(20.473,-3.412){0}{\usebox{\plotpoint}}
\put(422.46,430.00){\usebox{\plotpoint}}
\multiput(423,430)(20.473,-3.412){0}{\usebox{\plotpoint}}
\multiput(429,429)(20.756,0.000){0}{\usebox{\plotpoint}}
\multiput(435,429)(20.473,-3.412){0}{\usebox{\plotpoint}}
\put(443.02,427.66){\usebox{\plotpoint}}
\multiput(447,427)(20.473,-3.412){0}{\usebox{\plotpoint}}
\multiput(453,426)(20.547,-2.935){0}{\usebox{\plotpoint}}
\put(463.52,424.41){\usebox{\plotpoint}}
\multiput(466,424)(20.473,-3.412){0}{\usebox{\plotpoint}}
\multiput(472,423)(19.690,-6.563){0}{\usebox{\plotpoint}}
\put(483.75,420.04){\usebox{\plotpoint}}
\multiput(484,420)(19.690,-6.563){0}{\usebox{\plotpoint}}
\multiput(490,418)(20.547,-2.935){0}{\usebox{\plotpoint}}
\multiput(497,417)(19.690,-6.563){0}{\usebox{\plotpoint}}
\put(503.77,414.87){\usebox{\plotpoint}}
\multiput(509,414)(19.690,-6.563){0}{\usebox{\plotpoint}}
\multiput(515,412)(20.473,-3.412){0}{\usebox{\plotpoint}}
\put(523.89,410.04){\usebox{\plotpoint}}
\multiput(527,409)(20.473,-3.412){0}{\usebox{\plotpoint}}
\multiput(533,408)(19.957,-5.702){0}{\usebox{\plotpoint}}
\put(544.06,405.32){\usebox{\plotpoint}}
\multiput(546,405)(20.473,-3.412){0}{\usebox{\plotpoint}}
\multiput(552,404)(19.690,-6.563){0}{\usebox{\plotpoint}}
\multiput(558,402)(20.473,-3.412){0}{\usebox{\plotpoint}}
\put(564.28,400.91){\usebox{\plotpoint}}
\multiput(570,399)(20.547,-2.935){0}{\usebox{\plotpoint}}
\multiput(577,398)(20.473,-3.412){0}{\usebox{\plotpoint}}
\put(584.55,396.74){\usebox{\plotpoint}}
\multiput(589,396)(19.690,-6.563){0}{\usebox{\plotpoint}}
\multiput(595,394)(20.473,-3.412){0}{\usebox{\plotpoint}}
\put(604.79,392.37){\usebox{\plotpoint}}
\multiput(607,392)(20.547,-2.935){0}{\usebox{\plotpoint}}
\multiput(614,391)(20.473,-3.412){0}{\usebox{\plotpoint}}
\put(625.29,389.12){\usebox{\plotpoint}}
\multiput(626,389)(20.473,-3.412){0}{\usebox{\plotpoint}}
\multiput(632,388)(20.473,-3.412){0}{\usebox{\plotpoint}}
\multiput(638,387)(20.756,0.000){0}{\usebox{\plotpoint}}
\put(645.85,386.74){\usebox{\plotpoint}}
\multiput(651,386)(20.473,-3.412){0}{\usebox{\plotpoint}}
\multiput(657,385)(20.473,-3.412){0}{\usebox{\plotpoint}}
\put(666.39,384.00){\usebox{\plotpoint}}
\multiput(669,384)(20.473,-3.412){0}{\usebox{\plotpoint}}
\multiput(675,383)(20.473,-3.412){0}{\usebox{\plotpoint}}
\put(686.98,382.00){\usebox{\plotpoint}}
\multiput(688,382)(20.473,-3.412){0}{\usebox{\plotpoint}}
\multiput(694,381)(20.756,0.000){0}{\usebox{\plotpoint}}
\multiput(700,381)(20.473,-3.412){0}{\usebox{\plotpoint}}
\put(706,380){\usebox{\plotpoint}}
\sbox{\plotpoint}{\rule[-0.600pt]{1.200pt}{1.200pt}}%
\put(96,373){\usebox{\plotpoint}}
\put(176,371.01){\rule{1.445pt}{1.200pt}}
\multiput(176.00,370.51)(3.000,1.000){2}{\rule{0.723pt}{1.200pt}}
\put(96.0,373.0){\rule[-0.600pt]{19.272pt}{1.200pt}}
\put(201,372.01){\rule{1.445pt}{1.200pt}}
\multiput(201.00,371.51)(3.000,1.000){2}{\rule{0.723pt}{1.200pt}}
\put(182.0,374.0){\rule[-0.600pt]{4.577pt}{1.200pt}}
\put(219,373.01){\rule{1.445pt}{1.200pt}}
\multiput(219.00,372.51)(3.000,1.000){2}{\rule{0.723pt}{1.200pt}}
\put(207.0,375.0){\rule[-0.600pt]{2.891pt}{1.200pt}}
\put(232,374.01){\rule{1.445pt}{1.200pt}}
\multiput(232.00,373.51)(3.000,1.000){2}{\rule{0.723pt}{1.200pt}}
\put(225.0,376.0){\rule[-0.600pt]{1.686pt}{1.200pt}}
\put(244,375.01){\rule{1.445pt}{1.200pt}}
\multiput(244.00,374.51)(3.000,1.000){2}{\rule{0.723pt}{1.200pt}}
\put(250,376.01){\rule{1.445pt}{1.200pt}}
\multiput(250.00,375.51)(3.000,1.000){2}{\rule{0.723pt}{1.200pt}}
\put(256,377.01){\rule{1.445pt}{1.200pt}}
\multiput(256.00,376.51)(3.000,1.000){2}{\rule{0.723pt}{1.200pt}}
\put(262,378.01){\rule{1.686pt}{1.200pt}}
\multiput(262.00,377.51)(3.500,1.000){2}{\rule{0.843pt}{1.200pt}}
\put(269,379.01){\rule{1.445pt}{1.200pt}}
\multiput(269.00,378.51)(3.000,1.000){2}{\rule{0.723pt}{1.200pt}}
\put(275,380.51){\rule{1.445pt}{1.200pt}}
\multiput(275.00,379.51)(3.000,2.000){2}{\rule{0.723pt}{1.200pt}}
\put(281,382.01){\rule{1.445pt}{1.200pt}}
\multiput(281.00,381.51)(3.000,1.000){2}{\rule{0.723pt}{1.200pt}}
\put(287,383.51){\rule{1.445pt}{1.200pt}}
\multiput(287.00,382.51)(3.000,2.000){2}{\rule{0.723pt}{1.200pt}}
\put(293,385.51){\rule{1.445pt}{1.200pt}}
\multiput(293.00,384.51)(3.000,2.000){2}{\rule{0.723pt}{1.200pt}}
\put(299,387.51){\rule{1.445pt}{1.200pt}}
\multiput(299.00,386.51)(3.000,2.000){2}{\rule{0.723pt}{1.200pt}}
\put(305,389.51){\rule{1.686pt}{1.200pt}}
\multiput(305.00,388.51)(3.500,2.000){2}{\rule{0.843pt}{1.200pt}}
\put(312,392.01){\rule{1.445pt}{1.200pt}}
\multiput(312.00,390.51)(3.000,3.000){2}{\rule{0.723pt}{1.200pt}}
\put(318,395.01){\rule{1.445pt}{1.200pt}}
\multiput(318.00,393.51)(3.000,3.000){2}{\rule{0.723pt}{1.200pt}}
\put(324,398.01){\rule{1.445pt}{1.200pt}}
\multiput(324.00,396.51)(3.000,3.000){2}{\rule{0.723pt}{1.200pt}}
\put(330,401.01){\rule{1.445pt}{1.200pt}}
\multiput(330.00,399.51)(3.000,3.000){2}{\rule{0.723pt}{1.200pt}}
\put(336,404.01){\rule{1.445pt}{1.200pt}}
\multiput(336.00,402.51)(3.000,3.000){2}{\rule{0.723pt}{1.200pt}}
\put(342,407.51){\rule{1.686pt}{1.200pt}}
\multiput(342.00,405.51)(3.500,4.000){2}{\rule{0.843pt}{1.200pt}}
\put(349,411.01){\rule{1.445pt}{1.200pt}}
\multiput(349.00,409.51)(3.000,3.000){2}{\rule{0.723pt}{1.200pt}}
\put(355,414.01){\rule{1.445pt}{1.200pt}}
\multiput(355.00,412.51)(3.000,3.000){2}{\rule{0.723pt}{1.200pt}}
\put(361,417.51){\rule{1.445pt}{1.200pt}}
\multiput(361.00,415.51)(3.000,4.000){2}{\rule{0.723pt}{1.200pt}}
\put(367,420.51){\rule{1.445pt}{1.200pt}}
\multiput(367.00,419.51)(3.000,2.000){2}{\rule{0.723pt}{1.200pt}}
\put(373,423.01){\rule{1.445pt}{1.200pt}}
\multiput(373.00,421.51)(3.000,3.000){2}{\rule{0.723pt}{1.200pt}}
\put(379,425.51){\rule{1.686pt}{1.200pt}}
\multiput(379.00,424.51)(3.500,2.000){2}{\rule{0.843pt}{1.200pt}}
\put(386,427.01){\rule{1.445pt}{1.200pt}}
\multiput(386.00,426.51)(3.000,1.000){2}{\rule{0.723pt}{1.200pt}}
\put(392,428.01){\rule{1.445pt}{1.200pt}}
\multiput(392.00,427.51)(3.000,1.000){2}{\rule{0.723pt}{1.200pt}}
\put(238.0,377.0){\rule[-0.600pt]{1.445pt}{1.200pt}}
\put(404,428.01){\rule{1.445pt}{1.200pt}}
\multiput(404.00,428.51)(3.000,-1.000){2}{\rule{0.723pt}{1.200pt}}
\put(410,427.01){\rule{1.445pt}{1.200pt}}
\multiput(410.00,427.51)(3.000,-1.000){2}{\rule{0.723pt}{1.200pt}}
\put(416,425.51){\rule{1.686pt}{1.200pt}}
\multiput(416.00,426.51)(3.500,-2.000){2}{\rule{0.843pt}{1.200pt}}
\put(423,423.01){\rule{1.445pt}{1.200pt}}
\multiput(423.00,424.51)(3.000,-3.000){2}{\rule{0.723pt}{1.200pt}}
\put(429,420.51){\rule{1.445pt}{1.200pt}}
\multiput(429.00,421.51)(3.000,-2.000){2}{\rule{0.723pt}{1.200pt}}
\put(435,417.51){\rule{1.445pt}{1.200pt}}
\multiput(435.00,419.51)(3.000,-4.000){2}{\rule{0.723pt}{1.200pt}}
\put(441,414.01){\rule{1.445pt}{1.200pt}}
\multiput(441.00,415.51)(3.000,-3.000){2}{\rule{0.723pt}{1.200pt}}
\put(447,411.01){\rule{1.445pt}{1.200pt}}
\multiput(447.00,412.51)(3.000,-3.000){2}{\rule{0.723pt}{1.200pt}}
\put(453,407.51){\rule{1.686pt}{1.200pt}}
\multiput(453.00,409.51)(3.500,-4.000){2}{\rule{0.843pt}{1.200pt}}
\put(460,404.01){\rule{1.445pt}{1.200pt}}
\multiput(460.00,405.51)(3.000,-3.000){2}{\rule{0.723pt}{1.200pt}}
\put(466,401.01){\rule{1.445pt}{1.200pt}}
\multiput(466.00,402.51)(3.000,-3.000){2}{\rule{0.723pt}{1.200pt}}
\put(472,398.01){\rule{1.445pt}{1.200pt}}
\multiput(472.00,399.51)(3.000,-3.000){2}{\rule{0.723pt}{1.200pt}}
\put(478,395.01){\rule{1.445pt}{1.200pt}}
\multiput(478.00,396.51)(3.000,-3.000){2}{\rule{0.723pt}{1.200pt}}
\put(484,392.01){\rule{1.445pt}{1.200pt}}
\multiput(484.00,393.51)(3.000,-3.000){2}{\rule{0.723pt}{1.200pt}}
\put(490,389.51){\rule{1.686pt}{1.200pt}}
\multiput(490.00,390.51)(3.500,-2.000){2}{\rule{0.843pt}{1.200pt}}
\put(497,387.51){\rule{1.445pt}{1.200pt}}
\multiput(497.00,388.51)(3.000,-2.000){2}{\rule{0.723pt}{1.200pt}}
\put(503,385.51){\rule{1.445pt}{1.200pt}}
\multiput(503.00,386.51)(3.000,-2.000){2}{\rule{0.723pt}{1.200pt}}
\put(509,383.51){\rule{1.445pt}{1.200pt}}
\multiput(509.00,384.51)(3.000,-2.000){2}{\rule{0.723pt}{1.200pt}}
\put(515,382.01){\rule{1.445pt}{1.200pt}}
\multiput(515.00,382.51)(3.000,-1.000){2}{\rule{0.723pt}{1.200pt}}
\put(521,380.51){\rule{1.445pt}{1.200pt}}
\multiput(521.00,381.51)(3.000,-2.000){2}{\rule{0.723pt}{1.200pt}}
\put(527,379.01){\rule{1.445pt}{1.200pt}}
\multiput(527.00,379.51)(3.000,-1.000){2}{\rule{0.723pt}{1.200pt}}
\put(533,378.01){\rule{1.686pt}{1.200pt}}
\multiput(533.00,378.51)(3.500,-1.000){2}{\rule{0.843pt}{1.200pt}}
\put(540,377.01){\rule{1.445pt}{1.200pt}}
\multiput(540.00,377.51)(3.000,-1.000){2}{\rule{0.723pt}{1.200pt}}
\put(546,376.01){\rule{1.445pt}{1.200pt}}
\multiput(546.00,376.51)(3.000,-1.000){2}{\rule{0.723pt}{1.200pt}}
\put(552,375.01){\rule{1.445pt}{1.200pt}}
\multiput(552.00,375.51)(3.000,-1.000){2}{\rule{0.723pt}{1.200pt}}
\put(398.0,431.0){\rule[-0.600pt]{1.445pt}{1.200pt}}
\put(564,374.01){\rule{1.445pt}{1.200pt}}
\multiput(564.00,374.51)(3.000,-1.000){2}{\rule{0.723pt}{1.200pt}}
\put(558.0,377.0){\rule[-0.600pt]{1.445pt}{1.200pt}}
\put(577,373.01){\rule{1.445pt}{1.200pt}}
\multiput(577.00,373.51)(3.000,-1.000){2}{\rule{0.723pt}{1.200pt}}
\put(570.0,376.0){\rule[-0.600pt]{1.686pt}{1.200pt}}
\put(595,372.01){\rule{1.445pt}{1.200pt}}
\multiput(595.00,372.51)(3.000,-1.000){2}{\rule{0.723pt}{1.200pt}}
\put(583.0,375.0){\rule[-0.600pt]{2.891pt}{1.200pt}}
\put(620,371.01){\rule{1.445pt}{1.200pt}}
\multiput(620.00,371.51)(3.000,-1.000){2}{\rule{0.723pt}{1.200pt}}
\put(601.0,374.0){\rule[-0.600pt]{4.577pt}{1.200pt}}
\put(626.0,373.0){\rule[-0.600pt]{19.272pt}{1.200pt}}
\end{picture}

%% file: Nuclpls.tex
\setlength{\unitlength}{0.240900pt}
\ifx\plotpoint\undefined\newsavebox{\plotpoint}\fi
\sbox{\plotpoint}{\rule[-0.200pt]{0.400pt}{0.400pt}}%
\begin{picture}(750,450)(0,0)
\font\gnuplot=cmr10 at 12pt
\gnuplot
\sbox{\plotpoint}{\rule[-0.200pt]{0.400pt}{0.400pt}}%
\put(84,19){\makebox(0,0)[r]{-4}}
\put(76.0,19.0){\rule[-0.200pt]{4.818pt}{0.400pt}}
\put(84,90){\makebox(0,0)[r]{-3}}
\put(76.0,90.0){\rule[-0.200pt]{4.818pt}{0.400pt}}
\put(84,160){\makebox(0,0)[r]{-2}}
\put(76.0,160.0){\rule[-0.200pt]{4.818pt}{0.400pt}}
\put(84,231){\makebox(0,0)[r]{-1}}
\put(76.0,231.0){\rule[-0.200pt]{4.818pt}{0.400pt}}
\put(84,302){\makebox(0,0)[r]{0}}
\put(76.0,302.0){\rule[-0.200pt]{4.818pt}{0.400pt}}
\put(84,372){\makebox(0,0)[r]{1}}
\put(76.0,372.0){\rule[-0.200pt]{4.818pt}{0.400pt}}
\put(84,443){\makebox(0,0)[r]{2}}
\put(76.0,443.0){\rule[-0.200pt]{4.818pt}{0.400pt}}
\put(96.0,-1.0){\rule[-0.200pt]{0.400pt}{4.818pt}}
\put(218.0,-1.0){\rule[-0.200pt]{0.400pt}{4.818pt}}
\put(340.0,-1.0){\rule[-0.200pt]{0.400pt}{4.818pt}}
\put(462.0,-1.0){\rule[-0.200pt]{0.400pt}{4.818pt}}
\put(584.0,-1.0){\rule[-0.200pt]{0.400pt}{4.818pt}}
\put(706.0,-1.0){\rule[-0.200pt]{0.400pt}{4.818pt}}
\put(96.0,19.0){\rule[-0.200pt]{146.949pt}{0.400pt}}
\put(706.0,19.0){\rule[-0.200pt]{0.400pt}{102.142pt}}
\put(96.0,443.0){\rule[-0.200pt]{146.949pt}{0.400pt}}
\put(737,266){\makebox(0,0)[l]{$x$}}
\put(127,478){\makebox(0,0)[l]{$p_{\mathrm{nucl}}(x-\frac{1}{2})$}}
\put(218,-22){\makebox(0,0)[l]{0.2}}
\put(340,-22){\makebox(0,0)[l]{0.4}}
\put(462,-22){\makebox(0,0)[l]{0.6}}
\put(584,-22){\makebox(0,0)[l]{0.8}}
\put(706,-22){\makebox(0,0)[l]{1}}
\put(96,-22){\makebox(0,0)[l]{0}}
\put(96.0,19.0){\rule[-0.200pt]{0.400pt}{102.142pt}}
\put(96,302){\vector(1,0){641}}
\put(96,19){\vector(0,1){459}}
\put(96,348){\usebox{\plotpoint}}
\put(96,346.17){\rule{1.300pt}{0.400pt}}
\multiput(96.00,347.17)(3.302,-2.000){2}{\rule{0.650pt}{0.400pt}}
\put(102,344.67){\rule{1.445pt}{0.400pt}}
\multiput(102.00,345.17)(3.000,-1.000){2}{\rule{0.723pt}{0.400pt}}
\put(108,343.17){\rule{1.300pt}{0.400pt}}
\multiput(108.00,344.17)(3.302,-2.000){2}{\rule{0.650pt}{0.400pt}}
\put(114,341.67){\rule{1.686pt}{0.400pt}}
\multiput(114.00,342.17)(3.500,-1.000){2}{\rule{0.843pt}{0.400pt}}
\put(121,340.17){\rule{1.300pt}{0.400pt}}
\multiput(121.00,341.17)(3.302,-2.000){2}{\rule{0.650pt}{0.400pt}}
\put(127,338.17){\rule{1.300pt}{0.400pt}}
\multiput(127.00,339.17)(3.302,-2.000){2}{\rule{0.650pt}{0.400pt}}
\put(133,336.17){\rule{1.300pt}{0.400pt}}
\multiput(133.00,337.17)(3.302,-2.000){2}{\rule{0.650pt}{0.400pt}}
\put(139,334.17){\rule{1.300pt}{0.400pt}}
\multiput(139.00,335.17)(3.302,-2.000){2}{\rule{0.650pt}{0.400pt}}
\put(145,332.17){\rule{1.300pt}{0.400pt}}
\multiput(145.00,333.17)(3.302,-2.000){2}{\rule{0.650pt}{0.400pt}}
\put(151,330.17){\rule{1.500pt}{0.400pt}}
\multiput(151.00,331.17)(3.887,-2.000){2}{\rule{0.750pt}{0.400pt}}
\multiput(158.00,328.95)(1.132,-0.447){3}{\rule{0.900pt}{0.108pt}}
\multiput(158.00,329.17)(4.132,-3.000){2}{\rule{0.450pt}{0.400pt}}
\put(164,325.17){\rule{1.300pt}{0.400pt}}
\multiput(164.00,326.17)(3.302,-2.000){2}{\rule{0.650pt}{0.400pt}}
\multiput(170.00,323.95)(1.132,-0.447){3}{\rule{0.900pt}{0.108pt}}
\multiput(170.00,324.17)(4.132,-3.000){2}{\rule{0.450pt}{0.400pt}}
\multiput(176.00,320.95)(1.132,-0.447){3}{\rule{0.900pt}{0.108pt}}
\multiput(176.00,321.17)(4.132,-3.000){2}{\rule{0.450pt}{0.400pt}}
\multiput(182.00,317.95)(1.132,-0.447){3}{\rule{0.900pt}{0.108pt}}
\multiput(182.00,318.17)(4.132,-3.000){2}{\rule{0.450pt}{0.400pt}}
\multiput(188.00,314.95)(1.355,-0.447){3}{\rule{1.033pt}{0.108pt}}
\multiput(188.00,315.17)(4.855,-3.000){2}{\rule{0.517pt}{0.400pt}}
\multiput(195.00,311.94)(0.774,-0.468){5}{\rule{0.700pt}{0.113pt}}
\multiput(195.00,312.17)(4.547,-4.000){2}{\rule{0.350pt}{0.400pt}}
\multiput(201.00,307.94)(0.774,-0.468){5}{\rule{0.700pt}{0.113pt}}
\multiput(201.00,308.17)(4.547,-4.000){2}{\rule{0.350pt}{0.400pt}}
\multiput(207.00,303.94)(0.774,-0.468){5}{\rule{0.700pt}{0.113pt}}
\multiput(207.00,304.17)(4.547,-4.000){2}{\rule{0.350pt}{0.400pt}}
\multiput(213.00,299.94)(0.774,-0.468){5}{\rule{0.700pt}{0.113pt}}
\multiput(213.00,300.17)(4.547,-4.000){2}{\rule{0.350pt}{0.400pt}}
\multiput(219.00,295.93)(0.599,-0.477){7}{\rule{0.580pt}{0.115pt}}
\multiput(219.00,296.17)(4.796,-5.000){2}{\rule{0.290pt}{0.400pt}}
\multiput(225.00,290.93)(0.710,-0.477){7}{\rule{0.660pt}{0.115pt}}
\multiput(225.00,291.17)(5.630,-5.000){2}{\rule{0.330pt}{0.400pt}}
\multiput(232.00,285.93)(0.491,-0.482){9}{\rule{0.500pt}{0.116pt}}
\multiput(232.00,286.17)(4.962,-6.000){2}{\rule{0.250pt}{0.400pt}}
\multiput(238.00,279.93)(0.599,-0.477){7}{\rule{0.580pt}{0.115pt}}
\multiput(238.00,280.17)(4.796,-5.000){2}{\rule{0.290pt}{0.400pt}}
\multiput(244.59,273.65)(0.482,-0.581){9}{\rule{0.116pt}{0.567pt}}
\multiput(243.17,274.82)(6.000,-5.824){2}{\rule{0.400pt}{0.283pt}}
\multiput(250.00,267.93)(0.491,-0.482){9}{\rule{0.500pt}{0.116pt}}
\multiput(250.00,268.17)(4.962,-6.000){2}{\rule{0.250pt}{0.400pt}}
\multiput(256.59,260.65)(0.482,-0.581){9}{\rule{0.116pt}{0.567pt}}
\multiput(255.17,261.82)(6.000,-5.824){2}{\rule{0.400pt}{0.283pt}}
\multiput(262.59,253.69)(0.485,-0.569){11}{\rule{0.117pt}{0.557pt}}
\multiput(261.17,254.84)(7.000,-6.844){2}{\rule{0.400pt}{0.279pt}}
\multiput(269.59,245.37)(0.482,-0.671){9}{\rule{0.116pt}{0.633pt}}
\multiput(268.17,246.69)(6.000,-6.685){2}{\rule{0.400pt}{0.317pt}}
\multiput(275.59,237.09)(0.482,-0.762){9}{\rule{0.116pt}{0.700pt}}
\multiput(274.17,238.55)(6.000,-7.547){2}{\rule{0.400pt}{0.350pt}}
\multiput(281.59,228.09)(0.482,-0.762){9}{\rule{0.116pt}{0.700pt}}
\multiput(280.17,229.55)(6.000,-7.547){2}{\rule{0.400pt}{0.350pt}}
\multiput(287.59,218.82)(0.482,-0.852){9}{\rule{0.116pt}{0.767pt}}
\multiput(286.17,220.41)(6.000,-8.409){2}{\rule{0.400pt}{0.383pt}}
\multiput(293.59,208.54)(0.482,-0.943){9}{\rule{0.116pt}{0.833pt}}
\multiput(292.17,210.27)(6.000,-9.270){2}{\rule{0.400pt}{0.417pt}}
\multiput(299.59,197.54)(0.482,-0.943){9}{\rule{0.116pt}{0.833pt}}
\multiput(298.17,199.27)(6.000,-9.270){2}{\rule{0.400pt}{0.417pt}}
\multiput(305.59,186.74)(0.485,-0.874){11}{\rule{0.117pt}{0.786pt}}
\multiput(304.17,188.37)(7.000,-10.369){2}{\rule{0.400pt}{0.393pt}}
\multiput(312.59,174.26)(0.482,-1.033){9}{\rule{0.116pt}{0.900pt}}
\multiput(311.17,176.13)(6.000,-10.132){2}{\rule{0.400pt}{0.450pt}}
\multiput(318.59,161.99)(0.482,-1.123){9}{\rule{0.116pt}{0.967pt}}
\multiput(317.17,163.99)(6.000,-10.994){2}{\rule{0.400pt}{0.483pt}}
\multiput(324.59,148.99)(0.482,-1.123){9}{\rule{0.116pt}{0.967pt}}
\multiput(323.17,150.99)(6.000,-10.994){2}{\rule{0.400pt}{0.483pt}}
\multiput(330.59,135.99)(0.482,-1.123){9}{\rule{0.116pt}{0.967pt}}
\multiput(329.17,137.99)(6.000,-10.994){2}{\rule{0.400pt}{0.483pt}}
\multiput(336.59,122.99)(0.482,-1.123){9}{\rule{0.116pt}{0.967pt}}
\multiput(335.17,124.99)(6.000,-10.994){2}{\rule{0.400pt}{0.483pt}}
\multiput(342.59,110.26)(0.485,-1.026){11}{\rule{0.117pt}{0.900pt}}
\multiput(341.17,112.13)(7.000,-12.132){2}{\rule{0.400pt}{0.450pt}}
\multiput(349.59,95.99)(0.482,-1.123){9}{\rule{0.116pt}{0.967pt}}
\multiput(348.17,97.99)(6.000,-10.994){2}{\rule{0.400pt}{0.483pt}}
\multiput(355.59,83.26)(0.482,-1.033){9}{\rule{0.116pt}{0.900pt}}
\multiput(354.17,85.13)(6.000,-10.132){2}{\rule{0.400pt}{0.450pt}}
\multiput(361.59,71.54)(0.482,-0.943){9}{\rule{0.116pt}{0.833pt}}
\multiput(360.17,73.27)(6.000,-9.270){2}{\rule{0.400pt}{0.417pt}}
\multiput(367.59,60.82)(0.482,-0.852){9}{\rule{0.116pt}{0.767pt}}
\multiput(366.17,62.41)(6.000,-8.409){2}{\rule{0.400pt}{0.383pt}}
\multiput(373.59,51.09)(0.482,-0.762){9}{\rule{0.116pt}{0.700pt}}
\multiput(372.17,52.55)(6.000,-7.547){2}{\rule{0.400pt}{0.350pt}}
\multiput(379.00,43.93)(0.492,-0.485){11}{\rule{0.500pt}{0.117pt}}
\multiput(379.00,44.17)(5.962,-7.000){2}{\rule{0.250pt}{0.400pt}}
\multiput(386.00,36.94)(0.774,-0.468){5}{\rule{0.700pt}{0.113pt}}
\multiput(386.00,37.17)(4.547,-4.000){2}{\rule{0.350pt}{0.400pt}}
\multiput(392.00,32.95)(1.132,-0.447){3}{\rule{0.900pt}{0.108pt}}
\multiput(392.00,33.17)(4.132,-3.000){2}{\rule{0.450pt}{0.400pt}}
\multiput(404.00,31.61)(1.132,0.447){3}{\rule{0.900pt}{0.108pt}}
\multiput(404.00,30.17)(4.132,3.000){2}{\rule{0.450pt}{0.400pt}}
\multiput(410.00,34.60)(0.774,0.468){5}{\rule{0.700pt}{0.113pt}}
\multiput(410.00,33.17)(4.547,4.000){2}{\rule{0.350pt}{0.400pt}}
\multiput(416.00,38.59)(0.492,0.485){11}{\rule{0.500pt}{0.117pt}}
\multiput(416.00,37.17)(5.962,7.000){2}{\rule{0.250pt}{0.400pt}}
\multiput(423.59,45.00)(0.482,0.762){9}{\rule{0.116pt}{0.700pt}}
\multiput(422.17,45.00)(6.000,7.547){2}{\rule{0.400pt}{0.350pt}}
\multiput(429.59,54.00)(0.482,0.852){9}{\rule{0.116pt}{0.767pt}}
\multiput(428.17,54.00)(6.000,8.409){2}{\rule{0.400pt}{0.383pt}}
\multiput(435.59,64.00)(0.482,0.943){9}{\rule{0.116pt}{0.833pt}}
\multiput(434.17,64.00)(6.000,9.270){2}{\rule{0.400pt}{0.417pt}}
\multiput(441.59,75.00)(0.482,1.033){9}{\rule{0.116pt}{0.900pt}}
\multiput(440.17,75.00)(6.000,10.132){2}{\rule{0.400pt}{0.450pt}}
\multiput(447.59,87.00)(0.482,1.123){9}{\rule{0.116pt}{0.967pt}}
\multiput(446.17,87.00)(6.000,10.994){2}{\rule{0.400pt}{0.483pt}}
\multiput(453.59,100.00)(0.485,1.026){11}{\rule{0.117pt}{0.900pt}}
\multiput(452.17,100.00)(7.000,12.132){2}{\rule{0.400pt}{0.450pt}}
\multiput(460.59,114.00)(0.482,1.123){9}{\rule{0.116pt}{0.967pt}}
\multiput(459.17,114.00)(6.000,10.994){2}{\rule{0.400pt}{0.483pt}}
\multiput(466.59,127.00)(0.482,1.123){9}{\rule{0.116pt}{0.967pt}}
\multiput(465.17,127.00)(6.000,10.994){2}{\rule{0.400pt}{0.483pt}}
\multiput(472.59,140.00)(0.482,1.123){9}{\rule{0.116pt}{0.967pt}}
\multiput(471.17,140.00)(6.000,10.994){2}{\rule{0.400pt}{0.483pt}}
\multiput(478.59,153.00)(0.482,1.123){9}{\rule{0.116pt}{0.967pt}}
\multiput(477.17,153.00)(6.000,10.994){2}{\rule{0.400pt}{0.483pt}}
\multiput(484.59,166.00)(0.482,1.033){9}{\rule{0.116pt}{0.900pt}}
\multiput(483.17,166.00)(6.000,10.132){2}{\rule{0.400pt}{0.450pt}}
\multiput(490.59,178.00)(0.485,0.874){11}{\rule{0.117pt}{0.786pt}}
\multiput(489.17,178.00)(7.000,10.369){2}{\rule{0.400pt}{0.393pt}}
\multiput(497.59,190.00)(0.482,0.943){9}{\rule{0.116pt}{0.833pt}}
\multiput(496.17,190.00)(6.000,9.270){2}{\rule{0.400pt}{0.417pt}}
\multiput(503.59,201.00)(0.482,0.943){9}{\rule{0.116pt}{0.833pt}}
\multiput(502.17,201.00)(6.000,9.270){2}{\rule{0.400pt}{0.417pt}}
\multiput(509.59,212.00)(0.482,0.852){9}{\rule{0.116pt}{0.767pt}}
\multiput(508.17,212.00)(6.000,8.409){2}{\rule{0.400pt}{0.383pt}}
\multiput(515.59,222.00)(0.482,0.762){9}{\rule{0.116pt}{0.700pt}}
\multiput(514.17,222.00)(6.000,7.547){2}{\rule{0.400pt}{0.350pt}}
\multiput(521.59,231.00)(0.482,0.762){9}{\rule{0.116pt}{0.700pt}}
\multiput(520.17,231.00)(6.000,7.547){2}{\rule{0.400pt}{0.350pt}}
\multiput(527.59,240.00)(0.482,0.671){9}{\rule{0.116pt}{0.633pt}}
\multiput(526.17,240.00)(6.000,6.685){2}{\rule{0.400pt}{0.317pt}}
\multiput(533.59,248.00)(0.485,0.569){11}{\rule{0.117pt}{0.557pt}}
\multiput(532.17,248.00)(7.000,6.844){2}{\rule{0.400pt}{0.279pt}}
\multiput(540.59,256.00)(0.482,0.581){9}{\rule{0.116pt}{0.567pt}}
\multiput(539.17,256.00)(6.000,5.824){2}{\rule{0.400pt}{0.283pt}}
\multiput(546.00,263.59)(0.491,0.482){9}{\rule{0.500pt}{0.116pt}}
\multiput(546.00,262.17)(4.962,6.000){2}{\rule{0.250pt}{0.400pt}}
\multiput(552.59,269.00)(0.482,0.581){9}{\rule{0.116pt}{0.567pt}}
\multiput(551.17,269.00)(6.000,5.824){2}{\rule{0.400pt}{0.283pt}}
\multiput(558.00,276.59)(0.599,0.477){7}{\rule{0.580pt}{0.115pt}}
\multiput(558.00,275.17)(4.796,5.000){2}{\rule{0.290pt}{0.400pt}}
\multiput(564.00,281.59)(0.491,0.482){9}{\rule{0.500pt}{0.116pt}}
\multiput(564.00,280.17)(4.962,6.000){2}{\rule{0.250pt}{0.400pt}}
\multiput(570.00,287.59)(0.710,0.477){7}{\rule{0.660pt}{0.115pt}}
\multiput(570.00,286.17)(5.630,5.000){2}{\rule{0.330pt}{0.400pt}}
\multiput(577.00,292.59)(0.599,0.477){7}{\rule{0.580pt}{0.115pt}}
\multiput(577.00,291.17)(4.796,5.000){2}{\rule{0.290pt}{0.400pt}}
\multiput(583.00,297.60)(0.774,0.468){5}{\rule{0.700pt}{0.113pt}}
\multiput(583.00,296.17)(4.547,4.000){2}{\rule{0.350pt}{0.400pt}}
\multiput(589.00,301.60)(0.774,0.468){5}{\rule{0.700pt}{0.113pt}}
\multiput(589.00,300.17)(4.547,4.000){2}{\rule{0.350pt}{0.400pt}}
\multiput(595.00,305.60)(0.774,0.468){5}{\rule{0.700pt}{0.113pt}}
\multiput(595.00,304.17)(4.547,4.000){2}{\rule{0.350pt}{0.400pt}}
\multiput(601.00,309.60)(0.774,0.468){5}{\rule{0.700pt}{0.113pt}}
\multiput(601.00,308.17)(4.547,4.000){2}{\rule{0.350pt}{0.400pt}}
\multiput(607.00,313.61)(1.355,0.447){3}{\rule{1.033pt}{0.108pt}}
\multiput(607.00,312.17)(4.855,3.000){2}{\rule{0.517pt}{0.400pt}}
\multiput(614.00,316.61)(1.132,0.447){3}{\rule{0.900pt}{0.108pt}}
\multiput(614.00,315.17)(4.132,3.000){2}{\rule{0.450pt}{0.400pt}}
\multiput(620.00,319.61)(1.132,0.447){3}{\rule{0.900pt}{0.108pt}}
\multiput(620.00,318.17)(4.132,3.000){2}{\rule{0.450pt}{0.400pt}}
\multiput(626.00,322.61)(1.132,0.447){3}{\rule{0.900pt}{0.108pt}}
\multiput(626.00,321.17)(4.132,3.000){2}{\rule{0.450pt}{0.400pt}}
\put(632,325.17){\rule{1.300pt}{0.400pt}}
\multiput(632.00,324.17)(3.302,2.000){2}{\rule{0.650pt}{0.400pt}}
\multiput(638.00,327.61)(1.132,0.447){3}{\rule{0.900pt}{0.108pt}}
\multiput(638.00,326.17)(4.132,3.000){2}{\rule{0.450pt}{0.400pt}}
\put(644,330.17){\rule{1.500pt}{0.400pt}}
\multiput(644.00,329.17)(3.887,2.000){2}{\rule{0.750pt}{0.400pt}}
\put(651,332.17){\rule{1.300pt}{0.400pt}}
\multiput(651.00,331.17)(3.302,2.000){2}{\rule{0.650pt}{0.400pt}}
\put(657,334.17){\rule{1.300pt}{0.400pt}}
\multiput(657.00,333.17)(3.302,2.000){2}{\rule{0.650pt}{0.400pt}}
\put(663,336.17){\rule{1.300pt}{0.400pt}}
\multiput(663.00,335.17)(3.302,2.000){2}{\rule{0.650pt}{0.400pt}}
\put(669,338.17){\rule{1.300pt}{0.400pt}}
\multiput(669.00,337.17)(3.302,2.000){2}{\rule{0.650pt}{0.400pt}}
\put(675,340.17){\rule{1.300pt}{0.400pt}}
\multiput(675.00,339.17)(3.302,2.000){2}{\rule{0.650pt}{0.400pt}}
\put(681,341.67){\rule{1.686pt}{0.400pt}}
\multiput(681.00,341.17)(3.500,1.000){2}{\rule{0.843pt}{0.400pt}}
\put(688,343.17){\rule{1.300pt}{0.400pt}}
\multiput(688.00,342.17)(3.302,2.000){2}{\rule{0.650pt}{0.400pt}}
\put(694,344.67){\rule{1.445pt}{0.400pt}}
\multiput(694.00,344.17)(3.000,1.000){2}{\rule{0.723pt}{0.400pt}}
\put(700,346.17){\rule{1.300pt}{0.400pt}}
\multiput(700.00,345.17)(3.302,2.000){2}{\rule{0.650pt}{0.400pt}}
\put(398.0,31.0){\rule[-0.200pt]{1.445pt}{0.400pt}}
\sbox{\plotpoint}{\rule[-0.500pt]{1.000pt}{1.000pt}}%
\put(96,363){\usebox{\plotpoint}}
\put(96.00,363.00){\usebox{\plotpoint}}
\multiput(102,363)(20.473,-3.412){0}{\usebox{\plotpoint}}
\multiput(108,362)(20.473,-3.412){0}{\usebox{\plotpoint}}
\put(116.59,361.00){\usebox{\plotpoint}}
\multiput(121,361)(20.473,-3.412){0}{\usebox{\plotpoint}}
\multiput(127,360)(20.473,-3.412){0}{\usebox{\plotpoint}}
\put(137.12,358.31){\usebox{\plotpoint}}
\multiput(139,358)(20.473,-3.412){0}{\usebox{\plotpoint}}
\multiput(145,357)(20.473,-3.412){0}{\usebox{\plotpoint}}
\put(157.43,354.16){\usebox{\plotpoint}}
\multiput(158,354)(20.473,-3.412){0}{\usebox{\plotpoint}}
\multiput(164,353)(20.473,-3.412){0}{\usebox{\plotpoint}}
\multiput(170,352)(19.690,-6.563){0}{\usebox{\plotpoint}}
\put(177.59,349.47){\usebox{\plotpoint}}
\multiput(182,348)(20.473,-3.412){0}{\usebox{\plotpoint}}
\multiput(188,347)(19.957,-5.702){0}{\usebox{\plotpoint}}
\put(197.45,343.77){\usebox{\plotpoint}}
\multiput(201,342)(19.690,-6.563){0}{\usebox{\plotpoint}}
\multiput(207,340)(19.690,-6.563){0}{\usebox{\plotpoint}}
\put(216.70,336.15){\usebox{\plotpoint}}
\multiput(219,335)(18.564,-9.282){0}{\usebox{\plotpoint}}
\multiput(225,332)(18.021,-10.298){0}{\usebox{\plotpoint}}
\put(235.05,326.47){\usebox{\plotpoint}}
\multiput(238,325)(17.270,-11.513){0}{\usebox{\plotpoint}}
\multiput(244,321)(15.945,-13.287){0}{\usebox{\plotpoint}}
\put(252.03,314.65){\usebox{\plotpoint}}
\multiput(256,312)(14.676,-14.676){0}{\usebox{\plotpoint}}
\put(268.10,301.64){\usebox{\plotpoint}}
\multiput(269,301)(13.508,-15.759){0}{\usebox{\plotpoint}}
\multiput(275,294)(13.508,-15.759){0}{\usebox{\plotpoint}}
\put(281.73,286.03){\usebox{\plotpoint}}
\multiput(287,279)(12.453,-16.604){0}{\usebox{\plotpoint}}
\put(294.09,269.36){\usebox{\plotpoint}}
\put(304.70,251.55){\usebox{\plotpoint}}
\multiput(305,251)(11.143,-17.511){0}{\usebox{\plotpoint}}
\put(314.97,233.56){\usebox{\plotpoint}}
\put(323.33,214.56){\usebox{\plotpoint}}
\multiput(324,213)(8.176,-19.077){0}{\usebox{\plotpoint}}
\put(331.27,195.39){\usebox{\plotpoint}}
\put(338.18,175.82){\usebox{\plotpoint}}
\put(345.36,156.35){\usebox{\plotpoint}}
\put(352.39,136.83){\usebox{\plotpoint}}
\put(358.77,117.07){\usebox{\plotpoint}}
\put(365.02,97.28){\usebox{\plotpoint}}
\put(371.72,77.64){\usebox{\plotpoint}}
\put(378.93,58.18){\usebox{\plotpoint}}
\multiput(379,58)(9.840,-18.275){0}{\usebox{\plotpoint}}
\put(389.22,40.18){\usebox{\plotpoint}}
\multiput(392,36)(17.270,-11.513){0}{\usebox{\plotpoint}}
\multiput(398,32)(20.756,0.000){0}{\usebox{\plotpoint}}
\put(406.10,33.40){\usebox{\plotpoint}}
\multiput(410,36)(11.513,17.270){0}{\usebox{\plotpoint}}
\put(418.49,49.63){\usebox{\plotpoint}}
\put(426.95,68.53){\usebox{\plotpoint}}
\put(433.96,88.06){\usebox{\plotpoint}}
\put(440.31,107.82){\usebox{\plotpoint}}
\put(446.56,127.61){\usebox{\plotpoint}}
\multiput(447,129)(6.563,19.690){0}{\usebox{\plotpoint}}
\put(453.12,147.30){\usebox{\plotpoint}}
\put(460.59,166.67){\usebox{\plotpoint}}
\put(467.50,186.24){\usebox{\plotpoint}}
\put(474.85,205.64){\usebox{\plotpoint}}
\put(483.02,224.72){\usebox{\plotpoint}}
\multiput(484,227)(8.698,18.845){0}{\usebox{\plotpoint}}
\put(492.12,243.34){\usebox{\plotpoint}}
\put(502.59,261.25){\usebox{\plotpoint}}
\multiput(503,262)(11.513,17.270){0}{\usebox{\plotpoint}}
\put(514.45,278.26){\usebox{\plotpoint}}
\multiput(515,279)(12.453,16.604){0}{\usebox{\plotpoint}}
\multiput(521,287)(13.508,15.759){0}{\usebox{\plotpoint}}
\put(527.40,294.47){\usebox{\plotpoint}}
\multiput(533,301)(16.889,12.064){0}{\usebox{\plotpoint}}
\put(542.51,308.51){\usebox{\plotpoint}}
\multiput(546,312)(17.270,11.513){0}{\usebox{\plotpoint}}
\multiput(552,316)(15.945,13.287){0}{\usebox{\plotpoint}}
\put(558.66,321.44){\usebox{\plotpoint}}
\multiput(564,325)(18.564,9.282){0}{\usebox{\plotpoint}}
\put(576.63,331.79){\usebox{\plotpoint}}
\multiput(577,332)(18.564,9.282){0}{\usebox{\plotpoint}}
\multiput(583,335)(18.564,9.282){0}{\usebox{\plotpoint}}
\multiput(589,338)(19.690,6.563){0}{\usebox{\plotpoint}}
\put(595.56,340.19){\usebox{\plotpoint}}
\multiput(601,342)(18.564,9.282){0}{\usebox{\plotpoint}}
\multiput(607,345)(19.957,5.702){0}{\usebox{\plotpoint}}
\put(615.02,347.17){\usebox{\plotpoint}}
\multiput(620,348)(19.690,6.563){0}{\usebox{\plotpoint}}
\multiput(626,350)(19.690,6.563){0}{\usebox{\plotpoint}}
\put(635.01,352.50){\usebox{\plotpoint}}
\multiput(638,353)(20.473,3.412){0}{\usebox{\plotpoint}}
\multiput(644,354)(19.957,5.702){0}{\usebox{\plotpoint}}
\put(655.30,356.72){\usebox{\plotpoint}}
\multiput(657,357)(20.473,3.412){0}{\usebox{\plotpoint}}
\multiput(663,358)(20.473,3.412){0}{\usebox{\plotpoint}}
\multiput(669,359)(20.473,3.412){0}{\usebox{\plotpoint}}
\put(675.78,360.13){\usebox{\plotpoint}}
\multiput(681,361)(20.756,0.000){0}{\usebox{\plotpoint}}
\multiput(688,361)(20.473,3.412){0}{\usebox{\plotpoint}}
\put(696.35,362.39){\usebox{\plotpoint}}
\multiput(700,363)(20.756,0.000){0}{\usebox{\plotpoint}}
\put(706,363){\usebox{\plotpoint}}
\sbox{\plotpoint}{\rule[-0.600pt]{1.200pt}{1.200pt}}%
\put(96,372){\usebox{\plotpoint}}
\put(164,369.01){\rule{1.445pt}{1.200pt}}
\multiput(164.00,369.51)(3.000,-1.000){2}{\rule{0.723pt}{1.200pt}}
\put(96.0,372.0){\rule[-0.600pt]{16.381pt}{1.200pt}}
\put(195,368.01){\rule{1.445pt}{1.200pt}}
\multiput(195.00,368.51)(3.000,-1.000){2}{\rule{0.723pt}{1.200pt}}
\put(170.0,371.0){\rule[-0.600pt]{6.022pt}{1.200pt}}
\put(213,367.01){\rule{1.445pt}{1.200pt}}
\multiput(213.00,367.51)(3.000,-1.000){2}{\rule{0.723pt}{1.200pt}}
\put(201.0,370.0){\rule[-0.600pt]{2.891pt}{1.200pt}}
\put(225,366.01){\rule{1.686pt}{1.200pt}}
\multiput(225.00,366.51)(3.500,-1.000){2}{\rule{0.843pt}{1.200pt}}
\put(219.0,369.0){\rule[-0.600pt]{1.445pt}{1.200pt}}
\put(238,365.01){\rule{1.445pt}{1.200pt}}
\multiput(238.00,365.51)(3.000,-1.000){2}{\rule{0.723pt}{1.200pt}}
\put(244,364.01){\rule{1.445pt}{1.200pt}}
\multiput(244.00,364.51)(3.000,-1.000){2}{\rule{0.723pt}{1.200pt}}
\put(250,363.01){\rule{1.445pt}{1.200pt}}
\multiput(250.00,363.51)(3.000,-1.000){2}{\rule{0.723pt}{1.200pt}}
\put(256,362.01){\rule{1.445pt}{1.200pt}}
\multiput(256.00,362.51)(3.000,-1.000){2}{\rule{0.723pt}{1.200pt}}
\put(262,360.51){\rule{1.686pt}{1.200pt}}
\multiput(262.00,361.51)(3.500,-2.000){2}{\rule{0.843pt}{1.200pt}}
\put(269,359.01){\rule{1.445pt}{1.200pt}}
\multiput(269.00,359.51)(3.000,-1.000){2}{\rule{0.723pt}{1.200pt}}
\put(275,357.51){\rule{1.445pt}{1.200pt}}
\multiput(275.00,358.51)(3.000,-2.000){2}{\rule{0.723pt}{1.200pt}}
\put(281,355.51){\rule{1.445pt}{1.200pt}}
\multiput(281.00,356.51)(3.000,-2.000){2}{\rule{0.723pt}{1.200pt}}
\put(287,353.01){\rule{1.445pt}{1.200pt}}
\multiput(287.00,354.51)(3.000,-3.000){2}{\rule{0.723pt}{1.200pt}}
\put(293,350.01){\rule{1.445pt}{1.200pt}}
\multiput(293.00,351.51)(3.000,-3.000){2}{\rule{0.723pt}{1.200pt}}
\put(299,346.51){\rule{1.445pt}{1.200pt}}
\multiput(299.00,348.51)(3.000,-4.000){2}{\rule{0.723pt}{1.200pt}}
\put(305,342.01){\rule{1.686pt}{1.200pt}}
\multiput(305.00,344.51)(3.500,-5.000){2}{\rule{0.843pt}{1.200pt}}
\put(312,337.01){\rule{1.445pt}{1.200pt}}
\multiput(312.00,339.51)(3.000,-5.000){2}{\rule{0.723pt}{1.200pt}}
\put(318.51,330){\rule{1.200pt}{1.686pt}}
\multiput(315.51,333.50)(6.000,-3.500){2}{\rule{1.200pt}{0.843pt}}
\multiput(326.24,322.11)(0.509,-0.113){2}{\rule{0.123pt}{1.900pt}}
\multiput(321.51,326.06)(6.000,-4.056){2}{\rule{1.200pt}{0.950pt}}
\multiput(332.24,312.45)(0.509,-0.452){2}{\rule{0.123pt}{2.300pt}}
\multiput(327.51,317.23)(6.000,-5.226){2}{\rule{1.200pt}{1.150pt}}
\multiput(338.24,299.96)(0.509,-0.962){2}{\rule{0.123pt}{2.900pt}}
\multiput(333.51,305.98)(6.000,-6.981){2}{\rule{1.200pt}{1.450pt}}
\multiput(344.24,287.08)(0.505,-1.027){4}{\rule{0.122pt}{2.871pt}}
\multiput(339.51,293.04)(7.000,-9.040){2}{\rule{1.200pt}{1.436pt}}
\multiput(351.24,266.15)(0.509,-2.150){2}{\rule{0.123pt}{4.300pt}}
\multiput(346.51,275.08)(6.000,-11.075){2}{\rule{1.200pt}{2.150pt}}
\multiput(357.24,242.83)(0.509,-2.829){2}{\rule{0.123pt}{5.100pt}}
\multiput(352.51,253.41)(6.000,-13.415){2}{\rule{1.200pt}{2.550pt}}
\multiput(363.24,214.68)(0.509,-3.678){2}{\rule{0.123pt}{6.100pt}}
\multiput(358.51,227.34)(6.000,-16.339){2}{\rule{1.200pt}{3.050pt}}
\multiput(369.24,179.87)(0.509,-4.867){2}{\rule{0.123pt}{7.500pt}}
\multiput(364.51,195.43)(6.000,-20.433){2}{\rule{1.200pt}{3.750pt}}
\multiput(375.24,140.55)(0.509,-5.546){2}{\rule{0.123pt}{8.300pt}}
\multiput(370.51,157.77)(6.000,-22.773){2}{\rule{1.200pt}{4.150pt}}
\multiput(381.24,103.87)(0.505,-3.622){4}{\rule{0.122pt}{7.500pt}}
\multiput(376.51,119.43)(7.000,-26.433){2}{\rule{1.200pt}{3.750pt}}
\multiput(388.24,61.04)(0.509,-5.037){2}{\rule{0.123pt}{7.700pt}}
\multiput(383.51,77.02)(6.000,-21.018){2}{\rule{1.200pt}{3.850pt}}
\multiput(394.24,36.49)(0.509,-2.490){2}{\rule{0.123pt}{4.700pt}}
\multiput(389.51,46.24)(6.000,-12.245){2}{\rule{1.200pt}{2.350pt}}
\put(232.0,368.0){\rule[-0.600pt]{1.445pt}{1.200pt}}
\multiput(406.24,34.00)(0.509,2.490){2}{\rule{0.123pt}{4.700pt}}
\multiput(401.51,34.00)(6.000,12.245){2}{\rule{1.200pt}{2.350pt}}
\multiput(412.24,56.00)(0.509,5.037){2}{\rule{0.123pt}{7.700pt}}
\multiput(407.51,56.00)(6.000,21.018){2}{\rule{1.200pt}{3.850pt}}
\multiput(418.24,93.00)(0.505,3.622){4}{\rule{0.122pt}{7.500pt}}
\multiput(413.51,93.00)(7.000,26.433){2}{\rule{1.200pt}{3.750pt}}
\multiput(425.24,135.00)(0.509,5.546){2}{\rule{0.123pt}{8.300pt}}
\multiput(420.51,135.00)(6.000,22.773){2}{\rule{1.200pt}{4.150pt}}
\multiput(431.24,175.00)(0.509,4.867){2}{\rule{0.123pt}{7.500pt}}
\multiput(426.51,175.00)(6.000,20.433){2}{\rule{1.200pt}{3.750pt}}
\multiput(437.24,211.00)(0.509,3.678){2}{\rule{0.123pt}{6.100pt}}
\multiput(432.51,211.00)(6.000,16.339){2}{\rule{1.200pt}{3.050pt}}
\multiput(443.24,240.00)(0.509,2.829){2}{\rule{0.123pt}{5.100pt}}
\multiput(438.51,240.00)(6.000,13.415){2}{\rule{1.200pt}{2.550pt}}
\multiput(449.24,264.00)(0.509,2.150){2}{\rule{0.123pt}{4.300pt}}
\multiput(444.51,264.00)(6.000,11.075){2}{\rule{1.200pt}{2.150pt}}
\multiput(455.24,284.00)(0.505,1.027){4}{\rule{0.122pt}{2.871pt}}
\multiput(450.51,284.00)(7.000,9.040){2}{\rule{1.200pt}{1.436pt}}
\multiput(462.24,299.00)(0.509,0.962){2}{\rule{0.123pt}{2.900pt}}
\multiput(457.51,299.00)(6.000,6.981){2}{\rule{1.200pt}{1.450pt}}
\multiput(468.24,312.00)(0.509,0.452){2}{\rule{0.123pt}{2.300pt}}
\multiput(463.51,312.00)(6.000,5.226){2}{\rule{1.200pt}{1.150pt}}
\multiput(474.24,322.00)(0.509,0.113){2}{\rule{0.123pt}{1.900pt}}
\multiput(469.51,322.00)(6.000,4.056){2}{\rule{1.200pt}{0.950pt}}
\put(478.51,330){\rule{1.200pt}{1.686pt}}
\multiput(475.51,330.00)(6.000,3.500){2}{\rule{1.200pt}{0.843pt}}
\put(484,337.01){\rule{1.445pt}{1.200pt}}
\multiput(484.00,334.51)(3.000,5.000){2}{\rule{0.723pt}{1.200pt}}
\put(490,342.01){\rule{1.686pt}{1.200pt}}
\multiput(490.00,339.51)(3.500,5.000){2}{\rule{0.843pt}{1.200pt}}
\put(497,346.51){\rule{1.445pt}{1.200pt}}
\multiput(497.00,344.51)(3.000,4.000){2}{\rule{0.723pt}{1.200pt}}
\put(503,350.01){\rule{1.445pt}{1.200pt}}
\multiput(503.00,348.51)(3.000,3.000){2}{\rule{0.723pt}{1.200pt}}
\put(509,353.01){\rule{1.445pt}{1.200pt}}
\multiput(509.00,351.51)(3.000,3.000){2}{\rule{0.723pt}{1.200pt}}
\put(515,355.51){\rule{1.445pt}{1.200pt}}
\multiput(515.00,354.51)(3.000,2.000){2}{\rule{0.723pt}{1.200pt}}
\put(521,357.51){\rule{1.445pt}{1.200pt}}
\multiput(521.00,356.51)(3.000,2.000){2}{\rule{0.723pt}{1.200pt}}
\put(527,359.01){\rule{1.445pt}{1.200pt}}
\multiput(527.00,358.51)(3.000,1.000){2}{\rule{0.723pt}{1.200pt}}
\put(533,360.51){\rule{1.686pt}{1.200pt}}
\multiput(533.00,359.51)(3.500,2.000){2}{\rule{0.843pt}{1.200pt}}
\put(540,362.01){\rule{1.445pt}{1.200pt}}
\multiput(540.00,361.51)(3.000,1.000){2}{\rule{0.723pt}{1.200pt}}
\put(546,363.01){\rule{1.445pt}{1.200pt}}
\multiput(546.00,362.51)(3.000,1.000){2}{\rule{0.723pt}{1.200pt}}
\put(552,364.01){\rule{1.445pt}{1.200pt}}
\multiput(552.00,363.51)(3.000,1.000){2}{\rule{0.723pt}{1.200pt}}
\put(558,365.01){\rule{1.445pt}{1.200pt}}
\multiput(558.00,364.51)(3.000,1.000){2}{\rule{0.723pt}{1.200pt}}
\put(398.0,34.0){\rule[-0.600pt]{1.445pt}{1.200pt}}
\put(570,366.01){\rule{1.686pt}{1.200pt}}
\multiput(570.00,365.51)(3.500,1.000){2}{\rule{0.843pt}{1.200pt}}
\put(564.0,368.0){\rule[-0.600pt]{1.445pt}{1.200pt}}
\put(583,367.01){\rule{1.445pt}{1.200pt}}
\multiput(583.00,366.51)(3.000,1.000){2}{\rule{0.723pt}{1.200pt}}
\put(577.0,369.0){\rule[-0.600pt]{1.445pt}{1.200pt}}
\put(601,368.01){\rule{1.445pt}{1.200pt}}
\multiput(601.00,367.51)(3.000,1.000){2}{\rule{0.723pt}{1.200pt}}
\put(589.0,370.0){\rule[-0.600pt]{2.891pt}{1.200pt}}
\put(632,369.01){\rule{1.445pt}{1.200pt}}
\multiput(632.00,368.51)(3.000,1.000){2}{\rule{0.723pt}{1.200pt}}
\put(607.0,371.0){\rule[-0.600pt]{6.022pt}{1.200pt}}
\put(638.0,372.0){\rule[-0.600pt]{16.381pt}{1.200pt}}
\end{picture}

%% file: treg.tex
\setlength{\unitlength}{0.240900pt}
\ifx\plotpoint\undefined\newsavebox{\plotpoint}\fi
\begin{picture}(750,450)(0,0)
\font\gnuplot=cmr10 at 12pt
\gnuplot
\sbox{\plotpoint}{\rule[-0.200pt]{0.400pt}{0.400pt}}%
\put(84,19){\makebox(0,0)[r]{0.2}}
\put(76.0,19.0){\rule[-0.200pt]{4.818pt}{0.400pt}}
\put(84,90){\makebox(0,0)[r]{0.3}}
\put(76.0,90.0){\rule[-0.200pt]{4.818pt}{0.400pt}}
\put(84,160){\makebox(0,0)[r]{0.4}}
\put(76.0,160.0){\rule[-0.200pt]{4.818pt}{0.400pt}}
\put(84,231){\makebox(0,0)[r]{0.5}}
\put(76.0,231.0){\rule[-0.200pt]{4.818pt}{0.400pt}}
\put(84,302){\makebox(0,0)[r]{0.6}}
\put(76.0,302.0){\rule[-0.200pt]{4.818pt}{0.400pt}}
\put(84,372){\makebox(0,0)[r]{0.7}}
\put(76.0,372.0){\rule[-0.200pt]{4.818pt}{0.400pt}}
\put(84,443){\makebox(0,0)[r]{0.8}}
\put(76.0,443.0){\rule[-0.200pt]{4.818pt}{0.400pt}}
\put(183.0,-1.0){\rule[-0.200pt]{0.400pt}{4.818pt}}
\put(270.0,-1.0){\rule[-0.200pt]{0.400pt}{4.818pt}}
\put(357.0,-1.0){\rule[-0.200pt]{0.400pt}{4.818pt}}
\put(445.0,-1.0){\rule[-0.200pt]{0.400pt}{4.818pt}}
\put(532.0,-1.0){\rule[-0.200pt]{0.400pt}{4.818pt}}
\put(619.0,-1.0){\rule[-0.200pt]{0.400pt}{4.818pt}}
\put(706.0,-1.0){\rule[-0.200pt]{0.400pt}{4.818pt}}
\put(96.0,19.0){\rule[-0.200pt]{146.949pt}{0.400pt}}
\put(706.0,19.0){\rule[-0.200pt]{0.400pt}{102.142pt}}
\put(96.0,443.0){\rule[-0.200pt]{146.949pt}{0.400pt}}
\put(793,19){\makebox(0,0)[l]{$\mu$}}
\put(227,478){\makebox(0,0)[l]{$x_{\mathrm{term}}(\mu)$}}
\put(270,-51){\makebox(0,0)[l]{0.1}}
\put(445,-51){\makebox(0,0)[l]{0.3}}
\put(619,-51){\makebox(0,0)[l]{0.5}}
\put(183,-51){\makebox(0,0)[l]{0}}
\put(96.0,19.0){\rule[-0.200pt]{0.400pt}{102.142pt}}
\put(96,19){\vector(1,0){654}}
\put(183,19){\vector(0,1){459}}
\put(96,19){\rule{1pt}{1pt}}
\put(140,26){\rule{1pt}{1pt}}
\put(183,47){\rule{1pt}{1pt}}
\put(227,61){\rule{1pt}{1pt}}
\put(270,76){\rule{1pt}{1pt}}
\put(357,111){\rule{1pt}{1pt}}
\put(445,153){\rule{1pt}{1pt}}
\put(532,231){\rule{1pt}{1pt}}
\put(619,295){\rule{1pt}{1pt}}
\put(706,408){\rule{1pt}{1pt}}
\end{picture}